# A Cognitive Architecture for Machine Consciousness and Artificial Superintelligence:

*Thought Is Structured by the Iterative Updating of Working Memory*

**Jared Edward Reser** Ph.D., M.A., M.A.

University of Southern California

jared@jaredreser.com

www.aithought.com

www.jared-research.com

This model builds on over 15 years of research (Reser, 2011, 2012, 2013, 2016, 2022), and fundamental components of the model were peer-reviewed and published as:

**Reser, J. E. (2016). Incremental change in the set of coactive cortical assemblies enables mental continuity. Physiology and Behavior, 167, 222-237.**

# Contents

# A Cognitive Architecture for Machine Consciousness and Artificial Superintelligence:

*Thought Is Structured by the Iterative Updating of Working Memory*

**Abstract**

This article provides an analytical framework for how to simulate human-like thought processes within a computer. It describes how attention and memory should be structured, updated, and utilized to search for associative additions to the stream of thought. The focus is on replicating the dynamics of the mammalian working memory system, which features two forms of persistent activity: sustained firing (preserving information on the order of seconds) and synaptic potentiation (preserving information from minutes to hours). The article uses a series of figures to systematically demonstrate how the iterative updating of these working memory stores provides functional organization to behavior, cognition, and awareness.

In a machine learning implementation, these two memory stores should be updated continuously and in an iterative fashion. This means each state should preserve a proportion of the coactive representations from the state before it (where each representation is an ensemble of neural network nodes). This makes each state a revised iteration of the preceding state and causes successive configurations to overlap and blend with respect to the information they contain. Thus, the set of concepts in working memory will evolve gradually and incrementally over time. Transitions between states happen as persistent activity spreads activation energy throughout the hierarchical network, searching long-term memory for the most appropriate representation to be added to the global workspace. The result is a chain of associatively linked intermediate states capable of advancing toward a solution or goal. Iterative updating is conceptualized here as an information processing strategy, a model of working memory, a theory of consciousness, and an algorithm for designing and programming artificial intelligence (AI, AGI, and ASI).

**Note:** For readers with time constraints, a concise understanding of the core content can be attained in around ten minutes by examining the figures and accompanying captions.

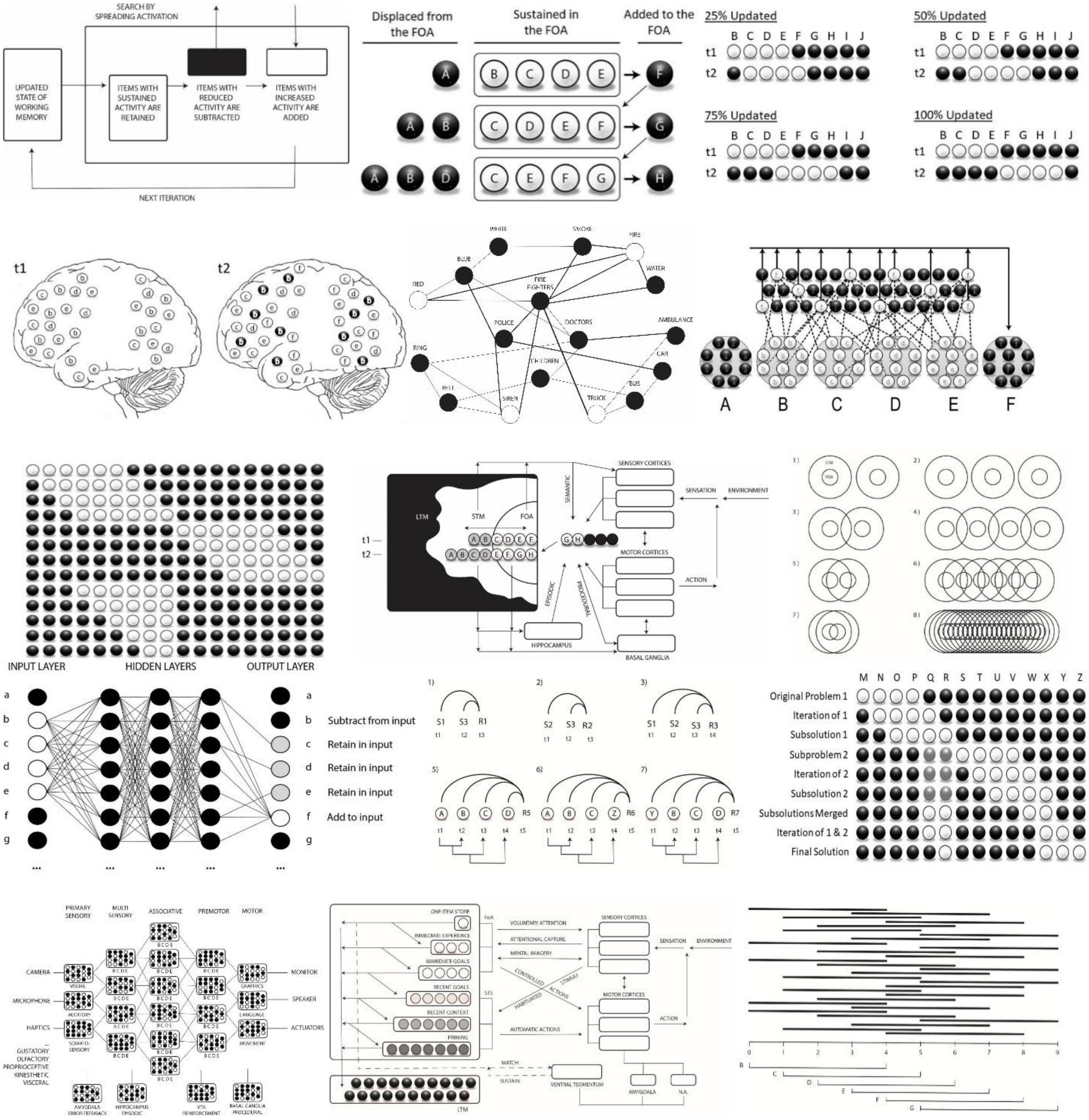

# **Part I:** Introduction

### 1.1 Machine Superintelligence Requires a Thought Process

> "It seems that the human mind has first to construct forms independently before we can find them in things… Knowledge cannot spring from experience alone, but only from a comparison of the inventions of the intellect with observed fact."
>
> Albert Einstein (1949)

The above quote suggests that for artificial intelligence (AI) to make sense of the world, it must go beyond training on data and think for itself, learning from its own internal processing. Incremental improvements to existing machine learning architectures will not yield knowledge creation or real understanding because they do not attempt to simulate thought or its reflective, analytical, or deliberative qualities. Although current artificial neural networks use various brain-inspired techniques such as attention, they do not use working memory the way mammals do. To do so would involve constructing a stream of thought by holding a set of representations coactive and continuously updating this set with the most pertinent associations. This results in an iterative system that embeds mental states within the states that came before them. Such a system would automate the generation and refinement of knowledge by constantly comparing, contextualizing, and generalizing information.

The present article identifies the elements of mammalian working memory that make this iterative process possible and describes how to organize them within a computer using contemporary machine learning technology. It also introduces several novel concepts, terms (Table 5), and illustrations to explain how an iterative cognitive cycle will permit a computer program to make the state-space transitions necessary to achieve general intellectual faculties. By simulating ongoing, self-directed, open-ended thought, as described here, an artificial agent could construct its own predictions and associations, simulate hypothetical situations, synthesize novel ideas, and thereby further scientific and technological progress. This framework can be used to build a novel AI system, offer organizational guidance to developing systems, or inspire enhancements to existing systems. Table 1 outlines some of the categorical traits of the model.

| General Features | Definition |
|---|---|
| **Connectionist** | Uses network-based, parallel, distributed processing |
| **Recurrent** | Processes sequential or time series data in search of temporal patterns |
| **Autoregressive** | Predicts future elements in a sequence based on previous elements |
| **Autonomous** | Self-governing activity independent of human training or feedback |
| **Continuous Learning** | Open-ended exploration and network refinement |
| **Cognition First** | Constant internal processing even in the absence of input or output |
| **Multimodal** | Various data types are combined and processed together |
| **Monolithic** | An architecture composed all in one piece |
| **Biomimetic** | Imitates biological (and neurological) properties |
| **Scalable** | Retaining effectiveness and efficiency when expanded in size or complexity |
| **Embodied** | Controls and receives feedback from a physical body or robot |
| **Neuro-symbolic** | Integrates parallel subsymbolic aspects with serial symbolic ones |
| **Unsupervised and Reinforcement Learning** | Learns without human feedback, and from reward, respectively |
| **Attention** | Uses mechanisms to weigh significance for inclusion in working memory |
| **Global Workspace** | Contains a central hub that integrates inputs from various subsystems |

**Table 1.** Comparative Overview of Features of an AI Based on the Iterative Updating Model

*This table summarizes some of the general characteristics of an intelligent system built in accordance with the architecture presented here.*

## 1.2 Iteration Defines the Workflow of Thought

AI research has yet to formalize and simulate the thinking process because psychology and neuroscience have completely ignored the crucial role of iteration. No contemporary models address iterative change in the contents of attention or working memory. In many discussions, updating of the information held in working memory is considered complete rather than partial, meaning that after being updated, the contents from the previous state are entirely replaced (e.g., Pina et al., 2018; Niklaus et al., 2019). In other discussions, information can be updated without complete replacement, but only such as when working memory holds three words and then accommodates a fourth in addition to the first three (e.g., Miller et al., 2018; Manohar et al., 2019). These views compartmentalize the thinking process, isolating current states from what came before them.

In contrast, the account presented here explores the hypothesis that partial updating occurs continuously. As representations are added, others are subtracted, and others from the previous state remain due to persistent neural activity (Figure 1). This cascading persistence allows successive states to share a proportion of their content in common, creating complex causal relationships between them (Reser, 2011, 2012). This iterative perspective illuminates how the gradually transforming collection of representations in working memory allows the thinking process to progress as updated states elaborate intelligently on the states that came before them (Reser, 2013, 2016, 2022).

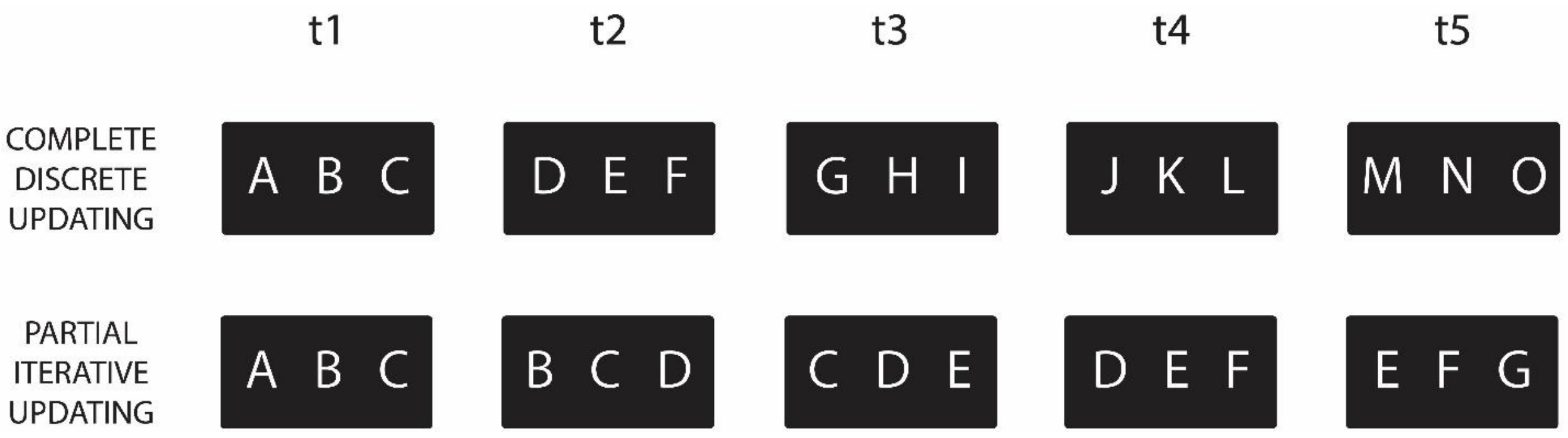

**Fig. 1.** Two Types of Working Memory Updating Compared

*Each row contains five rectangles labeled time one (t1) through time five (t5). Each rectangle corresponds to a state of working memory holding three items (uppercase letters). In the top row, successive rectangles do not hold any of the same items, indicating complete updating. In the second row, two items are shared between successive rectangles, indicating partial updating. This article contends that the iterative nature exhibited by the second row is a fundamental attribute of the thinking process.*

A familiar example of the concept of iteration is "iterative design." It is a method of developing commercial products through a cyclic process of prototyping, testing, and improving. With this method, designs are assessed through user feedback and enhanced in a cumulative fashion. Think of the installment histories of a popular product such as a cell phone, operating system, or car. The newest version of the product contains novel features but preserves many aspects of the previous version and even of versions before that. The workflow of human thought is interpreted here in a similar way (Figure 2). As mental representations in working memory are updated, the frame of reference is gradually replaced, and a thought about one scenario incrementally transitions into a thought about a related scenario. The result is a series of intermediate states capable of exploring a problem space by layering new information onto existing structures. This article will explore how this general process contributes to reasoning, mental modeling, executive processes, and consciousness.

This abstract, high-level model also offers an explanation for how the next iterative update to working memory is selected. The firing neurons that underlie the representations in working memory spread their combined excitatory and inhibitory effects to other cells throughout the cortex. Thus, the coactivation of the contents of working memory amounts to an associative search of long-term memory for applicable information (e.g., predictions, probabilities, and motor instructions). The nonactive (baseline) cells that receive the most spreading activation become active and comprise the representation(s) that will update working memory. Similarly, the representations that continue to receive activation energy are maintained in working memory. In contrast, those that receive reduced energy are subtracted from it. Performing subsymbolic search using a modified version of the previous search, and doing so repeatedly, amounts to a compounded form of search that ultimately enables the compounding of predictions and inferences.

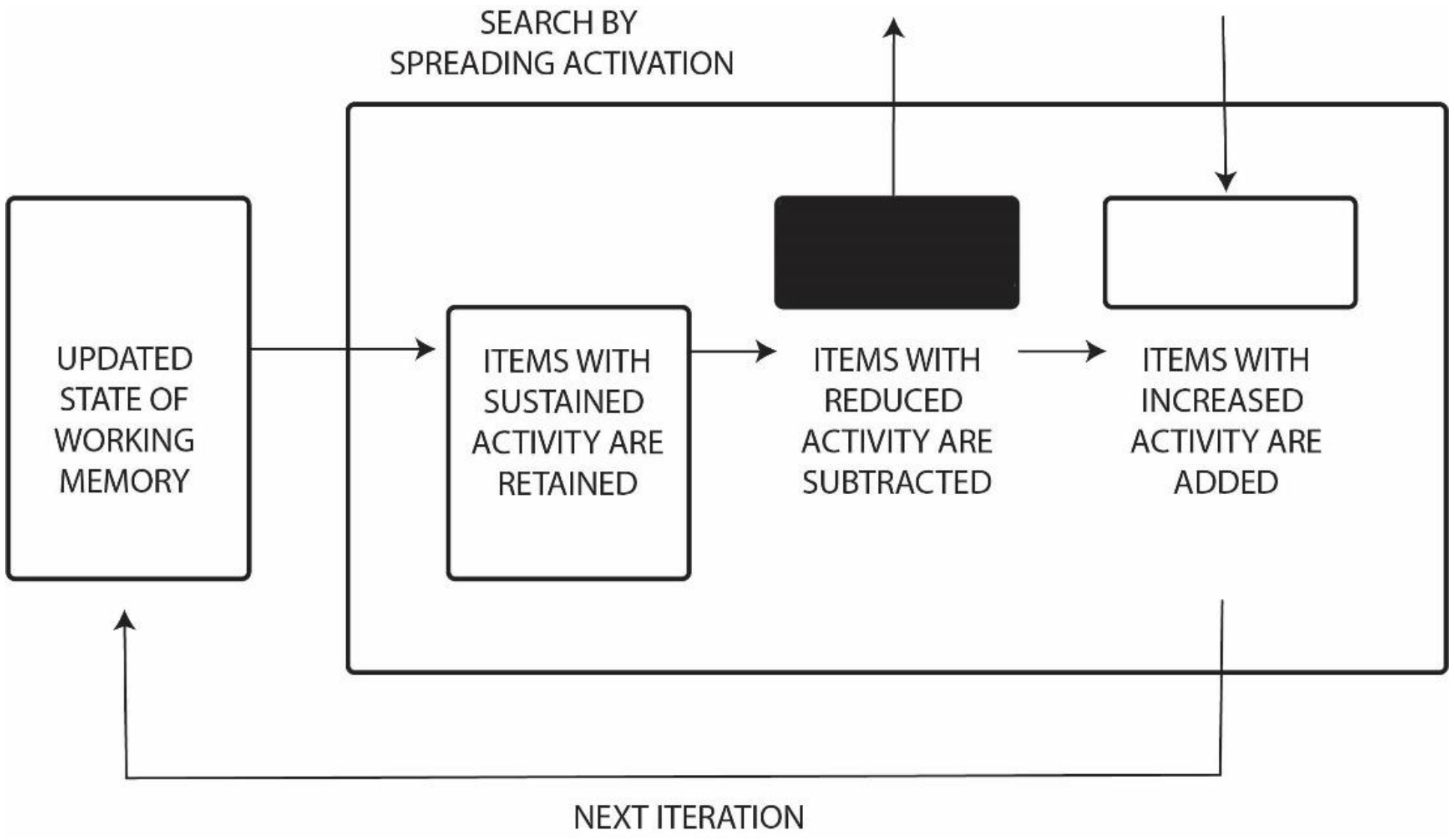

**Fig. 2.** Flowchart of Iterative Updating

*In an iterative process, a set of components is modified repetitively to generate a series of updated states. Each state is an iteration as well as the starting point for the next iteration. One way to accomplish iterative modification is to alter a given state by retaining pertinent elements and then subtracting and adding others. In the brain, the content to be added and subtracted is determined by spreading activation.*

Again, newly activated representations are added to the representations that have remained in working memory from the previous state. This updated set is used to conduct the next search. This cycle is then repeated in a loop to produce the thinking process. Thus, there is a direct structural correspondence between the turnover of persistent neural activity, the gradual updating of working memory, and the continuity of the stream of thought. Many of the major features of thought derived from introspection (Hamilton, 1860; Weger et al., 2018) are addressed by this hypothetical explanation, such as how mental context is conserved from one thought to the next, how one thought is associated with the next, and how it logically (or probabilistically) implies the next.

This article focuses on ongoing, internally generated activity within working memory and the emergent iterative pattern of information flow. This pattern, introduced in Figure 2, is elaborated on methodically through a series of over 50 figures that attempt to illustrate the "shape" of the thought process. Topics considered include the neural basis of items in working memory, variation in the rate of updating, the modeling of common psychological phenomena and how these can be implemented within artificial neural network models to enhance the performance of AI. This work builds on these issues while remaining consistent with prevailing knowledge and assimilating current theoretical approaches in brain and computer science. Part 2 reviews pertinent literature that forms the foundation of the iterative updating model. Parts 3 and 4 develop said model, and Part 5 details an AI implementation.

# **Part II:** Literature Review

### 2.1 Interactions Between Sensory Memory, Working Memory, and Long-term Memory

Working memory has been defined as the components of the mind that temporarily hold a limited amount of information in a heightened state of availability for use in ongoing information processing (Cowan, 2016). It involves holding ephemeral sensory and semantic information (e.g., objects, shapes, colors, locations, movement patterns, symbols, rules, concepts, numbers, and words) in attention until they are needed to execute an action or decision. It is one of multiple phases of memory and has been variously referred to as immediate memory and primary memory. It was conceptualized by William James (1842-1910) as the “trailing edge of the conscious present” and a major determinant of which portions of new information will be perceived and which of those will be analyzed (James, 1890). Working memory is thought to facilitate various operations, such as planning, language comprehension, reasoning, decision making, and problem solving (Baddeley, 2012).

The working memory store is constantly updated with new items, which then fade over the course of seconds or minutes (some more quickly than others). Updating allocates processing resources to important information coming from the senses (e.g., novelties, needs, or threats) or from internal states (e.g., intentions, plans, or schemas). Mental functions require the active maintenance of multiple items at once, along with systematic updating of these items (Baddeley, 2012). Active updating is necessary because working memory is capacity-limited and the importance of individual items changes as processing demands change (Myers et al., 2017).

Research on working memory has traditionally relied on behavioral investigations (such as memory tasks) to study interactions and dissociations between memory systems. Experimental studies on the topic are concerned with capacity limits, rehearsal, interference, suppression of irrelevant information, removal of unnecessary information, and other regular phenomena. Theorists have tried to capture these regularities using abstract models.

From the late 1950s to the 1960s, memory researchers (e.g., Atkinson & Shiffrin, 1968; Broadbent, 1958) developed models that conceptualized memory as being comprised of three interacting systems: (1) a sensory store that briefly holds and preprocesses sensory inputs, (2) an active short-term system capable of attending to this information over a time frame of seconds, and (3) a passive long-term system capable of maintaining information indefinitely (Fig. 3). Current models (including the present work) have retained many of these aspects.

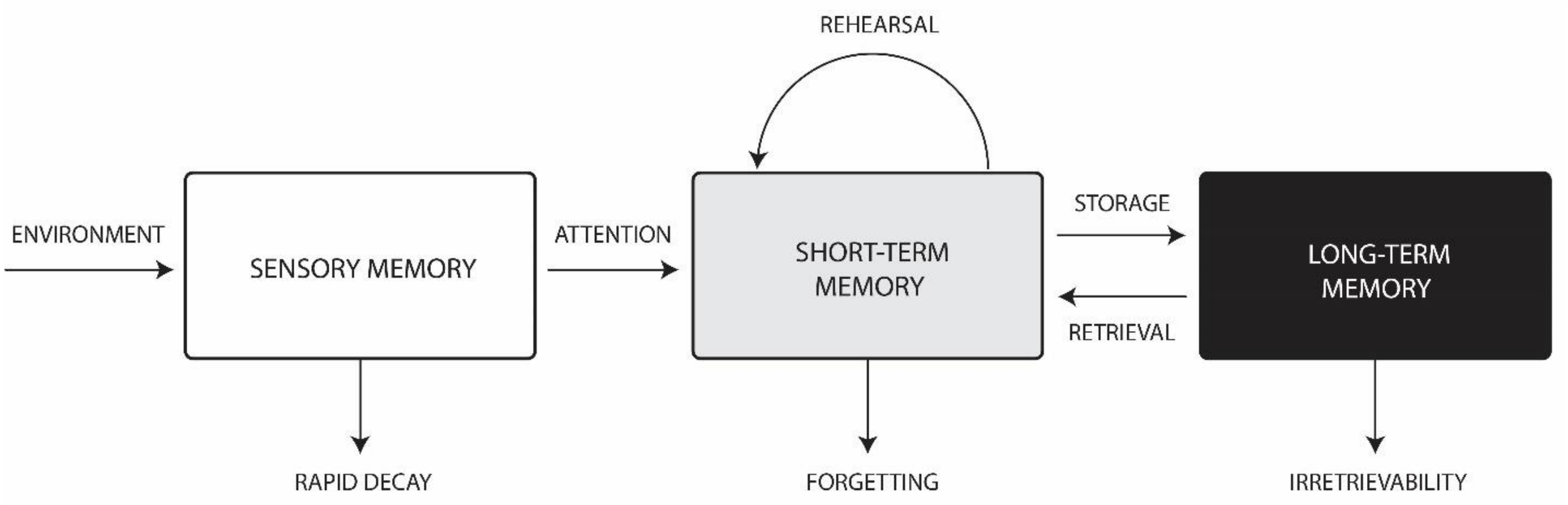

**Fig. 3.** Atkinson and Shiffrin's Multi-store Model (1968)

*This model depicts environmental stimuli received by the senses and held in sensory memory. If attended to, this stimulus information will enter short-term memory (i.e., working memory, shown in gray). If not rehearsed, it will be forgotten; if rehearsed, it will remain in short-term memory; and if sufficiently elaborated upon, it will be stored in long-term memory (shown in black), from which it can be retrieved later. Long-term memory is depicted in black here and throughout this article.*

The multi-store model has been expanded upon in several pivotal ways. Studies performed by Alan Baddeley and Graham Hitch (1974, 1986) using dual-task interference experiments indicated that the capacity limitations for visual and verbal working memory are independent, leading the authors to categorize these two modalities as separable. This distinction led to the authors' influential multicomponent model, which divided working memory into two domain-specific stores: the visuospatial sketchpad and the phonological buffer (Fig. 4). These stores work in concert to construct, sustain, and modify mental imagery.

Baddeley and Hitch also envisioned a dedicated supervisory subsystem, which they named the "central executive," that selects items for activity, shuttles information from one store to another, and makes other processing decisions. Because researchers have not determined how the central executive, visuospatial sketchpad, and phonological buffer operate, they remain areas of active research and theoretical inquiry.

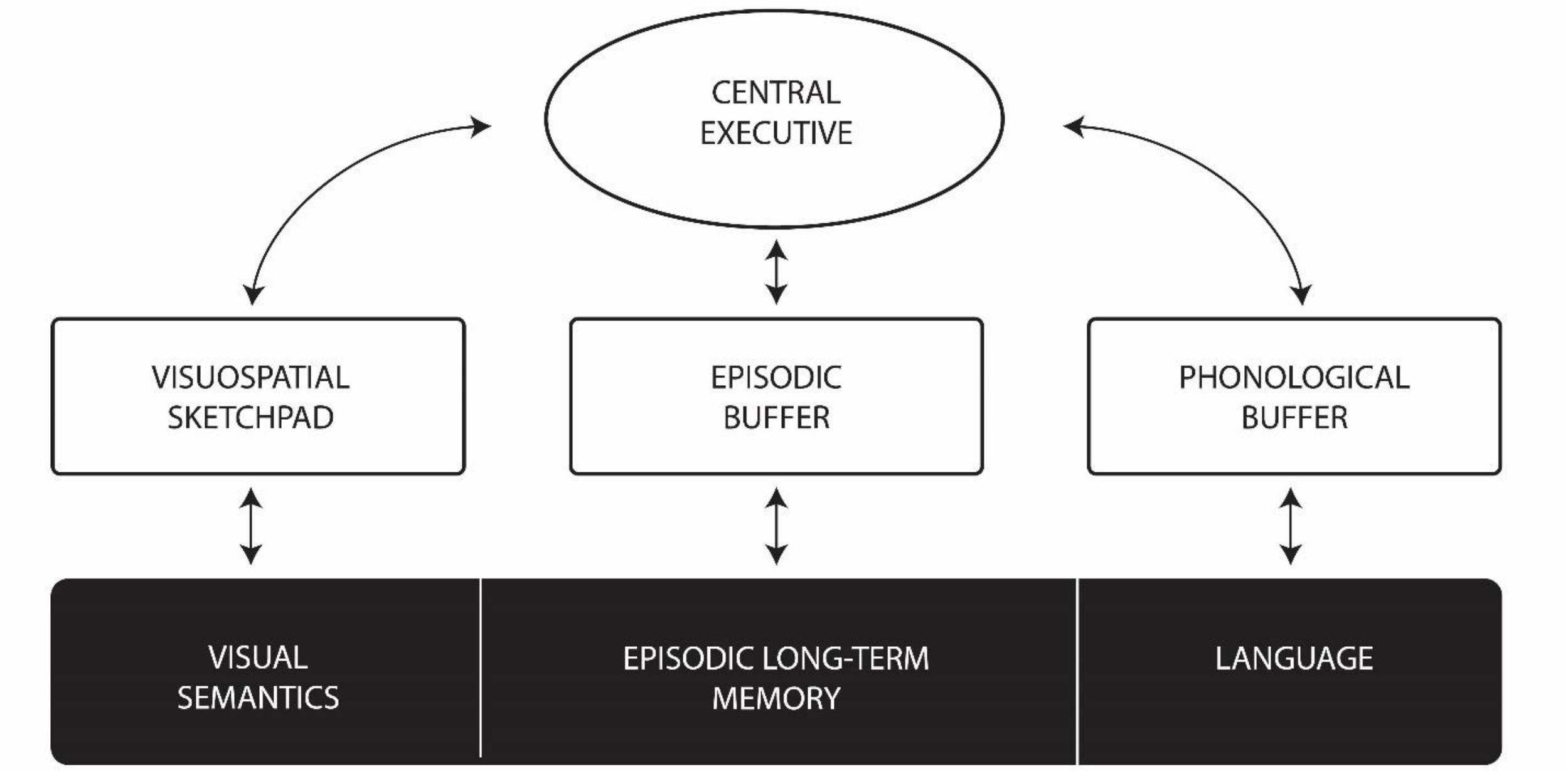

**Fig. 4.** Baddeley and Hitch's Multicomponent Model (1974)

*In this model, the short-term store from Atkinson and Shiffrin's model is split into four interacting components that together constitute working memory: the central executive, the visuospatial sketchpad, the phonological buffer, and the episodic buffer (Baddeley, 2000). These components interact with long-term memory, represented by the bottom rectangle.*

Since this time many cognitive architectures have been proposed modeling how humans store, retrieve and process information. These architectures aim to replicate how the human mind functions, typically by dividing cognitive functions into different modules often inspired by neuroscience and establishing how these components interact. Cognitive architectures are used in cognitive science and artificial intelligence to study and predict intelligent behavior.

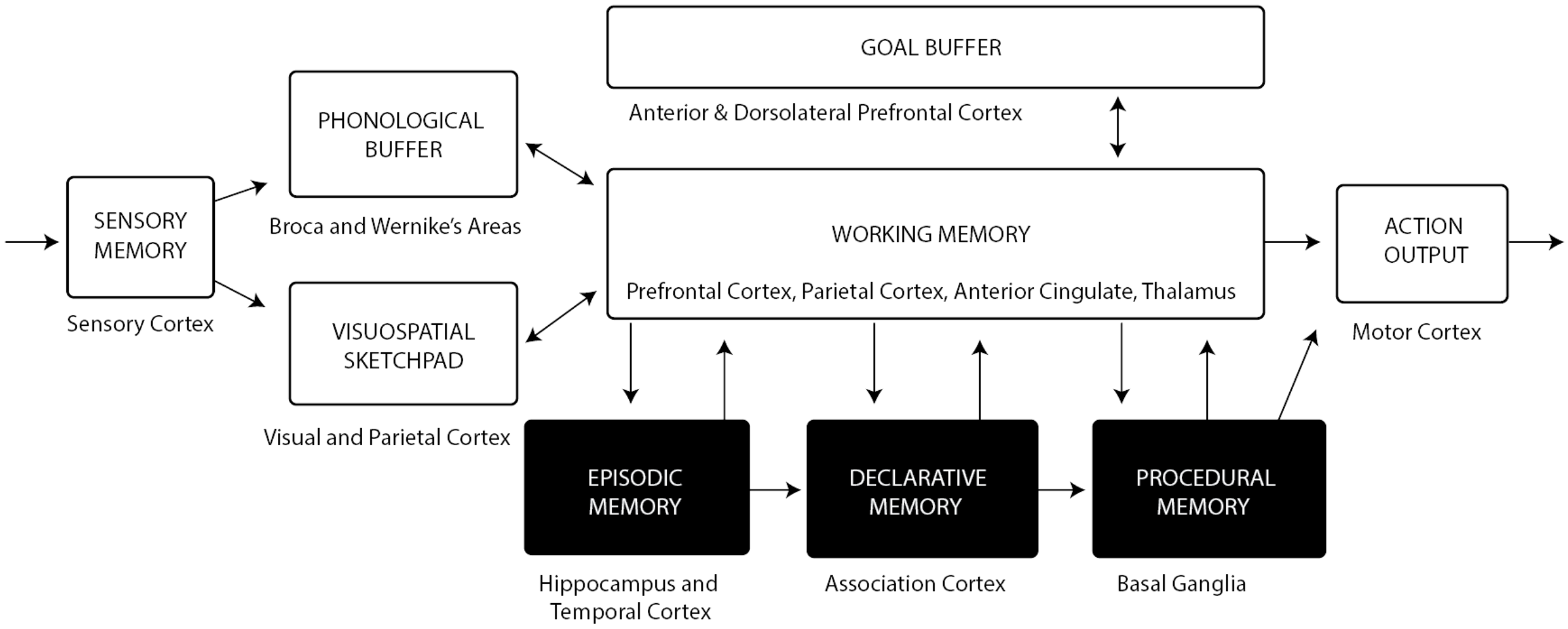

**Fig. 5** Prototypical Cognitive Architecture

*This figure incorporates aspects of various popular cognitive architectures and indicates the neural substrates thought to support these memory functions.*

Bernard Baars developed the functional framework model, which combines the multi-store model (Fig.3) with the multicomponent model (Fig. 4) (Baars, 2007). This framework, adapted in Figure 6, integrates other cognitive constructs such as attention, consciousness, and planning. It also draws further subdivisions within long-term memory.

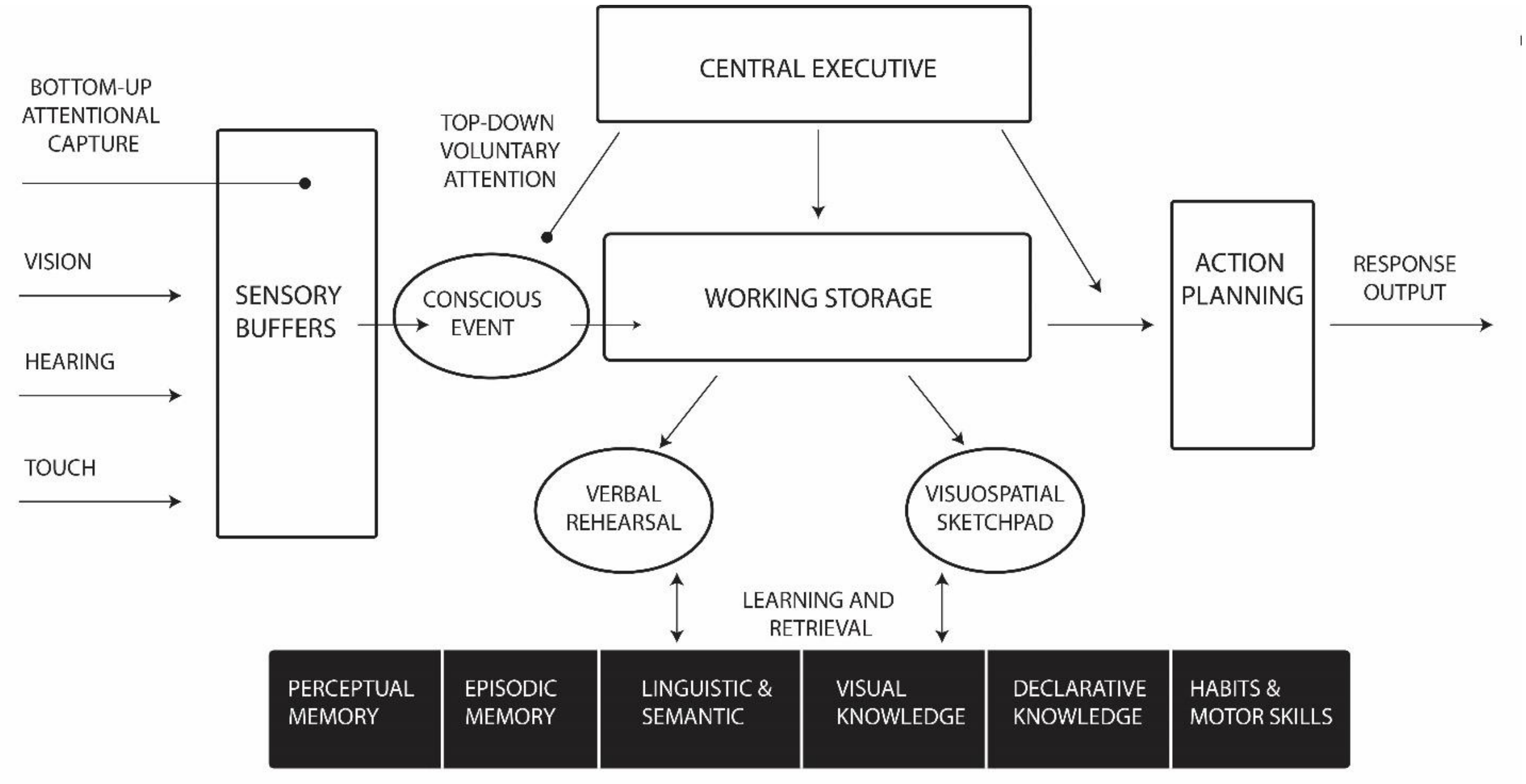

**Fig. 6.** Baars and Gage's Functional Framework (2007)

*Working memory activates long-term memories, knowledge, and skills, which are shown in the box at the bottom. Spontaneous (bottom-up) attention and voluntary (top-down) attention are symbolized as vectors.*

In 1988, Bernard Baars introduced the global workspace model (Fig. 7). Therein, active contents in working memory are broadcast throughout the brain, stimulating unconscious long-term memories. These long-term memories then compete to enter the global workspace. This type of organization is known as a "blackboard" architecture and can be traced back to Newel and Simon (1961). Many present-day computer science, neural, and psychological models assume a fleeting but centralized working memory capacity that acts as a common workspace where long-term memories become coactive and are exposed to one another (e.g., Dehaene, 2020; Ryan et al., 2019; Glushchenko et al., 2018).

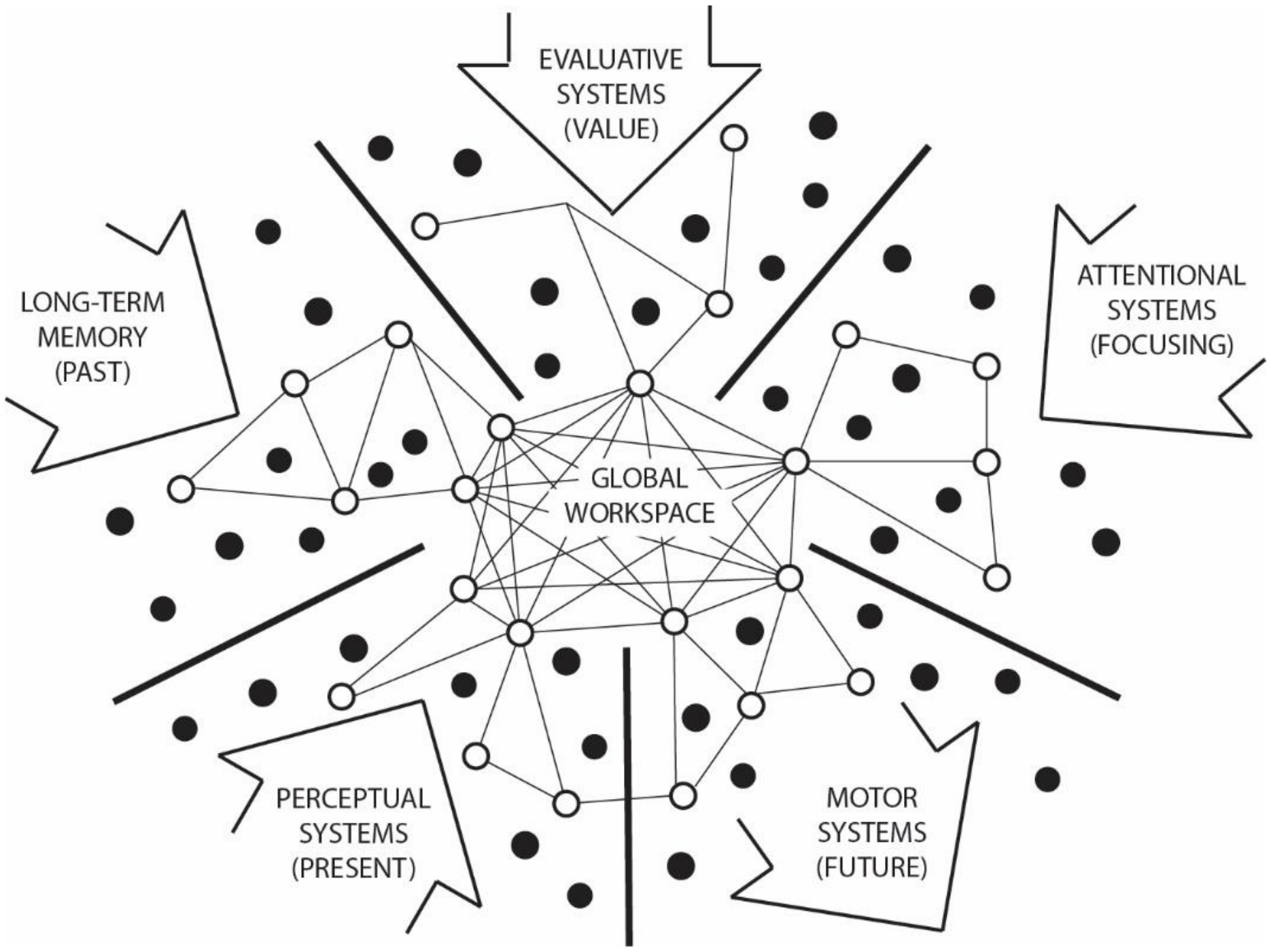

**Fig. 7.** Baar's Global Workspace Model (1988)

*The brain has several lower order modules that are generally isolated from each other. Unconscious processes and environmental stimuli processed by these modules compete for access to the global workspace. The most salient inputs enter the workspace where they are integrated, stored temporarily, and made conscious. These results are then broadcast back to the rest of the brain.*

Contemporary AI systems using artificial neural networks feature many of these models' properties. They store long-term memories by distributing them through the synaptic weights between neurons. They allow inputs to traverse the network, exciting the most closely associated neurons and neural representations, pulling them into a form of global short-term memory. Some also utilize forms of attention that potentiate certain representations for extended periods amounting to a kind of working memory. However, no current models arrange working memory elements the way the human brain does.

### 2.2 The Focus of Attention Is Embedded within the Short-term Memory Store

Early models attempting to explain how long-term memory is transferred into working memory were influenced by computer science. They envisioned long-term memories being copied and transferred from long-term storage to a separate processing substrate (i.e., from the hard drive to random-access memory (RAM) to the central processing unit (CPU)). In a departure from this conception, several theorists (e.g., Cowan, 1984; Norman, 1968; Treisman, 1964) conceived that information is encoded into working memory when existing units of long-term memory are activated and attended to without being copied or transported. Today, this is commonly referred to as activated long-term memory.

Brain imaging studies support this view and provide evidence that units of long-term memory reside in the exact locations involved in processing this information during non-working memory scenarios (D'Esposito & Postle, 2015). These findings suggest that information is not copied and transferred between dedicated registers, but activated right where it is (in situ) (Chein & Fiez, 2010; Moscovitch et al., 2007). The concept may apply equally to artificial neural networks. Thus, although neurons are stationary, as long as they remain active, they continue to broadcast their encoded information to the neurons they project to.

Nelson Cowan's embedded processes model (1988) reconciles the main features of the multi-store and multicomponent models with the concept of activated long-term memory. In Cowan's model, the short-term memory store is comprised of units of long-term memory that are activated above baseline levels, such as memories that have been primed. This activation can last from seconds to hours. Thus, the short-term store of working memory is simply an active subset of the long-term store it is "embedded" within (Cowan, 1999).

The other key component of Cowan's model is the focus of attention (FoA). The FoA holds consciously attended units of information and is embedded within the short-term store (Fig. 8). Units in the FoA comprise an even more active subset of the short-term store. Their elevated activity lasts from milliseconds to several seconds. Cowan and others consider the short-term store and the FoA together as constituting working memory (Cowan, 2005).

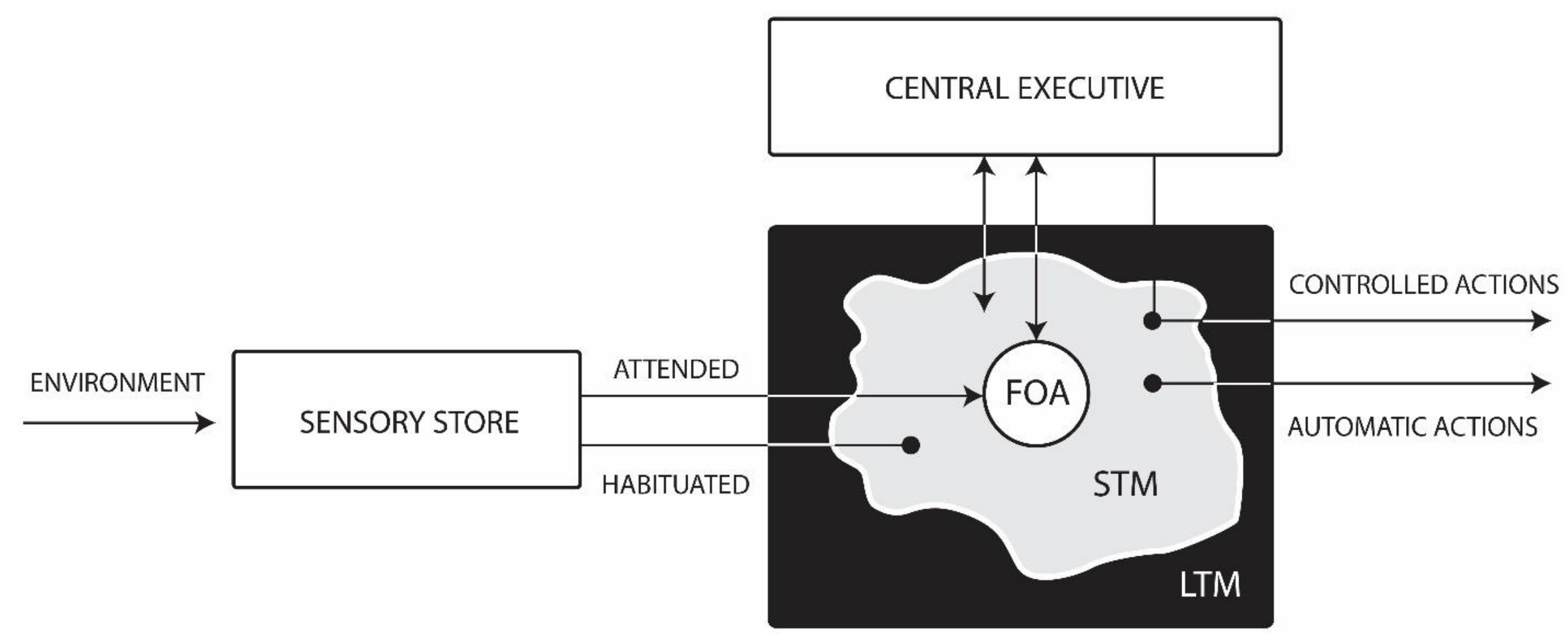

**Fig. 8.** Cowan's Embedded Processes Model (1988)

*According to this model, short-term storage is an activated subset of long-term storage. Similarly, the focus of attention (FoA) is an attended subset of short-term storage. These stores interact with a store for sensory memory and a central executive.*

During perception, task-relevant features from the sensory store are used to update the FoA. When attention shifts to other information, these items pass into the short-term store (Nyberg & Eriksson, 2016). However, information demoted from the FoA to the short-term store can still influence automatic actions and be readily reactivated into the FoA (Manohar et al., 2019). If not reactivated, this information returns to inert long-term memory (through the processes of decay, inhibition, interference, or contamination) (Cowan, 2009). Some items that enter working memory are demoted almost immediately, whereas others remain active for sustained periods (Cowan, 2011). This feature, along with features of the other models discussed thus far, forms critical assumptions about updating subsumed by the present model. Figure 9 provides a summary of the forms of human memory.

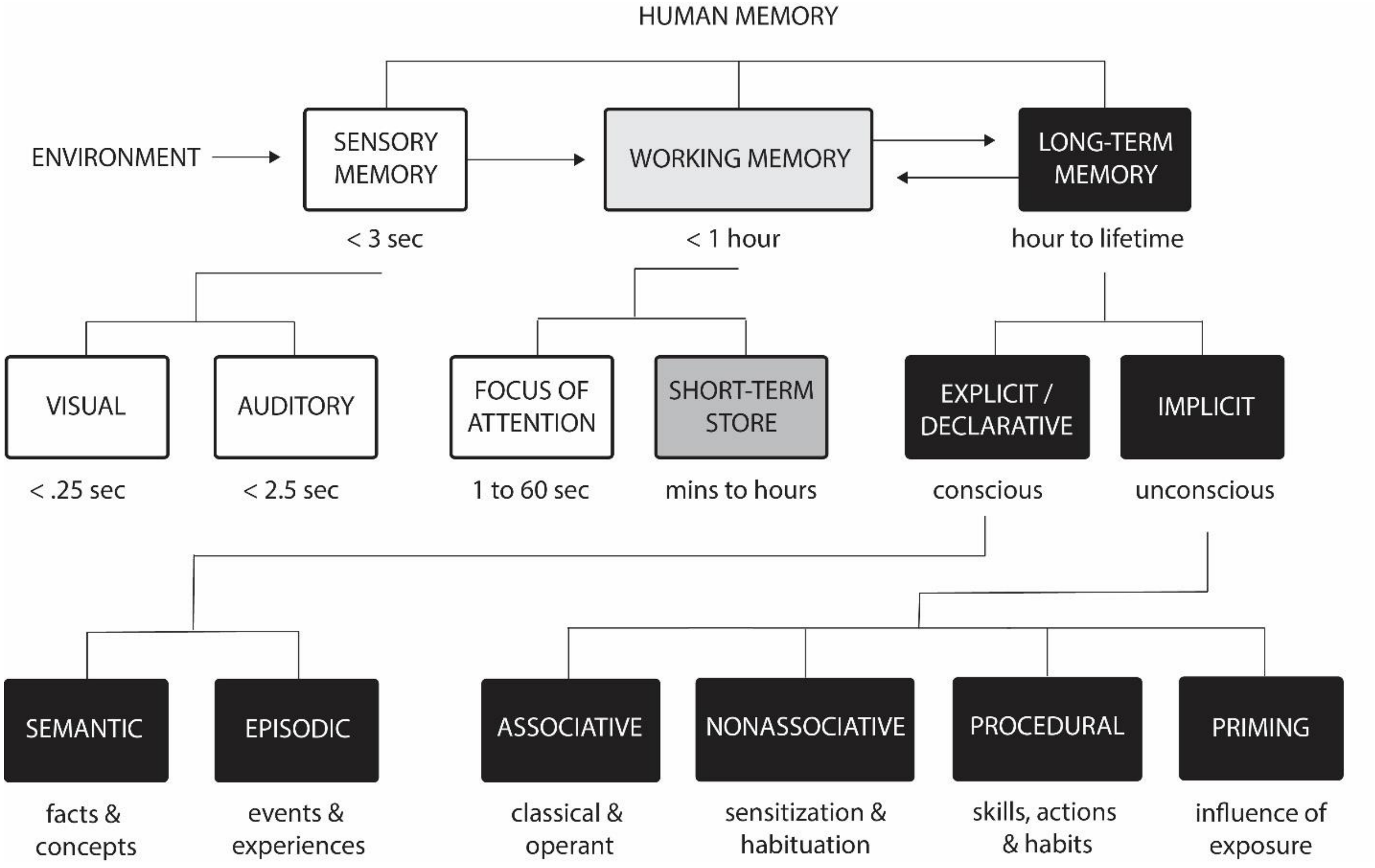

**Fig. 9.** Forms of Human Memory

*Human memory can be divided into three phases, each of which can be decomposed into other forms.*

The first of four quotes in this article from the prescient psychologist and philosopher William James (1842-1910) complements Cowan's conception of an FoA store, which interacts with and is embedded inside a short-term store:

> "My present field of consciousness is a centre surrounded by a fringe that shades insensibly into a subconscious more.... The centre works in one way while the margins work in another, and presently overpower the centre and are central themselves. What we conceptually identify ourselves with and say we are thinking of at any time is the centre; but our full self is the whole field, with all those indefinitely radiating subconscious possibilities of increase that we can only feel without conceiving, and can hardly begin to analyze."

William James (1909, p. 288)

### 2.3 Sustained Firing Maintains Information in the Focus of Attention

Understanding how the brain enables working memory should be paramount when designing working memory for computers. The neurophysiological basis of the persistent activity responsible for working memory is an active research area although research in the last two decades has been enlightening. Single-cell recordings of neurons in primates reveal that information retention occurs via a cellular phenomenon known as sustained firing. Glutamatergic pyramidal neurons in the prefrontal cortex (PFC), parietal cortex, and other association cortices are specialized for sustained activity, allowing the cells to generate action potentials at elevated rates for several seconds at a time (Funahashi, 2007; Fuster, 2015). Sustained firing is thought to maintain the signal of information that the neuron encodes. A neuron in the PFC with a background firing rate of 10 Hz (typical for cortical cells) might increase its firing rate to 20 Hz when utilizing sustained firing to preserve mnemonic information temporarily.

One of the earliest of these studies provides an illustrative example. In 1973, Joaquin Fuster recorded the sustained electrical activity of PFC neurons in monkeys performing a delayed matching task. In the task, a macaque monkey watches the experimenter place food under one of two identical cups. A shutter is then lowered for a variable delay period, so the cups are not visible. After the delay, the shutter is raised, and the monkey is given one attempt to collect the food. Through training, the animal learns to choose the correct cup on the first attempt. Completing the task requires the animal to hold the location of the food in working memory during the delay period. Presumably, the monkey must sustain either a retrospective sensory representation of the food's location or a prospective representation of the motor plan needed to retrieve it.

Using implanted electrodes, Fuster could record from neurons in the PFC that fired throughout the delay period. He found that the sustained firing subsided once the monkey responded, suggesting that the observed neuronal activity represented the food's location while the cup was out of sight. This landmark study revealed the brain's mechanism for keeping meaningful representations active without external input. It also suggested the presence of a dynamically updated pool of coactive neurons underlying thought and behavior. It is important to mention that processes besides sustained firing may be responsible for maintenance in the FoA (e.g., dynamic coding or activity states distributed across neuronal populations (Stokes, 2015; Jacob et al., 2018)), yet these can be expected to exhibit iterative updating as described here.

Subsequent primate research has found that the duration of sustained firing predicts whether items will be remembered. When this delay-period activity is weak, the likelihood of forgetting is greater (Funahashi et al., 1993). Moreover, lesioning of the prefrontal and association cortices (which contain neurons with the greatest capacity for sustained firing) significantly impairs performance in these tasks. Consistent with this animal work, functional magnetic resonance imaging (fMRI) studies in humans show that activity in prefrontal and association areas persists during the delay period of similar working memory tasks. In fact, the magnitude of this activation positively correlates with the number of items subjects are instructed to hold in memory (Rypma et al., 2002).

Patricia Goldman-Rakic (1987, 1990, 1995) was the first to suggest that the phenomenon of sustained firing in the PFC is responsible for the retention interval exhibited by working memory. Further work by Fuster (2009), Goldman-Rakic (1995), and others has shown that neuronal microcircuits within the PFC maintain information in working memory via recurrent, excitatory glutamatergic networks of pyramidal cells (Baddeley & Hitch, 1994; Miller & Cohen, 2001). Many researchers now believe that sustained firing is critical in maintaining working memory. The evidence backing this assumption is provided by studies reporting positive correlations between sustained firing and working memory performance. For example, both human and animal subjects can retain information in mind as long as sustained firing persists (Rypma et al., 2002). This has been found using extracellular, electroencephalographic, and hemodynamic approaches (D'Esposito & Postle, 2015).

Sustained firing in the PFC and parietal cortex is now assumed to underlie the capacity to internally maintain and update the contents of the FoA (Braver & Cohen, 2000; Sarter et al., 2001). As a result, working memory, executive processing, and cognitive control are now widely thought to rely on the maintenance of activity in multimodal association areas that correspond to goal-relevant features and patterns (Baddeley, 2007; Fuster, 2002a; Moscovich, 1992; Postle, 2007). Sustained rates of action potentials allow responses throughout the brain to be modulated by prior history over multiple timescales, from milliseconds to tens of seconds.

Cognitive neuroscience and models of working memory suggested that maintaining a dynamic and context-sensitive internal state was essential for any system attempting to process temporal patterns, reason over sequences, or otherwise engage in tasks that require memory. The sustained activity of neurons inspired the idea that artificial systems might also need a way to retain information over time so that previous states could influence current processing. This notion was embedded in the architecture of recurrent neural networks (RNNs), long short-term memory systems (LSTMs), and gated recurrent units (GRUs), and was instrumental in making them well-suited for tasks like language modeling, speech recognition, and time series analysis. However, it turned out that sustained firing was only half of the story.

### 2.4 Synaptic Potentiation Maintains Information in the Short-term Store

fMRI studies have suggested that the information represented by sustained firing corresponds only to the FoA, not the short-term store (Lewis-Peacock et al., 2012). This is because neuronal activity corresponding to items that have exited the FoA quickly drops to baseline firing rates.

Nevertheless, information about the items may be rapidly and reliably recalled after a brief delay. It is thought that the passive retention of information in the short-term store but outside the FoA may be mediated by a different "activity-silent" neural mechanism, such as changes in synaptic potentiation (short-term synaptic plasticity) (LaRocque et al., 2014; Rose, 2016). The evidence supporting this is strong (Silvanto, 2017; Nairne, 2002). For example, synaptic strength can be temporarily modified by transient increases in the concentration of presynaptic calcium ions or by GluR1-dependent short-term potentiation (Silvanto, 2017). The information potentiated by these changes in synaptic weighting can be converted back into active neural firing if the memory is reactivated by a contextual retrieval cue (Nairne, 2002).

Thus, the maintenance of information in working memory is achieved by at least two neural phenomena operating in parallel that correspond to distinct states of prioritization: sustained firing, which maintains information in the FoA, and synaptic potentiation, which maintains information in the short-term store. Both mechanisms contribute to the initialization of learning (long-term potentiation), including RNA synthesis, protein synthesis, and morphological synaptic changes that underlie the formation and consolidation of new long-term memories (Debanne, 2019). Table 2 summarizes the general properties of the four phases of human memory.

| | Sensory Memory | Focus of Attention | Short-term Store | Long-term Memory |
|---|---|---|---|---|
| **Description** | Low-level, early sensory processing | Attended, conscious content | Temporary store of recent content | Inactive but retrievable content |
| **Brain Correlate** | Firing neurons | Sustained firing | Synaptic potentiation | Long-term potentiation |
| **Capacity** | Numerous features | 1 to 9 items | Hours of memories | A lifetime of memories |
| **Duration** | Visual: .25 s. Auditory: 2.5 s. | Up to a minute | Minutes to hours | Days to lifetime |
| **Origin** | Pre-attentive | Attentive | Recent attention | Memorization |
| **Location** | Sensory cortex | Association cortex | Cerebral cortex | Cerebral cortex |
| **Maintenance** | Not possible | Continued attention | Rehearsal | Repetition, mnemonics |
| **Departure** | Decay, highly volatile | Replacement | Forgetting | Irretrievability |
| **Depiction in Figures** | N/A | 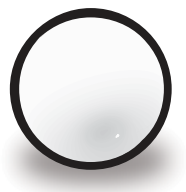 | 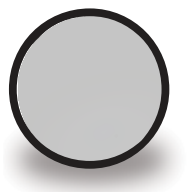 | 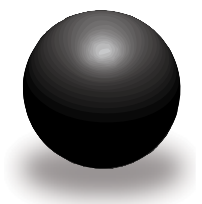 |

**Table 2.** General Characteristics of Four Forms of Memory

*This table summarizes some of the major comparisons between four different forms of memory. The details addressed in this table are not definitive and are active areas of debate and research. (see Baddeley et al., 2018; Christophel et al., 2017; Shipstead et al., 2015; Brydges et al., 2018; Cowan, 2017; Eriksson et al., 2015; Chia et al., 2018; Constantinidis et al., 2018).*

Modern artificial neural networks do not generally use direct analogs of sustained firing or synaptic potentiation, although some use forms of persistent activity such as recurrence and self-attention. As Figure 10 illustrates, a recurrent neuron, like those found in a recurrent neural network, reroutes its output back to its input. This allows it to hold the memory of an internal state, which can affect subsequent ones. These machine learning nodes can use this recurrent feature to selectively store, update, or forget information based on their input and previous state. As we will see, this functionality permits them to uncover patterns in time (long-range dependencies from sequential inputs). Recurrent neurons (or those orchestrated by self-attention), programmed properly, should permit artificial neural networks to simulate sustained firing and synaptic potentiation, as well as attain the cognitive properties discussed in upcoming sections, including the capacity to support a working memory that is iteratively updated.

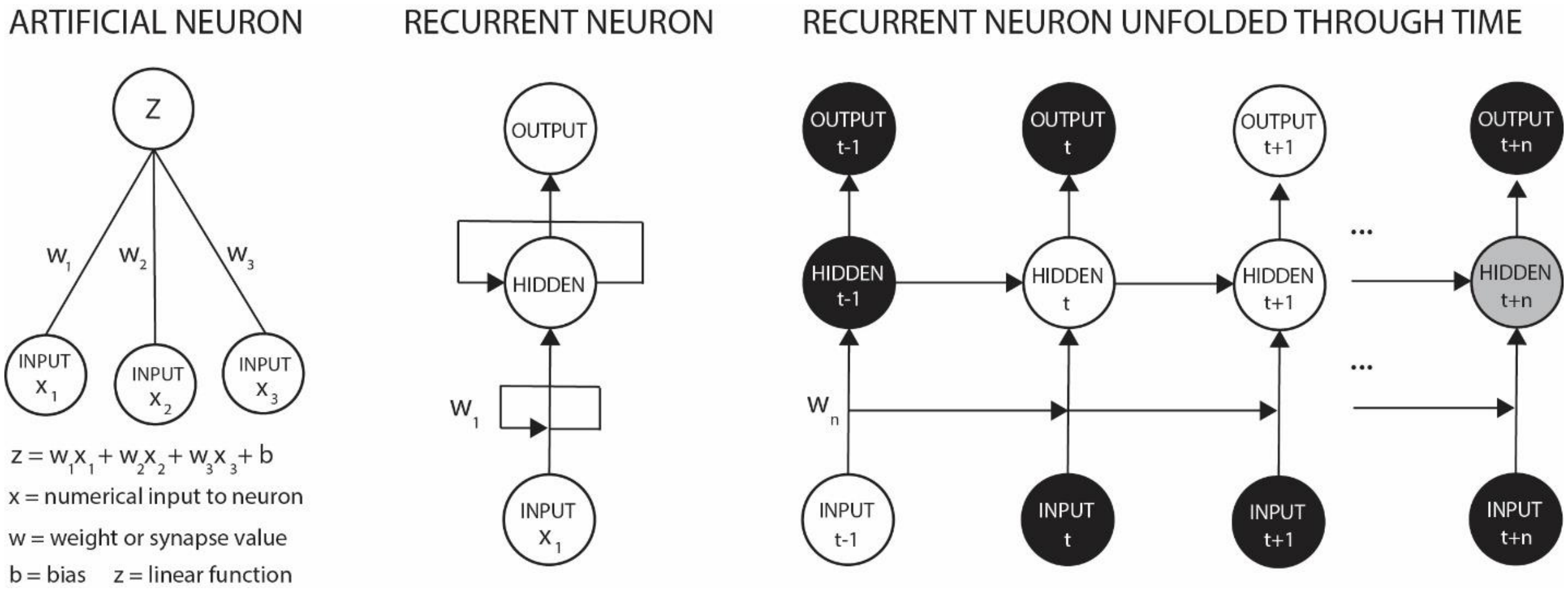

**Fig. 10.** Recurrency in Artificial Neural Networks Can Simulate Persistent Neural Activity

*A. A simplified version of an artificial neuron used in computer science, with inputs and weights labeled. B. An artificial recurrent neuron in a hidden layer, passing information from an input layer to an output layer. This hypothetical neuron (inspired by Table 2) exhibits recurrency in its cell body which is later converted to recurrency in its weights. Thus, it demonstrates an analog of both sustained firing and synaptic potentiation. C. The activity of this hidden neuron is unfolded through time as it goes from inactive to active to partially active.*

Persistent neural activity in the form of sustained firing and synaptic potentiation is time-limited. It runs out. When some neurons exit persistent activity and others enter it, working memory updating occurs. The updating of persistent activity provides the conceptual basis used by this article to view the previous six models from the perspective of iteration. The following section will discuss how this perspective provides insight into the process of thought.

# **Part III:** Working Memory Is Updated Iteratively

### 3.1 Persistent Activity Causes Successive States to Overlap

In the preceding sections, we delved into the diverse array of models that define our understanding of working memory. However, a conspicuous gap emerges in this landscape: the omission of iterative updating as a core mechanism. This is not a minor oversight because iteration is a crucial element in working memory dynamics. In this section, we will explore how iterative updating occurs in synchrony at both the neural and psychological levels, bridging these domains and offering new insights that extend current theoretical approaches.

Even though modern working memory research does not acknowledge that content is updated iteratively, the nature of persistent activity strongly implies that iterative updating is pervasive. Allow me to use another analogy to explain why. Take the human population of Earth, for instance. In the next year, many people will pass away, others will be born, yet most will remain living. In the same sense, in one second, some of the brain's neurons will stop exhibiting persistent activity, some new neurons will enter persistent activity, yet most will remain in it. The people and neurons that persist can influence subsequent states. In the same way that there could be no intergenerational knowledge transfer (culture) on a planet where generations do not overlap, there can be no thinking in a brain where spans of neural activity do not overlap.

The study of sustained firing has shown that the neocortex contains many neurons in persistent coactivity at any instant in time (Goldman-Rakic, 1995). Nevertheless, these coactive neurons could not have all started firing at the same time, nor could they all stop firing at the same time. This is similar to how people on Earth are not all born on the same day. Because sustained firing has been shown to occur for different durations in different neurons (Fuster, 2008, 2002b), their spans of activity must be staggered and must only partially overlap with one another rather than completely coincide (Reser, 2016), as portrayed in Figure 11. This effect may play a role in stitching together discrete states to create the seamless quality of mental life.

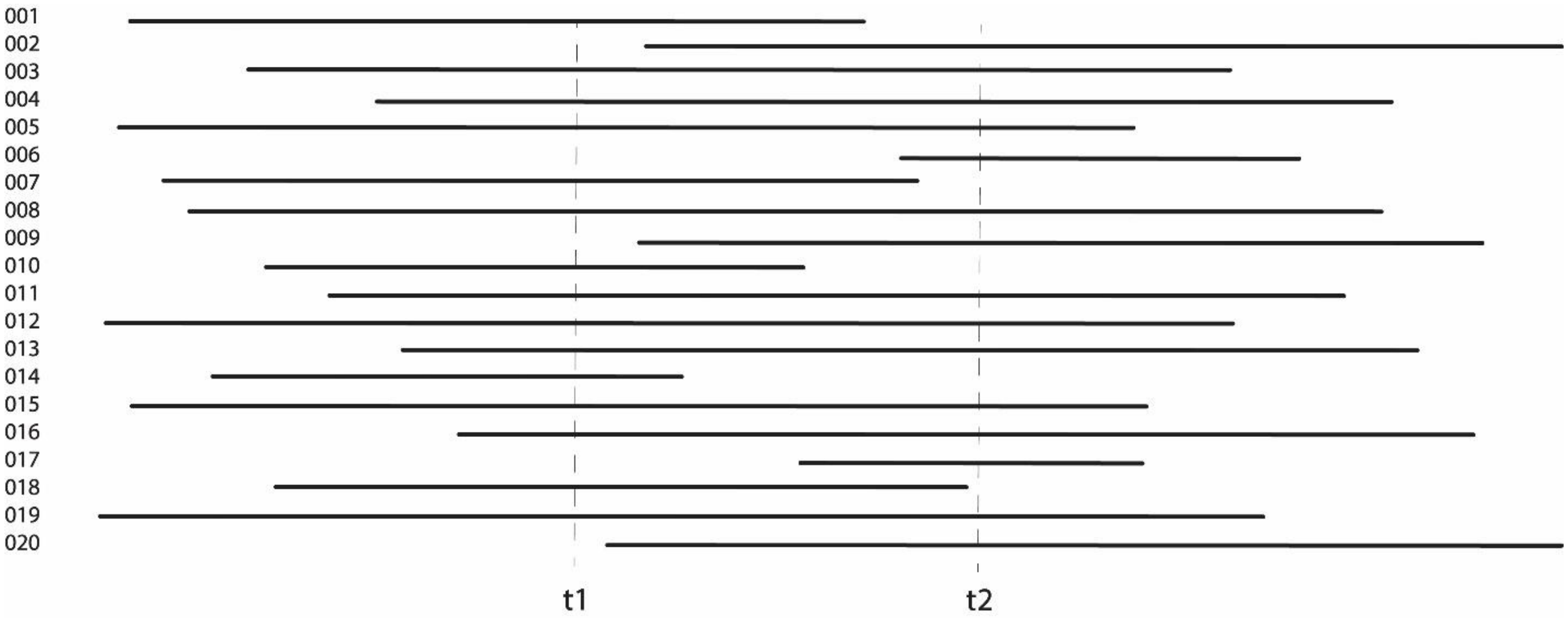

**Fig. 11.** A Set of 20 Neurons Exhibiting Sustained Firing

*The sustained firing of 20 hypothetical neurons is shown here. The x-axis represents time. The spans of individual neurons overlap but are staggered and asynchronous. T1 and t2 are marked at the bottom. The two time periods share 10 of 20 active neurons in common. This figure exemplifies the transitional reconfiguration that distinguishes iterative updating.*

Because updates to the set of neurons in sustained firing must involve partial rather than complete replacement, these dynamics indicate an ongoing pattern of iteration and recursion. Iteration involves the application of a computational procedure to the results of a previous application. Iteration's sister algorithm, recursion, is the reapplication of a rule, definition, or procedure to successive results. A recursive function references itself. Self-referential routines are common in mathematics and computer science but mostly unconsidered in psychology (Reser, 2016). The terms "iteration" and "recursion" uniquely capture different aspects of the present model, and both are used here depending on context.

The principles of iteration and recursion as they pertain to the present model are illustrated in Figure 12. At time 1 (t1), neuron "a" has stopped firing. Neurons b, c, d, and e exhibit sustained coactivity. By time 2 (t2), neuron "b" has stopped firing, while c, d, and e continue to fire, and "f" begins to fire. In time 2, c, d, and e recur. The figure depicts iteration because the set of coactive neurons at time 2 (c, d, e, and f) includes a subset (c, d, and e) of the coactive neurons at time 1. In computer programming, the goal of iteration is to obtain successively closer approximations to the solution of a problem. In later sections, this article will advocate that the algorithm of thought utilizes iteration for the same purpose.

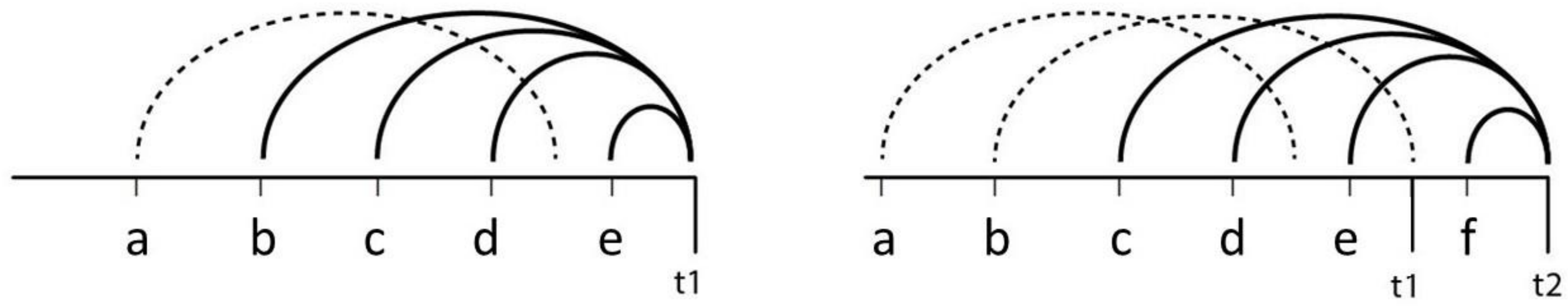

**Fig. 12.** Depiction of Iteration in Neurons Exhibiting Sustained Firing

*Each arc, designated by a lowercase letter, represents the time span during which a neuron exhibiting sustained firing remains active. The x-axis represents time. Dashed arcs represent neurons that have stopped firing, whereas full arcs denote neurons that are still active. The neurons shown here demonstrate iteration and recursion.*

Given that at any point in time, there are likely to be thousands of neurons engaged in sustained firing, we should expect the type of iterative pattern seen in Figure 12 to be ubiquitous. Furthermore, if examined on the order of hundreds of milliseconds, we should expect activity in the brain to be densely iterative. Iteration within the FoA causes consecutive brain states to be interrelated and autocorrelated as a function of the delay between them. Because a delineable subset of the active cells that characterize one brain state remain active in the next, each state is recursively nested within the one that precedes it. This allows the brain to record and keep track of its interactions with the environment (stateful) so that each interaction does not have to be handled based only on the information available at present (stateless).

It is asserted here that iterative updating should be considered inherent in any brain with neurons exhibiting persistent activity and that animals utilize it as a fundamental means of information processing. In particular, working memory may harness iteration in a way that allows potentially related representations to accumulate and coactivate despite delays between their initial appearances. This ensures that relevant processing products are temporarily sustained until a full suite of contextually related items is compiled to be used in aggregate to inform behavior. Such compilations could also permit the contextualization of each new impression and the capture of variable-length dependencies.

**Iterative Updating in Stores Other Than the Focus of Attention**

The short-term store should also be capable of iterative updating. Short-term synaptic potentiation (or any other neurophysiological mechanism) responsible for the maintenance of short-term memory can be expected to exhibit the same iterative properties as the sustained firing responsible for the FoA. This expectation derives from the fact that the sum of synaptically potentiated neurons is equivalent to a pool that is constantly being added to as new neurons are potentiated and subtracted from as other neurons lose their time-limited potentiation. Thus, both the short-term store and the FoA demonstrate iterative functionality, albeit on different time scales.

Aside from the FoA and short-term store, information found in other forms of temporary storage may also work iteratively. For example, neural binding may represent a third tier of temporary information storage, which might be embedded in the FoA and take place at even shorter intervals. Neural binding involves synchronized oscillations of network activity that form and dissipate on the order of milliseconds (Opitz, 2010). It is thought that this synchronization integrates different forms of information into cohesive conscious experiences (Pina et al., 2018). However, binding will be left out of the present discussion as it is unclear whether state changes in binding are complete or partial. Consequently, this article's focus on iterative updating may not apply to it.

Similarly, dynamic coding, closed recurrent loops, persistent attractor networks, and reentrant neural oscillations may or may not involve iterative updating. While Oberauer's one-item attentional store (2002) has found considerable empirical support (Niklaus et al., 2019), it could not work iteratively because (as just one item) it is either wholly updated or not updated at all.

The incremental updating expected at the neurophysiological level may be isomorphic with and provide a substrate for the incremental updating experienced on the psychological level. For example, a given line of thought does not change all at once but rather makes additive transitions that are grounded by content that remains unchanged. The subset of neurons that continue to exhibit persistent activity over the course of these incremental changes should be expected to embody the persisting subject of mental analysis. Stated differently, **neurons with the longest-lasting activation likely correspond to the underlying topic of thought that remains as other contextual features come and go.** This creates coherence and continuity between distinct epochs (Reser, 2016), as depicted in Figure 13. The present article posits that without the continuity made possible by iterative updating, thought as we know it cannot arise and will not be available to machines.

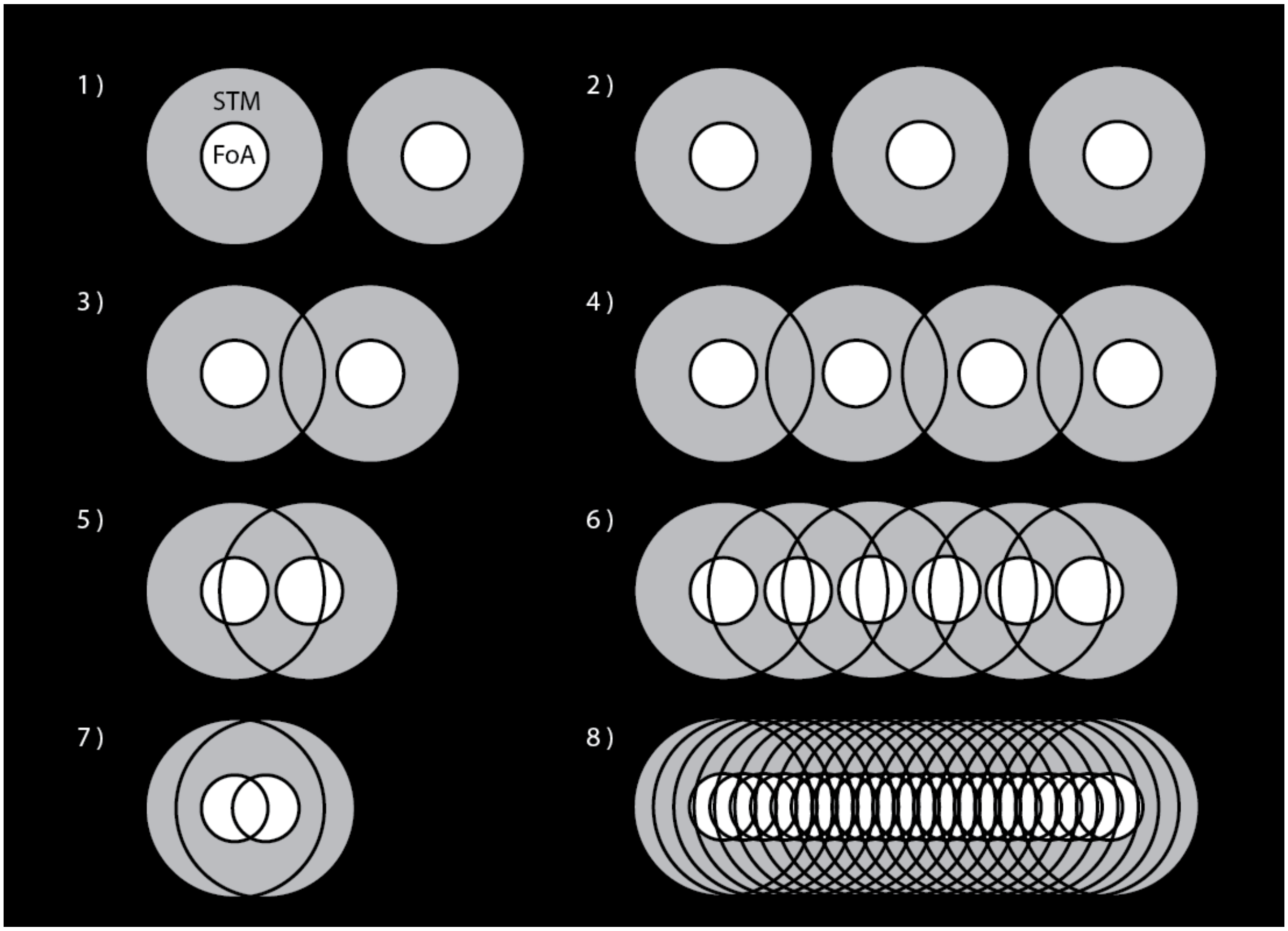

**Fig. 13.** Venn Diagrams of Information Shared Between Successive States of Working Memory

*These Venn diagrams depict informational overlap between successive states of working memory. The horizontal axis represents time. The small, white circles represent information within the FoA, while the large, grey circles represent information within the short-term store. Diagrams 1 and 2 show no overlap between states from different periods; 3 and 4 show overlap in the short-term store only; 5 and 6 show the short-term store of one state overlapping with the FoA of the neighboring state; and 7 and 8 show the FoA of separate states overlapping, suggesting attentive continuity. Given the time constraints noted in Table 2, it may be plausible that Diagrams 1 and 2 roughly depict sampling of cortical activity hours apart, 3 and 4 depict sampling several minutes apart, 5 and 6 depict sampling every minute, and 7 and 8 depict sampling every second.*

Diagram 1 of Figure 13 depicts two states of working memory whose contents do not overlap. We can assume that these states are from separate thoughts. Diagram 7 depicts two states whose contents overlap significantly. It is intended to represent a fractional transition in the thought process, such as two points in a line of reasoning. The overlapping informational content of the small circles shown in Diagram 7 indicates that the two states share neurons in common that exhibit sustained firing. The overlap of the large circles represents the sharing of potentiated synapses. Thus, the diagrams shown in Figure 13 depict updating as continuous change in active neurons and synapses. However, as the rest of this article will explore, partial change to the FoA may be more realistically depicted as iterative updates in discrete cognitive items.

| Term | Definition |
|---|---|
| **Iteration** | Repetition of a computational procedure applied to the product of a previous state repeated to approach the solution of a problem |
| **Working Memory** | A limited-capacity store dedicated to maintaining selected representations for use in further cognitive processing |
| **Working Memory Updating** | Changes to the contents of working memory occurring as processing proceeds through time |
| **Iterative Updating** | A shift in the contents of working memory that occurs during updating as some representations are added, others are removed, and still others are repeated |

**Table 3.** Definition of Key Terms

### 3.2 The Iterative Updating of Representations Allows Context to Shift

Figure 14 depicts an FoA store that holds four mental representations at a time. We will refer to these representations as items (also known in psychology as chunks). In this example, one discrete item is replaced at each point in time. Thus, it could be conceptualized as a "sliding store." The depiction of the FoA store as limited to four items is derived from an extensive literature review by Cowan (2001, 2005), which demonstrates that adults are generally able to recollect four items (plus or minus one) in situations when they cannot carry out chunking, rehearsal, or other memory strategies to aid them. This capacity of four items generally holds whether the items are numbers, words in a list, or visual objects in an array. The figures could alternatively feature seven items rather than four after less restrictively controlled research by George Miller (1956). While discussing the capacity of the FoA, Cowan remarked,

> "When people must recall items from a category in long-term memory, such as states of the United States, they do so in spurts of about three items on average. It is as if a bucket of short-term memory is filled from the well of long-term memory and must be emptied before it is refilled."
>
> Nelson Cowan (2009, p. 327)

Yet, perhaps this bucket does not need to be emptied to be refilled. Perhaps it can be emptied and filled simultaneously. While naming states and repeating numbers may not necessitate this, rational thought may.

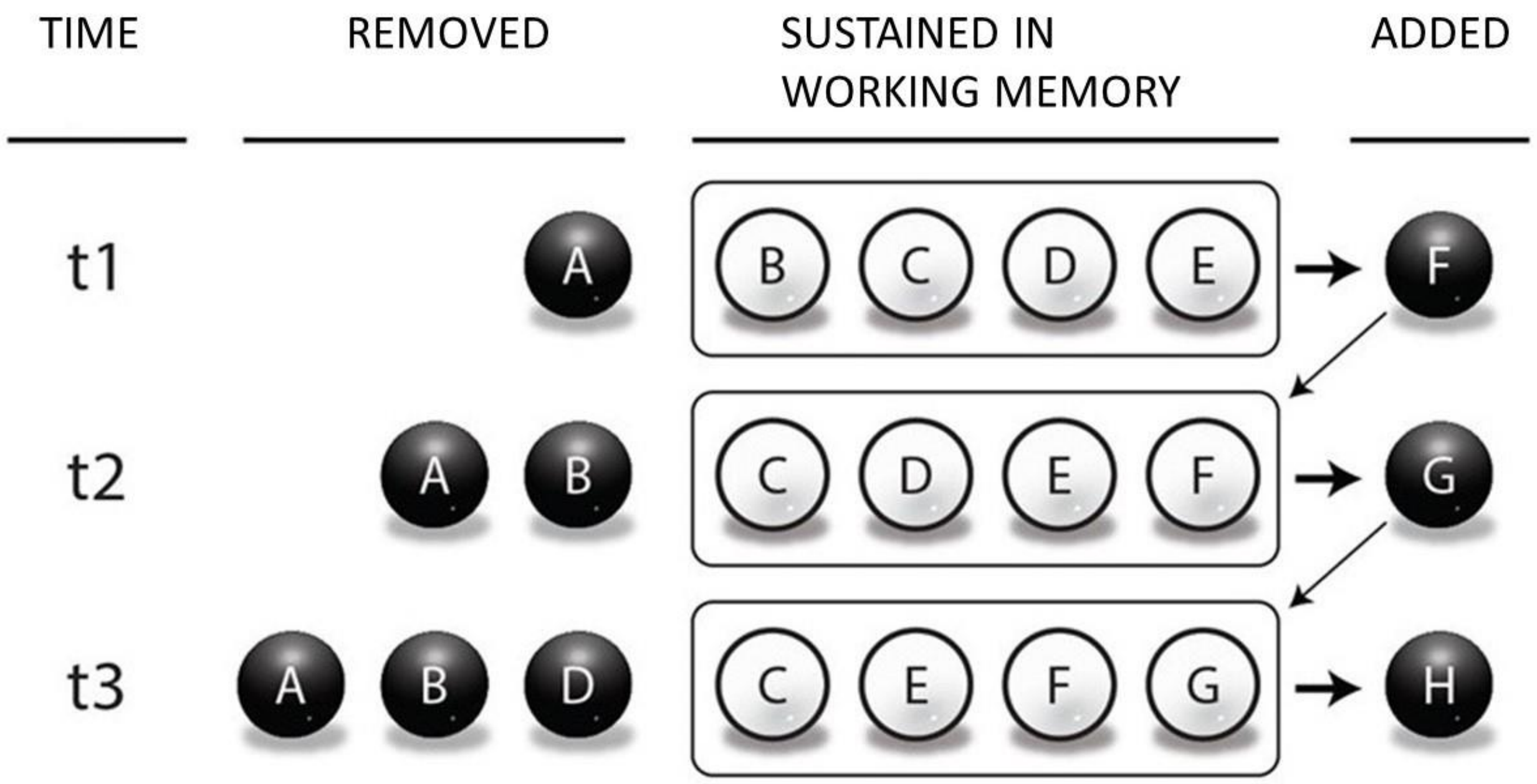

**Fig. 14.** Abstract Schematic of Iterative Updating in the FoA

*As with the following figures in this article, Figure 14 is an emblematic abstraction that uses a state-space model in discrete time. White spheres indicate active mental concepts (items), while black spheres indicate inactive ones. At time 1, item A has just been deactivated, while B, C, D, and E are coactive. This echoes the pattern of activity shown in Figure 12, except that the uppercase letters here represent items, whereas the lowercase letters in Figure 12 represent neurons. While coactive, these items (B, C, D, and E) spread their activation energy, resulting in the convergence of activity onto a new item, F. At time 2, B has been deactivated; C, D, and E remain active; and F has become active.*

At time 2, three items from time 1 (C, D, and E) remain active and are combined with the update located in time 1 (F). This new set of items is then used to search for the next update (G). Items C and E demonstrate reiteration because they exhibit uninterrupted activity from time 1 through time 3. The longer these items are coactive, the more likely they will become associated and possibly "chunk" or merge into a single item, "CE". While C and E remain active, their underlying neural circuits can be expected to impose sustained, top-down information processing biases on the targets they project to throughout the thalamocortical hierarchy. Items sustained enduringly in this way should be expected to influence the overarching theme of ongoing thought.

Imagine that item B represents your psychological concept of brownies, C represents your friend Cameron, D represents shopping, and E represents a grocery store. With these representations active in your FoA, you may form a mental image of your friend Cameron shopping for brownies at a grocery store. This scenario may cause you to remember Cameron's preference for drinking milk when he eats brownies. Thus, your next thought may be about your friend shopping in the same store for milk. Some contextual factors (the place, person, and activity) remained the same even though another (the object being shopped for) changed. This kind of narrative about the same place and person could take several seconds and many rounds of iteration to play out. This example illustrates how iteration enables continuity by allowing context to shift incrementally, which, this paper contends, is a central hallmark of the thought process. Figure 15 offers a different narrative example.

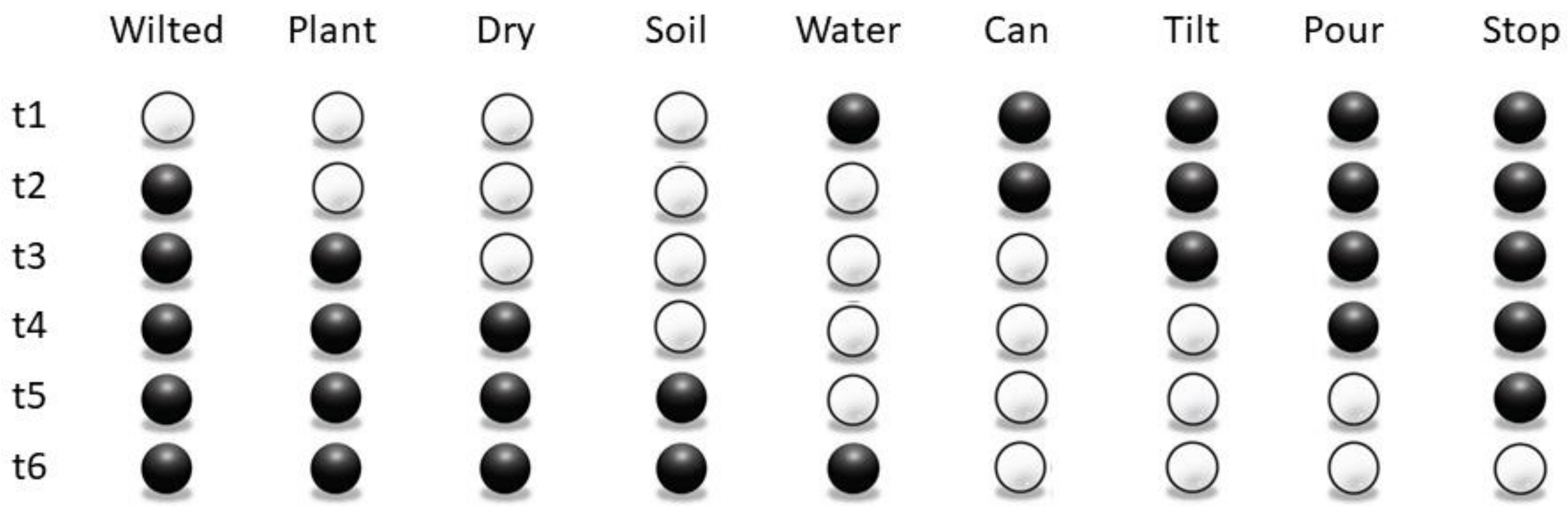

**Fig. 15.** Working Memory Updating Mediates Intelligent Transitions Between Mental States

*This series illustrates the iterative working memory activity of a person who thinks about watering a plant. The person imagines a wilting plant with dry soil. This set of coactivates in working memory spreads neural activity, converging on the concept of water. This new set, in turn, induces the ideation of using a watering can. The person then imagines tilting and then pouring the water from the watering can until they stop watering the plant.*

Figure 16 expands on the schematic from Figure 14, exemplifying how working memory capacity can vary. It is worth mentioning that large language models using the transformer architecture hold thousands of tokens in a rudimentary form of attention and update these coactive sets of tokens iteratively when forming predictions. An attentional window that can hold thousands of tokens may be a good reproduction of the short-term store but may be too large and likely to hold unrelated or incongruous elements to reproduce the FoA. Perhaps AI would benefit from having an analog of the brain's FoA which could maintain (and update) a mere handful of items in order to formulate a coherent and meaningful mental scene.

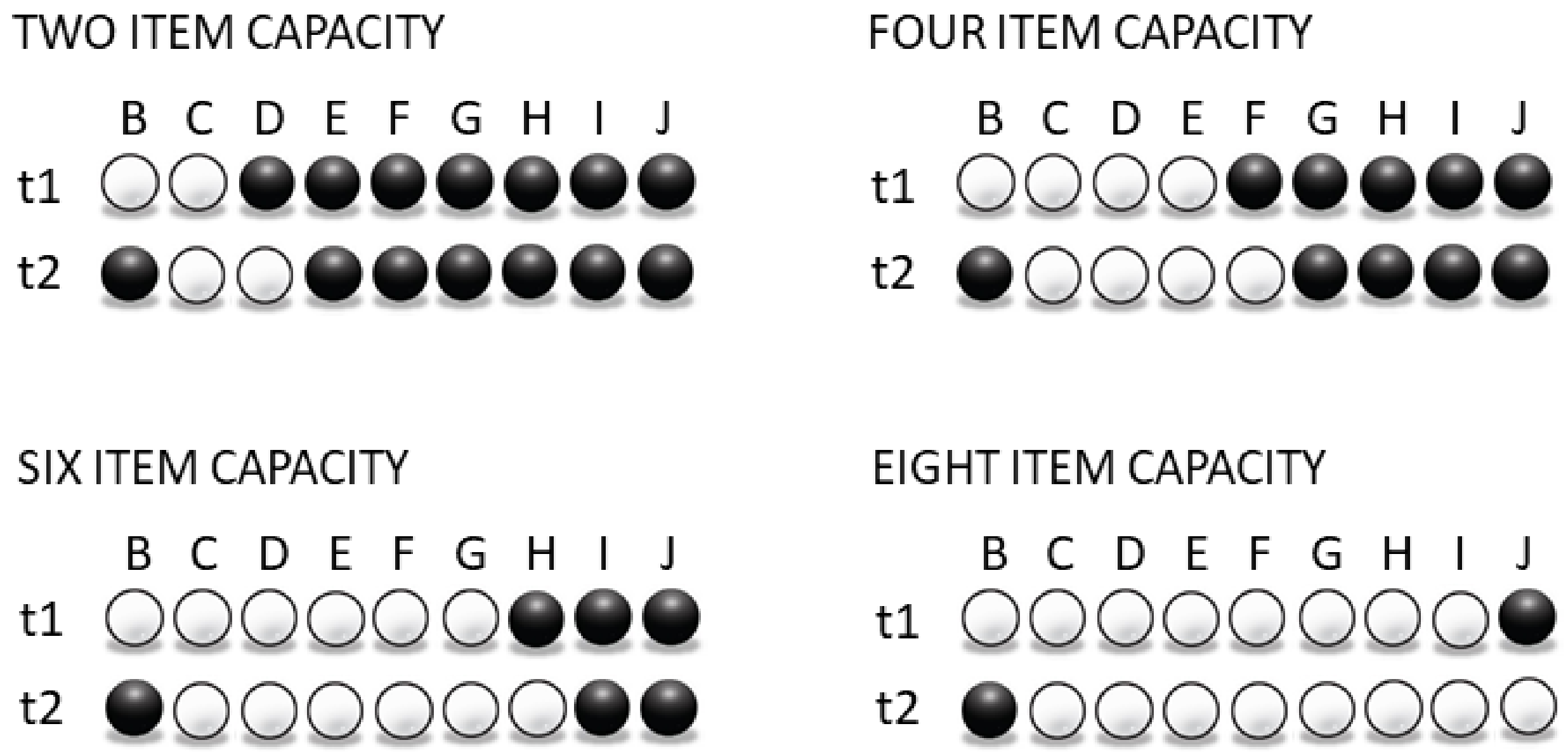

**Fig. 16.** The Capacity Limit of Working Memory Varies

*Four different examples of working memory capacity, from two to eight. The capacity of working memory can vary from task to task, person to person, and, presumably, from species to species.*

You may have noticed that at time 3 in Figure 14, item D exited the FoA before C (out of alphabetical order). This indicates that the order of entry does not determine the order of exit and, more specifically, that the items in the FoA for the longest time are not necessarily the first to be replaced. Thus, updating order can vary.

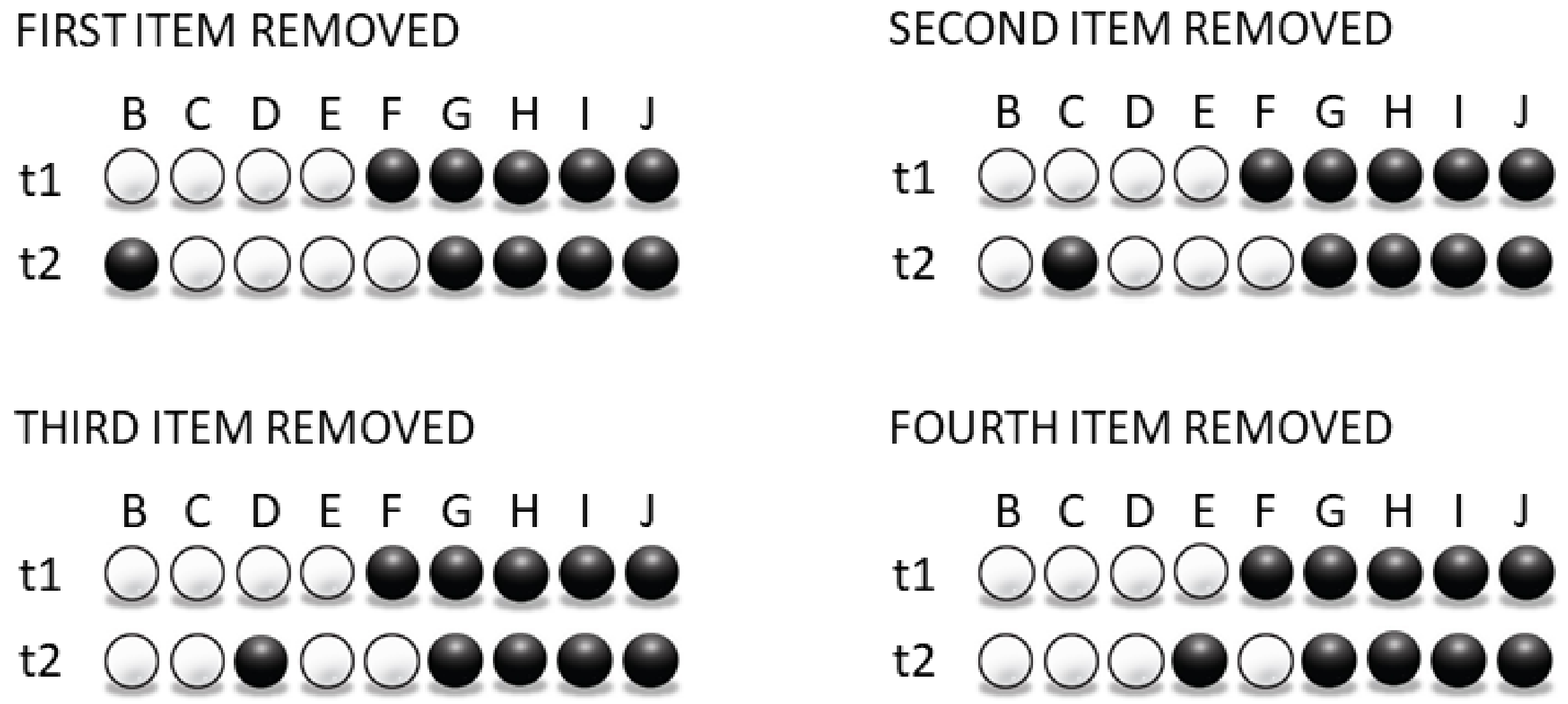

**Fig. 17.** Four Updating Replacement Schedules

*The item removed (evicted) can be the oldest entry in working memory (first in, first out), the newest entry (last in, first out), or anything in between. The least informative or relevant item should be the one selected for replacement.*

Figure 18 exemplifies how the contents of working memory can correspond to either external stimuli or internal concepts. It is probably fair to say that in all mammals, but in very few computer programs, external and internal representations coactivate and interact in real time.

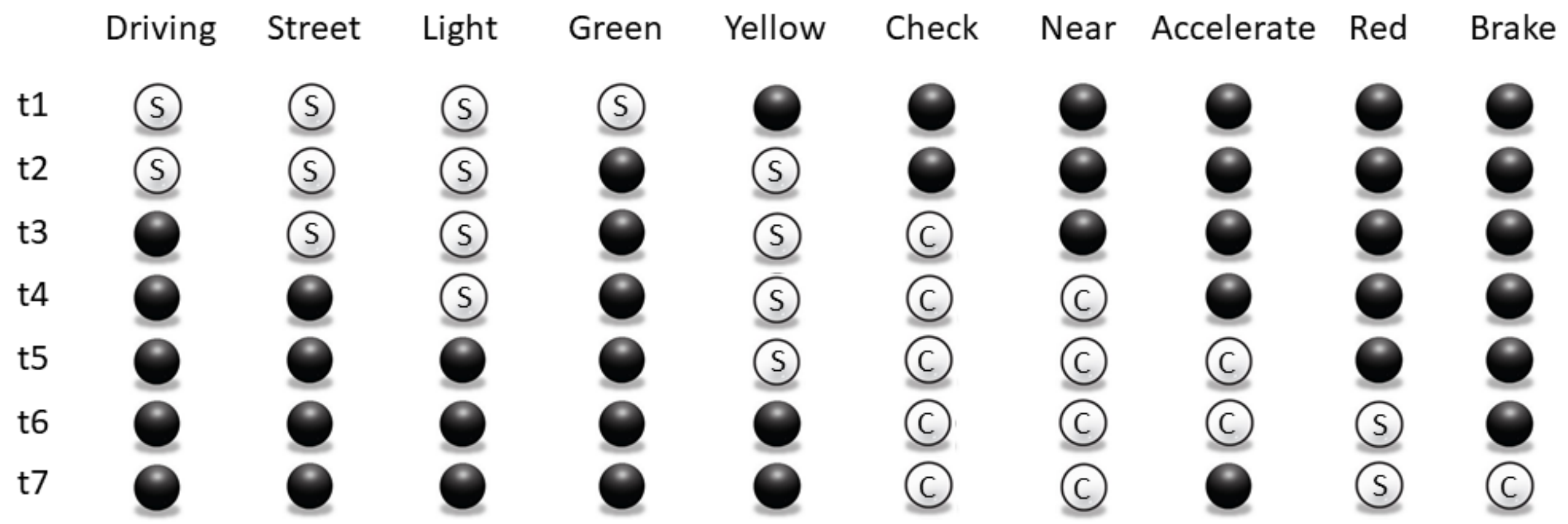

**Fig. 18.** Working Memory Responds to and Navigates Interactions with the World

*Items marked with an "S" represent external stimuli, while items marked with a "C" represent internally selected concepts. This series illustrates the working memory history of a driver responding to a green light that turns yellow. The yellow light in this context prompts the driver to check their distance to the intersection. When they find that the intersection is near, they accelerate. However, when the light turns red, this cues the driver to brake.*

### 3.3 The Rate of Iterative Updating Varies with Demand

Although Figures 14-18 depict working memory updating one unit at a time, this varies according to processing demands. For instance, when an individual pursues a new train of thought, initiates a different task, or is exposed to a novel or unexpected stimulus, their attention shifts entirely from its previous focus. When this happens, the content of the FoA can change completely. In this scenario, attentional resources are reallocated to the new context, and rather than a graduated transition, an abrupt transition occurs without iteration. Figure 19 depicts various transitions in the FoA.

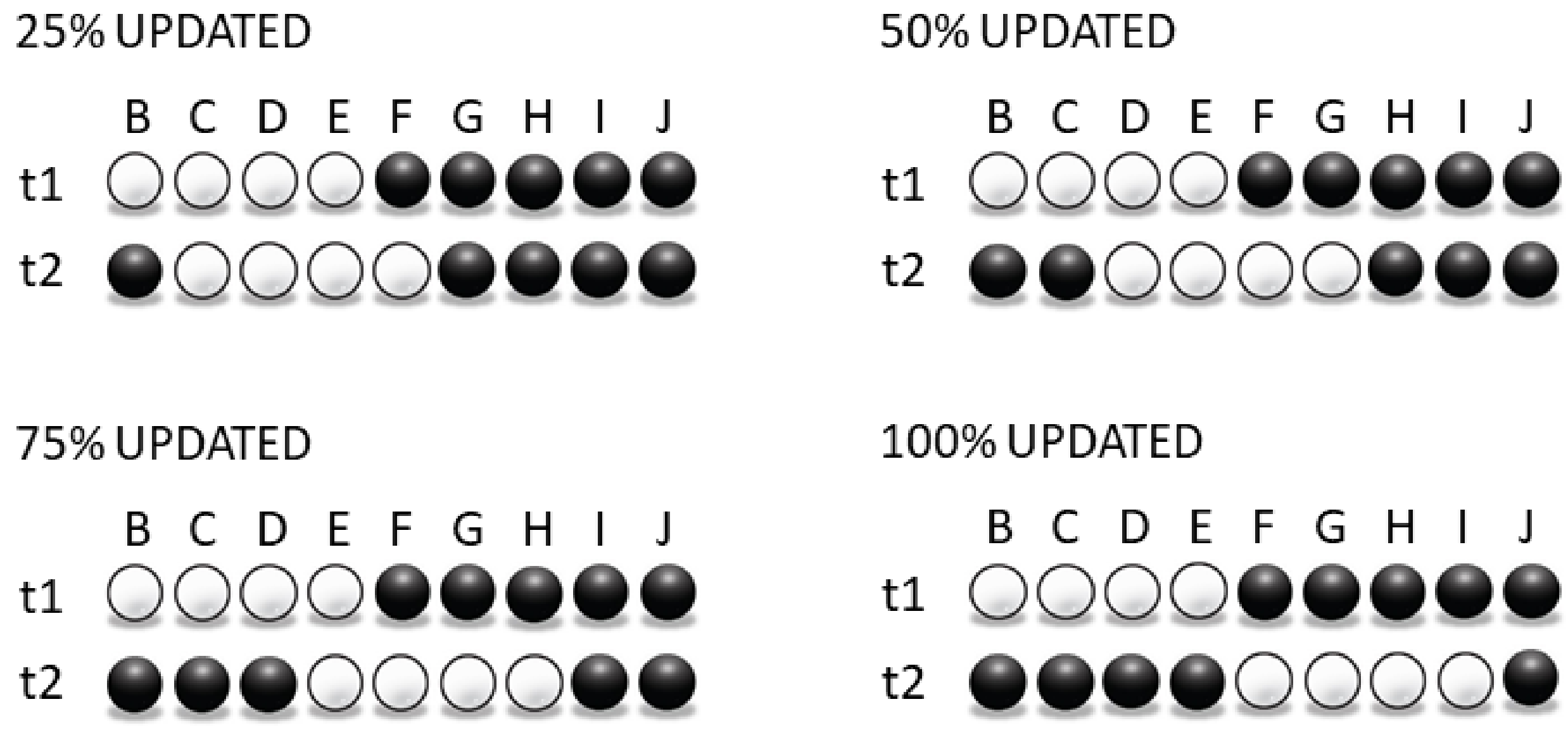

**Fig. 19.** Four Possible State Transitions in the FoA

*In the first diagram, there are four active items at time 1, which are marked as white spheres. At time 2, one of these four items has been replaced, so that one white sphere (B) becomes black (inactive) and a different black sphere (F) becomes white (active). Thus, 25% (1 ÷ 4) of the items have been updated between time 1 and time 2 without any change in the total number of active items. For clarity, most other figures in this article feature this single-item updating. However, in a store with four items, updating can occur in three other ways. The other diagrams in this figure depict 50%, 75%, and 100% updating.*

Note that abrupt, noniterative updating is not possible in the short-term store. This is due to the slower nature of turnover in synaptic potentiation. Because the number of active representations is much higher and they subside much more slowly (minutes) than in the FoA (seconds), the short-term store will continue to exhibit substantial iterative overlap, even during complete shifts in focal attention. Thus, the rate of updating from one period to the next is expected to remain relatively stable in the short-term store. In contrast, the rate of updating in the FoA is expected to fluctuate markedly under different processing requirements.

We should expect the average percentage of updating within the FoA per unit time to be lower in animals with larger, more complex brains. During mammalian evolution, association cortices were greatly enlarged relative to sensory cortices (Striedter, 2005). This development increased the number of neurons capable of sustained firing, as well as their maximum duration (Sousa et al., 2017), despite increased metabolic costs (Mongillo et al., 2008). For primates, and humans in particular, the presence of highly developed association areas likely leads to (1) more and longer sustained activity, (2) extended coactivity of items, (3) a lower percentage of updating per second, and (4) a corresponding higher degree of continuity between iterations.

In animals, a lower percentage of iterative updating might correlate with greater working memory capacity and higher fluid and general intelligence. This can be conceptualized as a longer working memory half-life. The concept of a half-life could be used to quantify the persistence of information in both the FoA and the short-term store, where generally, the shorter the half-life of activity in working memory, the shorter the attention span. For instance, the half-life for the diagram in Figure 19 exhibiting 25% updating is two time intervals, whereas the half-life for 50% updating is only one time interval.

Figure 20 addresses the decay rate using an FoA capacity of seven items. The first diagram illustrates how neural activity in small-brained animals primarily models the present and adjusts this model with bits from the recent past. The second diagram illustrates how neural activity in large-brained mammals models the recent past and adjusts it with bits from the present. In diagram 2, each new state carries more in common with the one before it, refining individual associations and enhancing the overall predictive capacity of the system.

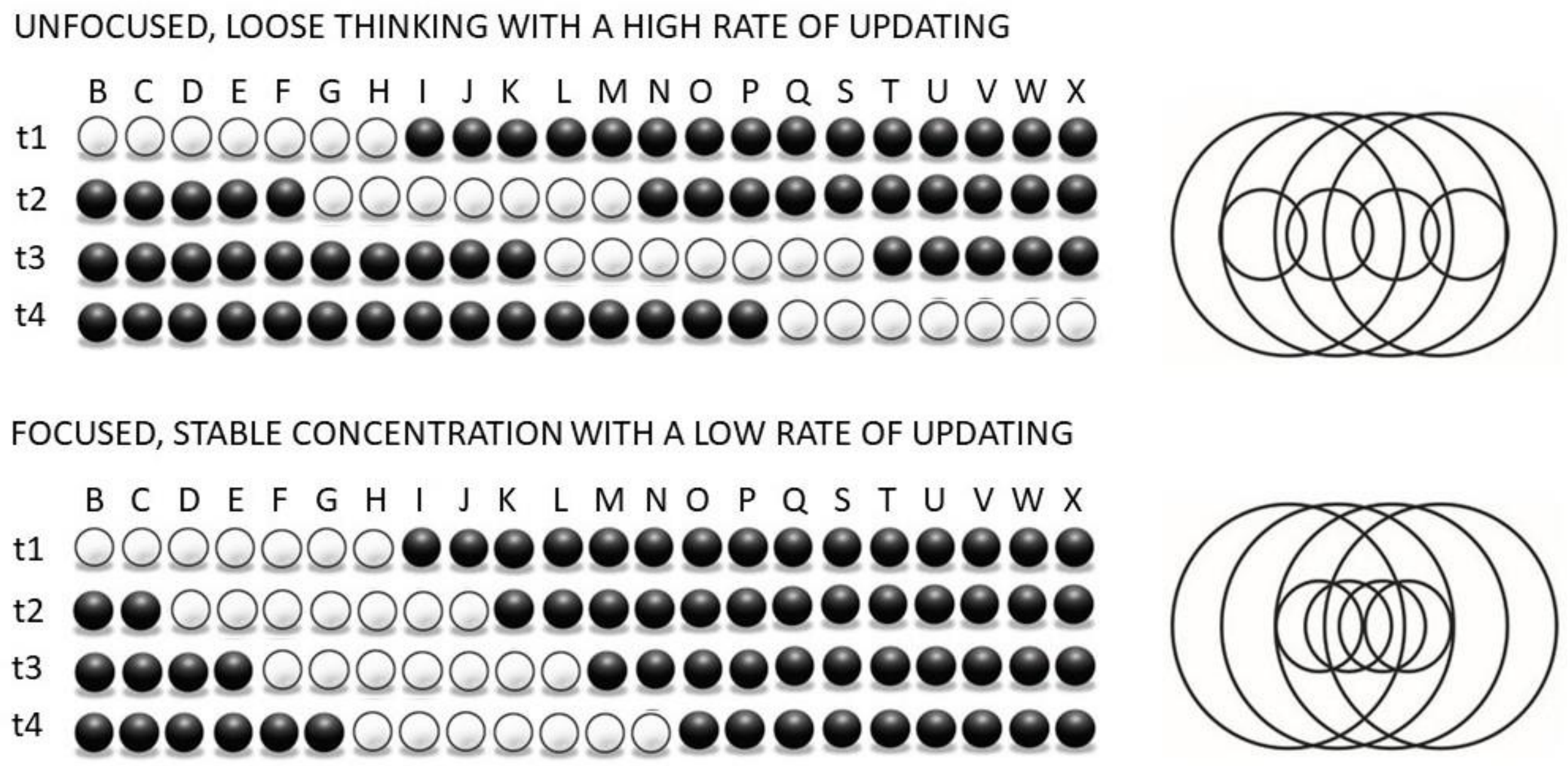

**Fig. 20.** Two Rates of Updating Carried Out Over Four Time Periods

*In the first scenario, 71% (5 ÷ 7) updating is carried out over four time periods. In the second scenario, 29% (2 ÷ 7) updating is carried out. This comparison delineates the difference between unfocused, minimally overlapping thought (loose iterative coupling) and highly focused, closely overlapping thought (tight iterative coupling). To better illustrate this point, the capacity of the FoA is depicted here as seven items after Miller (1956). The Venn diagrams to the right illustrate the percentage of iterative updating in the FoA using the style of Figure 13.*

The top diagram in Figure 20 covers a wider breadth of information, proceeds at a faster rate, and is likely more responsive and accommodating to external stimuli. However, it may be associated with an attention deficit, distractibility, and superficial associations. The bottom diagram is probably more conducive to concentrated attention, effortful/elaborative processing, and structured systematization of knowledge. This is because the search for the next state will be informed by a larger number of conserved parameters. Contrarily, in Diagram 1, more than half of the initial parameters are excluded after only one time interval because they could not be maintained. Thus, the next search performed loses precision and specificity.

For example, it should be more difficult to solve a mathematical word problem in one's head using the updating strategy depicted in Diagram 1 relative to that in Diagram 2 because too many of the problem's crucial elements would be forgotten prematurely and thus would not be available to contribute spreading activity in the search for a solution.

These two diagrams may represent the distinction not only between information processing in "lower" and "higher" animals but also between implicit and explicit processing in a single animal. Diagram 1 may be illustrative of implicit or system one processing (i.e., Kahneman's "thinking fast" (2011)) and its impulsive, heuristic, intuitive approach. Diagram 2 may illustrate explicit or system two processing (i.e., "thinking slow") in which a problem is encountered that requires multiple processing steps, recruitment of executive attention, the prefrontal cortex, and the prolonged maintenance of intermediate results. In machine learning this thinking slow approach is related to concepts like reflection, chain of thought, and intermediate reasoning tokens.

Figure 21 is meant to convey that implicit and explicit processing exist on a continuum and that implicit processing may transition into explicit when dopaminergic centers are engaged by demand, novelty, surprise, curiosity, anticipated reward, or error feedback, increasing the duration of sustained firing in the neurons that represent prioritized contextual variables.

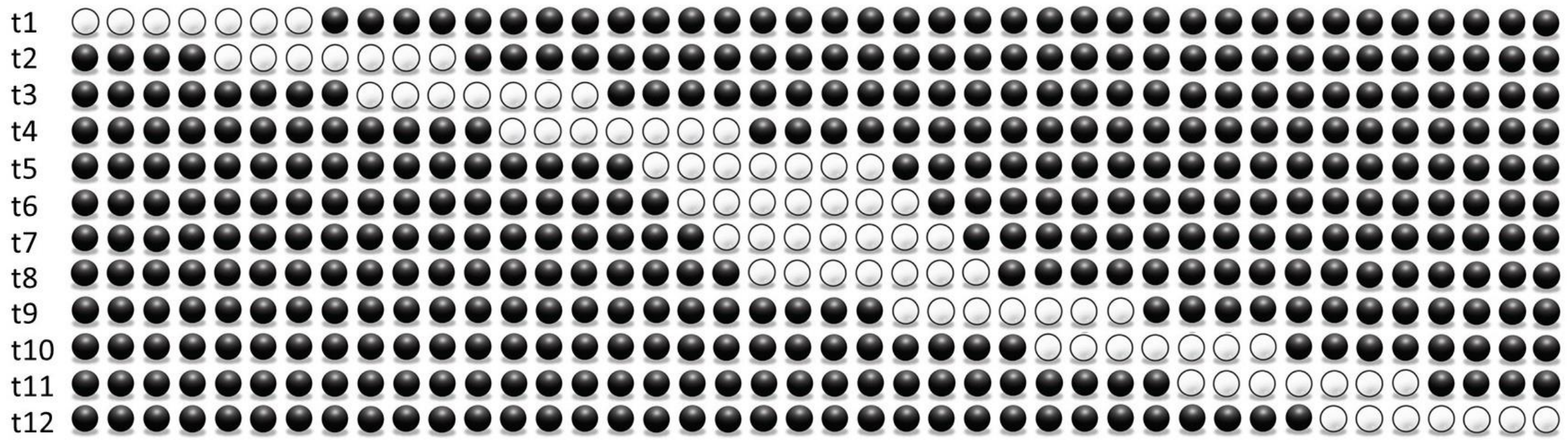

**Fig. 21.** Dopamine May Reduce the Rate of Iterative Updating in Working Memory

*A set of seven items is held in working memory. 57% (4 ÷ 7) updating is carried out over four time periods. At t5, the rate of updating is reduced to 14% (1 ÷ 7). This might happen when a person encounters a novel set of stimuli that causes the brain to release dopamine and shift from implicit processing (system one) to explicit attentive processing (system two). The activity of the items from t5 is sustained, and the concepts are anchored upon giving them more processing priority so that greater focus can be brought to bear on them. By t9, 57% updating resumes.*

As Figure 21 illustrates, it may be the case that the rate of iterative updating decreases during a thought but then increases during the transition between thoughts. The first diagram in Figure 22 features a larger number (4 vs. 2) of individual instances of continuity (i.e., discrete thoughts) compared to the second diagram. The transitions between thoughts could be conceptualized as intermittent noniterative updating. As a cognitive strategy, the processing found in the second diagram is probably more conducive to staying on topic, depth of processing, comprehending complicated scenarios, and solving complex problems. Presumably, however, animals alternate between these two strategies depending on the situation. To demonstrate a capacity for flexible thought, an AI system should have this ability, along with the abilities presented in the last six figures.

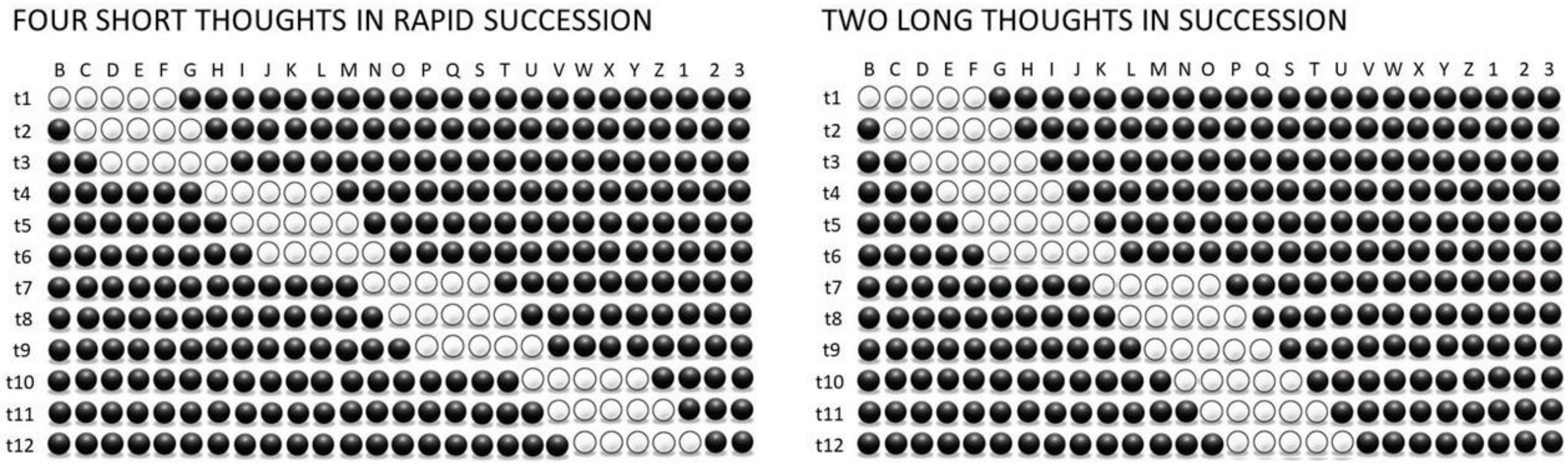

**Fig. 22.** Intermittent Noniterative Updating Marks a Boundary Between Thoughts

*In both diagrams, most of the updating occurs at a rate of 20% (1 ÷ 5). However, the first diagram shows three intermittent updates of 80% (4 ÷ 5). The second shows only one intermittent update of 80%. This comparison delineates the difference between four brief thoughts occurring in quick succession and two more prolonged thoughts. The first strategy would result in small islands of associative connections among long-term memory items. The second strategy would result in longer sequences of iterated associations and, consequently, less fragmented learning.*

### 3.4 Iterative Updating Gives Rise to Mental Continuity

Continuity is defined as the uninterrupted and consistent operation of something over a period of time. According to this model, continuity of thought involves a process in which a set of mental representations demonstrates gradual replacement across a series of processing states (Reser, 2016). Continuous, partial updating makes each mental state a reframed version of the last. This reframing process results in an updated group of conditions, modulating rather than replacing the conceptual blend created by the previous set of coactive items. The manner in which iteration permits relevant information from the past to conjoin and assimilate with information from the present may provide the connective tissue for the continuous nature of reflective thought and phenomenal consciousness. This property, portrayed in most of this article's figures, may be necessary but insufficient for machine consciousness.

A few analogies may clarify the nature of iterative continuity. When it demonstrates continuity, we should expect the attentional "spotlight" to move by degrees (e.g., the panning of a video camera) rather than abruptly (e.g., the saccade of an eye). The components within the spotlight vary smoothly. It is like the carousel function used in computer graphical interfaces where a collection of visible objects is updated as individual elements of the collection rotate into and out of view.

Mental continuity is also similar to the morphing technique used in computer animation, where an image is transformed fluidly into another by maintaining certain features but changing others in small, gradual steps. Corresponding points on the before and after images are usually anchored and incrementally transfigured from one to the other in a process called "crossfading." It is also like the changes taking place within the set of interlocking teeth of two gears. As the gears turn and a new tooth is added to this set, a different tooth is subtracted, yet other teeth remain interdigitated. In literary terms, continuity constitutes the "through line," connecting theme, or invisible thread that binds elements of a mental experience together. Mental continuity is an evolutionary process and, like natural selection, involves non-random retention and elimination of candidate structures leading to incremental modifications to a population.

In an earlier version of the present model (Reser, 2016), the subset of neurons demonstrating sustained firing over a series of states (represented by C, D, and E in Figure 14) was said to exhibit "state-spanning coactivity" (SSC). Over time, the set of coactive neurons shifts, creating "incremental change in state-spanning coactivity" (icSSC). The content of working memory is effectively in SSC, and as it progresses over time, the content exhibits icSSC. The iterative process of icSSC may provide continuity, not only to working memory but also to other constructs, such as attention, awareness, thought, and subjective conscious experience.

There are a few published articles that utilize iteration in describing various psychological phenomena (e.g., Shastri et al., 1999; Howard & Kahana, 2002; Hummel & Holyoak, 2003; Botvinick & Plaut, 2006; Kounatidou et al., 2018). However, these models are not applied to modeling continuity in brain activity or consciousness. Although modern research on these topics appears to be nonexistent (Reser, 2016), William James (1842-1910) addressed the continuous nature of consciousness in his writings. In a lecture from 1909 entitled "The Continuity of Experience," James spoke about the "units of our immediately felt life," describing how these units blend together to form a continuous sheet of experience:

> "It is like the log carried first by William and Henry, then by William, Henry, and John, then by Henry and John, then by John and Peter, and so on. All real units of experience overlap. Let a row of equidistant dots on a sheet of paper symbolize the concepts by which we intellectualize the world. Let a ruler long enough to cover at least three dots stand for our sensible experience. Then the conceived changes of the sensible experience can be symbolized by sliding the ruler along the line of dots. One concept after another will apply to it, one after another drop away, but it will always cover at least two of them, and no dots less than three will ever adequately cover it."
>
> William James (1909, p. 287)

The above quote evinces that James had envisioned an iterative model of consciousness over a hundred years ago. Moreover, his minimum of three "dots" coincides with Cowan's four (plus or minus one) items of working memory. The next section adds detail to the present account of the neural basis of the items in working memory and describes how active neurons search long-term memory for the next update.

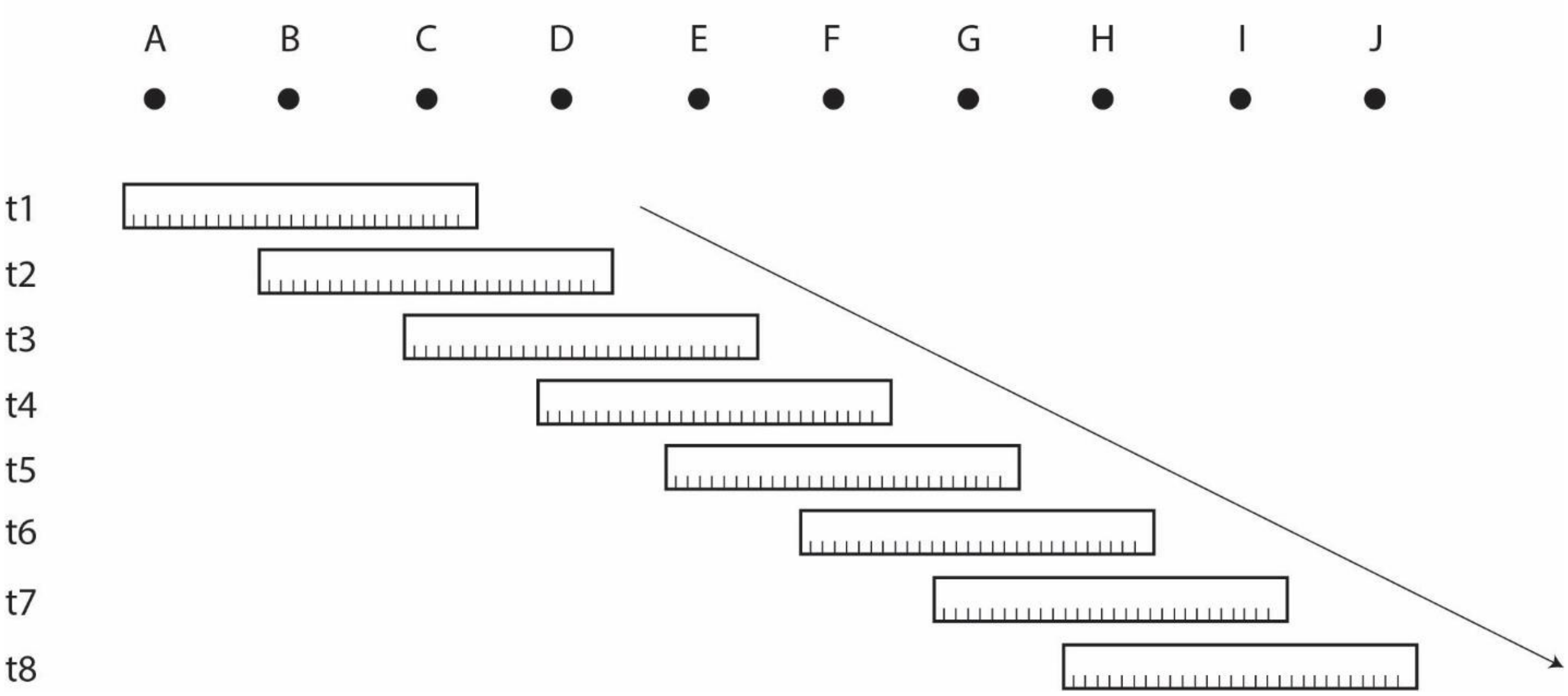

**Fig. 23.** A Representation of William James's Sliding Ruler

*This figure is meant to convey William James's ruler analogy for the overlapping units of conscious experience. The ruler encompasses a set of dots. As the ruler slides down a line of equidistant dots, the set it contains is updated iteratively.*

# **Part IV:** Implications of the Model

### 4.1 Iterative Updating Provides Structure to Associative Search

Donald Hebb (1949) first posited that a group of cells firing simultaneously could represent a memory fragment in the mind for as long as the neurons remained active. He called these groups of coactive cells "assemblies." Following Hebb's lead, many neuroscientists today describe cortical architecture as essentially a network of hierarchically organized pattern-recognizing assemblies (Gurney, 2009; Meyer & Damasio, 2009; Johnson-Laird, 1998; von der Malsburg, 1999). To recognize a complex entity, the network uses hierarchical pattern completion to locate and activate the group of assemblies that best represents the statistical function of the entity's constituent features (Hawkins, 2004; Kurzweil, 2012).

On this groundwork and that of the preceding sections, the present model proposes that the engram for an item of working memory consists of a large set (ensemble) of cell assemblies located in multimodal cortical association areas (where cells encode complex conjunctive patterns). The ensemble is symbolic, and its assemblies, like the neurons that compose them, are subsymbolic. An active ensemble of cells is not a stable, immutable group but a fuzzy set that varies every time the concept it encodes is activated (Reser, 2016). Also, the set of assemblies that make up the ensemble are likely to vary as the other contents of working memory change. Thus, an individual item in the FoA would correspond to an ensemble, a distinct subset of the total set of assemblies active in that instant. Recent studies have suggested that items may be formed from alternative processes such as dynamic population codes or low-dimensional subspace representations (Panichello & Buschman, 2021); however, we can expect even these would be updated iteratively.

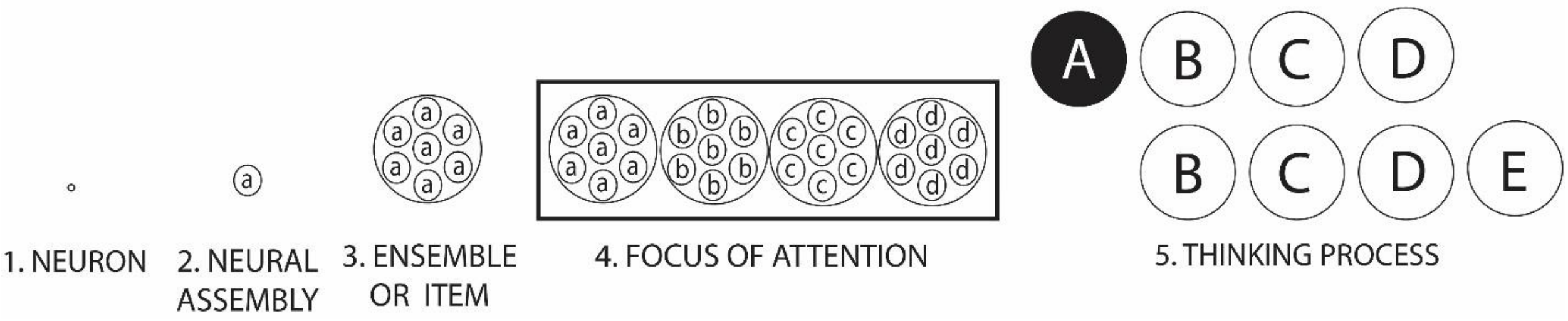

**Fig. 24.** Concepts of Interest at Different Levels of Abstraction

*1. A single neuron. 2. A neural assembly (indicated by a lower-case letter) composed of many nearby (local) neurons with similar receptive fields. An assembly (possibly a cortical minicolumn) is equivalent to a subsymbolic feature. 3. A neural ensemble composed of many nonlocal assemblies. An ensemble is an engram for an item, concept, or mental representation. 4. Four ensembles within the focus of attention of working memory. 5. Items (indicated by upper-case letters) in the focus of attention undergoing an iterative update.*

The assemblies constituting an ensemble would be densely interconnected and have strong interactions between them. They would also have the tendency to be added to (or subtracted from) working memory as a discrete group, as shown in Figure 25.

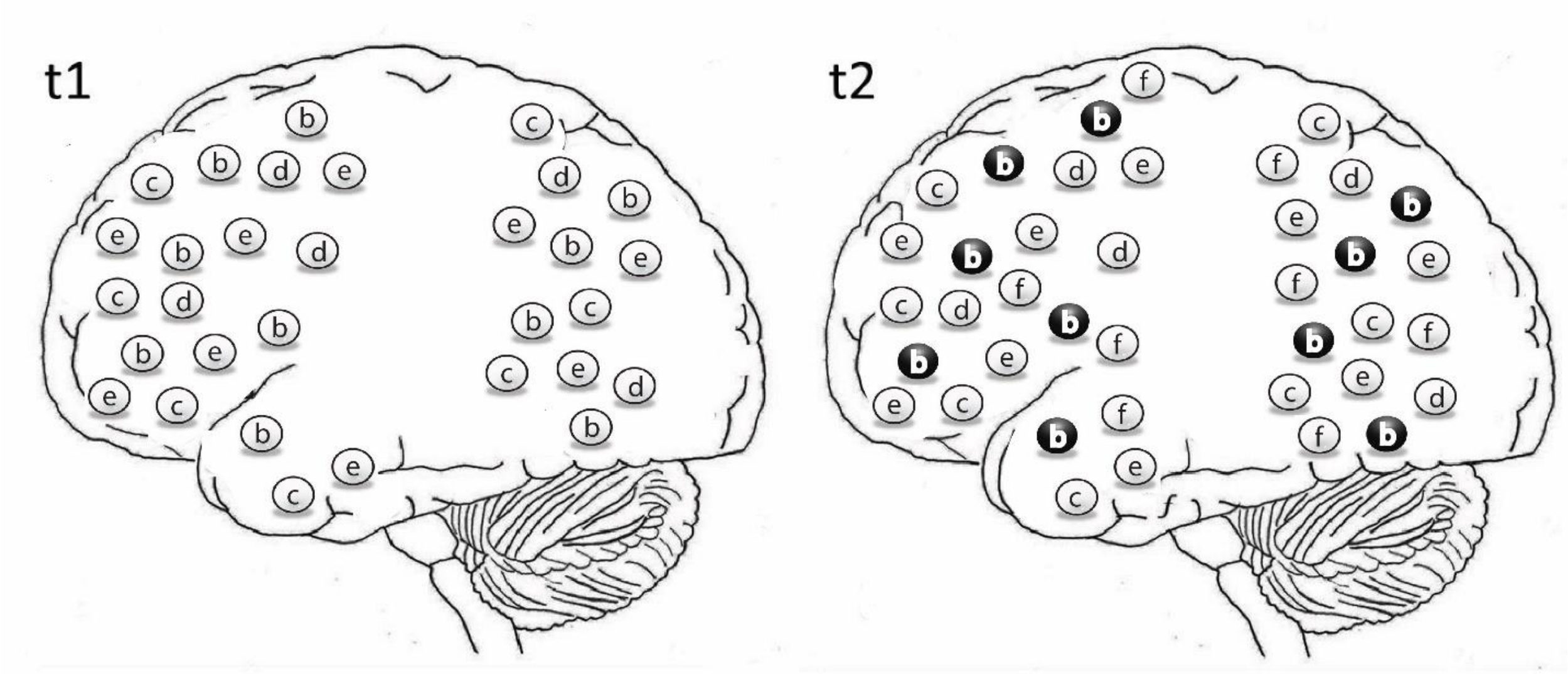

**Fig. 25.** Two Successive Instances of Coactive Assemblies in the FoA

*The ensembles for items B, C, D, E, and F are each composed of many assemblies of neurons active in association areas, represented by lowercase letters b, c, d, e, and f, respectively. At time 1, assemblies b, c, d, and e are active. At time 2, the assemblies for b have deactivated, while those for f have become active. Thus, time 2 is an iterated update of time 1.*

The primate neocortex can hold a number of contextually related items coactive for several seconds at a time. This model proposes that these items are used to perform a global search function by spreading the combined electrochemical activation energy of their neural assemblies throughout the thalamocortical network. This activation energy converges on and activates inactive items in long-term memory that are highly associated with the current state of activity. This is similar to the case where being exposed to the words “course,” “current,” “wet,” and “bank” might result in the involuntary activation of the brain’s representation for the word “river.” Hence, this model views each instantaneous state of active items in working memory as both a solution to the previous state’s search and a set of parameters for the next search.

This description of search is compatible with spreading activation theory. According to that theory, the capacity for search in associative networks is derived from activation energy (in the form of action potentials) produced by active neural assemblies (Anderson, 1983). Some of this energy is excitatory, and some is inhibitory. Activation energy from active assemblies spreads in parallel to inactive assemblies that are structurally connected to (i.e., associated with) the active ones due to a history of Hebbian plasticity (Collins & Loftus, 1975).

This activation energy propagates among assemblies through axons and dendrites and follows the weighted links of synapses. Ultimately, multiple alternative pathways originating from active assemblies converge on several of the same inactive items in long-term memory. The number of items converged on may be exceptionally large; however, not all of these can enter the FoA. The item(s) receiving the most excitatory energy is activated, becoming an iterative update to the FoA. We should expect this to be the concept most strongly psychologically associated with the items that converged upon it.

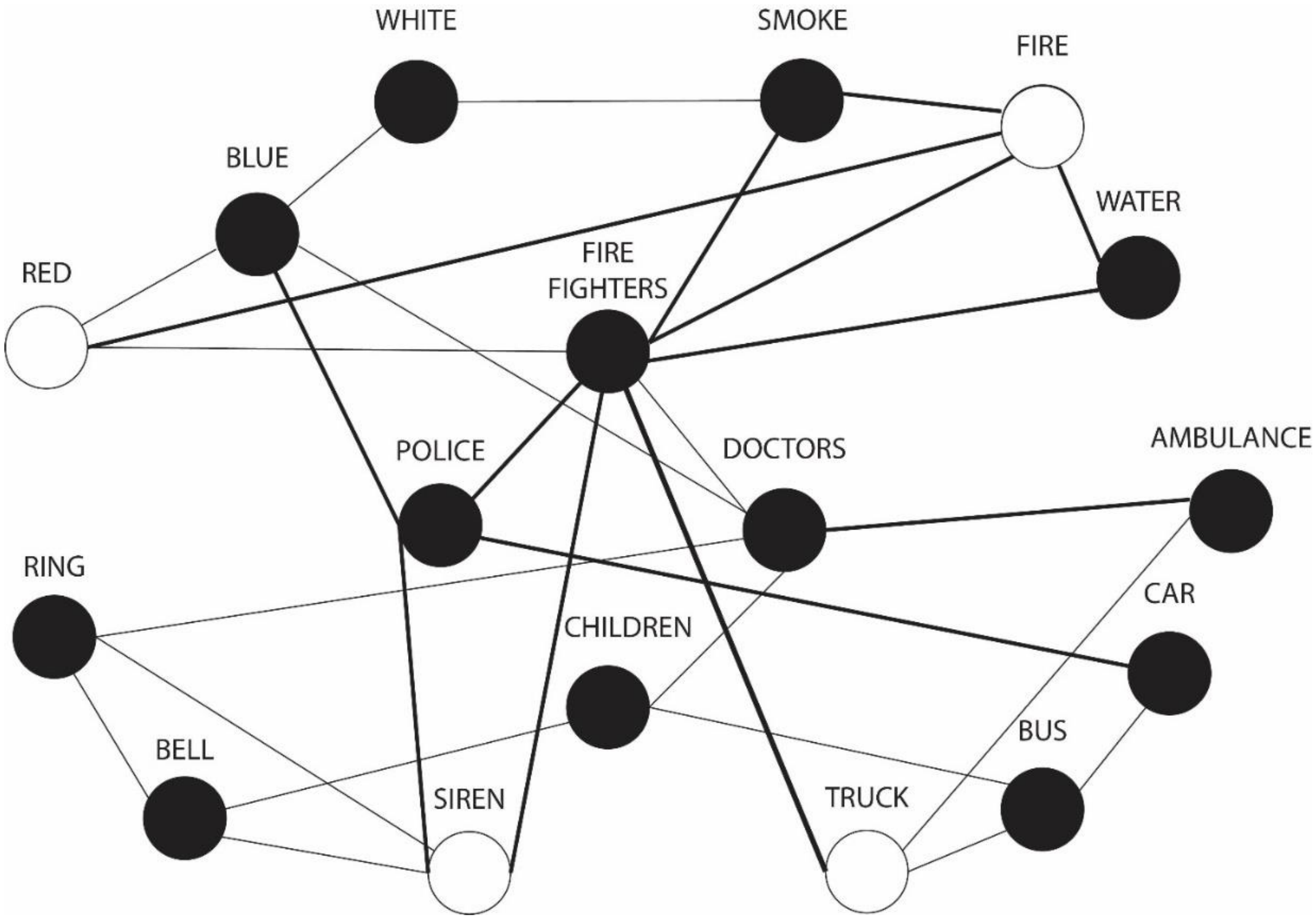

**Fig. 26.** A Semantic Web Showing Associative Connections Between Concepts

*Seventeen concepts (ensembles) are represented by circular nodes in a long-term memory network. The existence of an associative connection is indicated by a line, and its strength by the thickness of the line. The four items "red," "truck," "siren," and "fire" are currently coactive in this network, and thus spreading activation is likely to select "firefighters" as the next iterative update.*

Studies of semantic priming show that either conscious or subliminal exposure to a brief stimulus can temporarily increase the implicit availability of many associated concepts within long-term memory (Bargh & Chartrand, 2000). For instance, in a lexical decision task, merely priming the word "water" will speed up the recognition of various related words such as "fluid," "splash," "liquid," and "drink" (Schvaneveldt & Meyer, 1973). The standard interpretation of these findings is that activating the engram for "water" unconsciously spreads to the engrams for many semantically related words. This activation is rapid, automatic, irrepressive, and is theorized to be due to spreading activation in associative networks (Reisberg, 2010).

It may be reasonable to assume that updating working memory with a new item has a similar priming effect on spreading activation. This new item, added to the residual items, acts as an additional semantic retrieval cue, uniquely altering the field of items receiving activation. By assuming that updates to working memory are selected by its current contents, one can explain why new associations are marked by high contextual relevance and specificity. Combining this assessment with the earlier claims regarding iteration results in a system suited for producing a parade of complementary impressions, views, notions, and ideas.

### 4.2 Multiassociative Search Spreads the Combined Activation Energy of Multiple Items

"Associationism" is a longstanding philosophical position advocating that mental states determine their successor states by psychological associations between their contents. According to associationism, the sequence of ideas a person produces is essentially a matter of the preexisting links between stored associative memories (Shanks, 2010). William James believed one thought could induce another through a logical, correlative connection (1890). The face validity of associationism stems from the commonplace notion that one thought "suggests" the next.

In his discussion of "the succession of memories," Plato identifies three principles of association: similarity (resemblance), contiguity (in time and place), and contrast (difference). Numerous other principles capable of linking mental states were added to this list by the nineteenth century, including simultaneity, affinity, reinstatement of the remainder, cause and effect, reason and consequence, means and end, and premise and conclusion (Hamilton, 1860). When any of these forms of association occur, they may simply involve iterative updates, selected by spreading activation, joining a global workspace of persistent items. Spreading activity operating in this way may help us reconcile incongruous items by locating the most compatible update. For example, it may help us find solutions to problems such as the trivia prompt, "The name of a planet, a flowering plant, and a Roman goddess," where each of the clues contribute independently to unconscious neural convergence onto the ensemble representing the construct of "venus" (Reser, 2016).

The associationism school of thought primarily focused on a single logical relationship between one thought and another. This may provide only a limited explanation. The model presented here can be read as a version of associationism that escapes this limitation by assuming that all the neurons currently involved in working memory search cooperatively and probabilistically for the succeeding association. Thus, contiguous states are not only interrelated but are also interdependent. This cooperative search function may occur regardless of when the neurons started firing and irrespective of the item to which they belong.

Reser (2016) proposed that the selection of new items to be added to working memory might derive from the pooling of assembly activity in the cortical workspace. This unconscious, autonomous process, here termed "multiassociative search," operates as detailed in Table 4 below.

| Term | The Multiassociative Search Algorithm |
|---|---|
| | As excitatory and inhibitory activation energy from assemblies representing the items currently in working memory spreads, |
| **1** | items that continue to receive sufficient activation energy remain active, and |
| **2** | items that receive sufficiently reduced activation energy lose activity, and |
| **3** | inactive items that receive sufficient activation energy become active, and |
| **4** | the newest additions to the focus of attention spread activation energy to some items in working memory more than others, differentially increasing those items' capacity to contribute to the selection of the next item, and |
| **5** | the newest additions to the focus of attention spread activation energy to the items in working memory altering the composition of the fuzzy set of assemblies that compose each item, adjusting the meaning of each item based on context, or |
| **6** | when activation energy drops below a given threshold, all items in the FoA are deactivated and a new set of items from the most potentiated previous thread is made coactive and iteration resumes. |

**Table 4.** The Steps of in Multiassociative Search

*Processes involved in multiassociative search whereby spreading activation disseminates through the neural network resulting in the iterative updating of working memory and the thought cycle.*

The item(s) receiving sufficient activation energy (through spatial and temporal summation) from both the present constellation of coactive assemblies (FoA) and potentiated synapses (STM) may be recalled autoassociatively (i.e., an active subset of the item's assemblies is sufficient to activate the rest of the item). This nonlinear, stochastic process should be taken to be responsible not only for finding and activating the next item(s) (Fig. 27) but also for determining the percentage of items updated in the FoA (Figs. 19, 20, and 21).

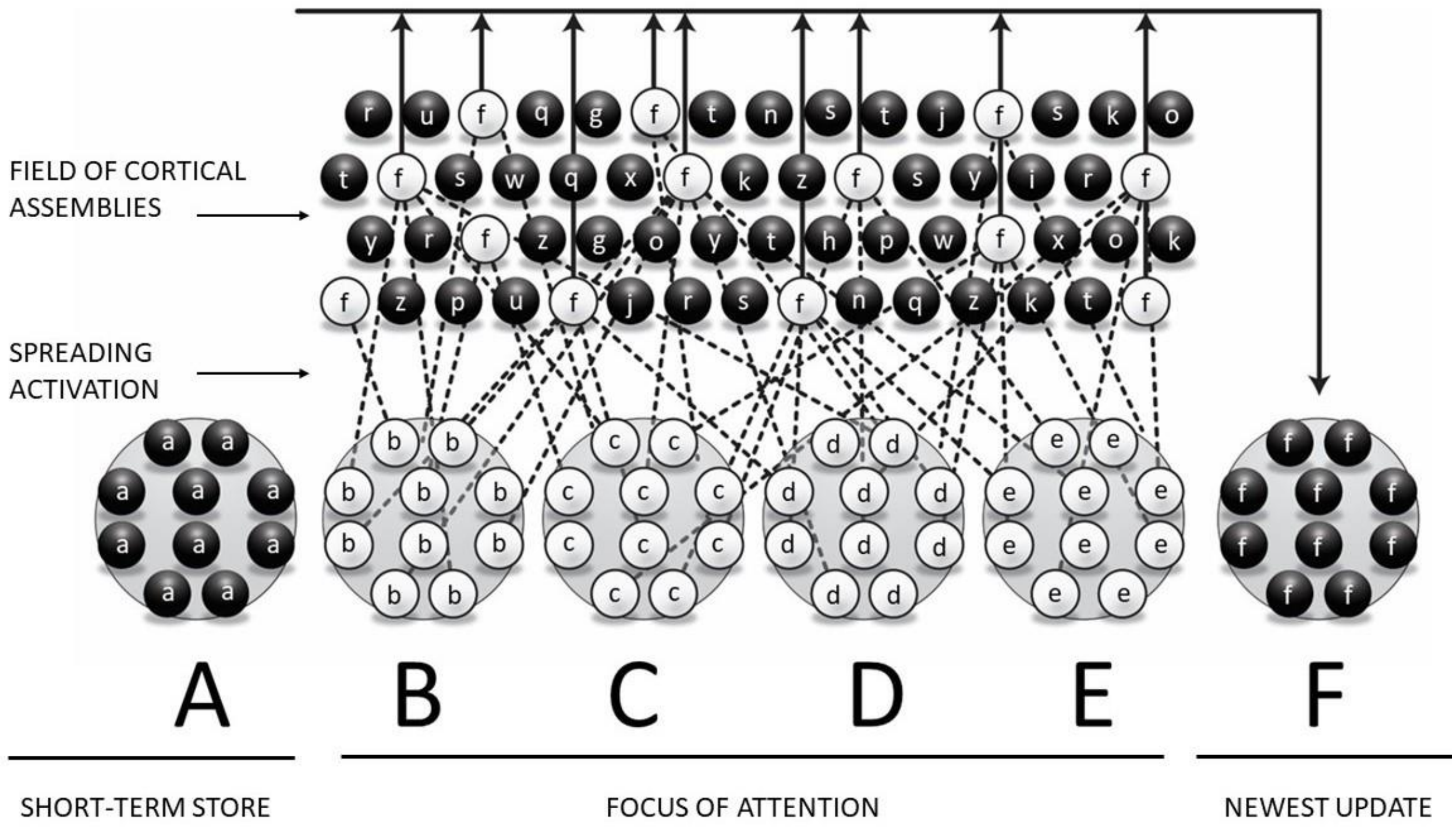

**Fig. 27.** A Schematic for Multiassociative Search

*Spreading activity from each of the assemblies (lowercase letters) of the four items (uppercase letters) in the FoA (B, C, D, and E) propagates throughout the cortex (represented by the field of assemblies above the items). This activates new assemblies constituting the newest item (F), which will be added to the FoA in the next state. The assemblies that constitute items B, C, D, and E are each individually associated with a very large number of potential items, but as a unique group, they are most closely associated with item F.*

This procedure is executed unceasingly during waking consciousness. It takes numerous neural assemblies that have never been coactive before and uses their collective spreading activation in parallel to select the most applicable iterative update. At every moment, the set of assemblies in coactivity is unprecedented. However, the set of items in coactivity may not be. When the set of items coactive in the FoA has been coactive at some point in the past, the spreading activity either converges on the same item that was selected the last time (recall) or it may converge on an altogether different item (novel inference). Regardless of which way this occurs, the process transforms the latent information inherent in the original set into new, manifest information by forcing it to interact with inert long-term memory. Each set of coactive items and the links between them can potentially be recorded to memory. Thus, every search creates new associative learning in the network, forming new boundaries and affinities between groups of items, and incrementally improving the animal's future behavior and model of reality. Thus, multiassociative searching gives rise to multiassociative learning.

These concepts explain how long-term (non-hippocampal) semantic memory might be updated. New memories don't replace old ones; rather, they retune the connectional strengths between groups of items. For instance, in Figure 27, the associative relationship between F and B is strengthened, but mostly in the presence of items C, D, and E. As items demonstrate coactivity within working memory, we should expect their assemblies to exhibit a Hebbian propensity to wire together, forming statistical codependencies that support the learning process. Reoccurring examples of coactivity would lead to the formation of heavily encoded associations (Asok et al., 2019), which would persist as procedural and semantic knowledge.

Beyond the first three steps of multiassociative search above, step four provides for more flexible and dynamic attentional control. Essentially, in the context of the newest item, some items in working memory are activated even more because their associative connections deem them more relevant. In the transformer architecture something similar is accomplished by self-attention where query-key similarity calculations redistribute the attention weights across the sequence allowing some tokens to contribute more strongly to determining the probability distribution used to predict the next token.

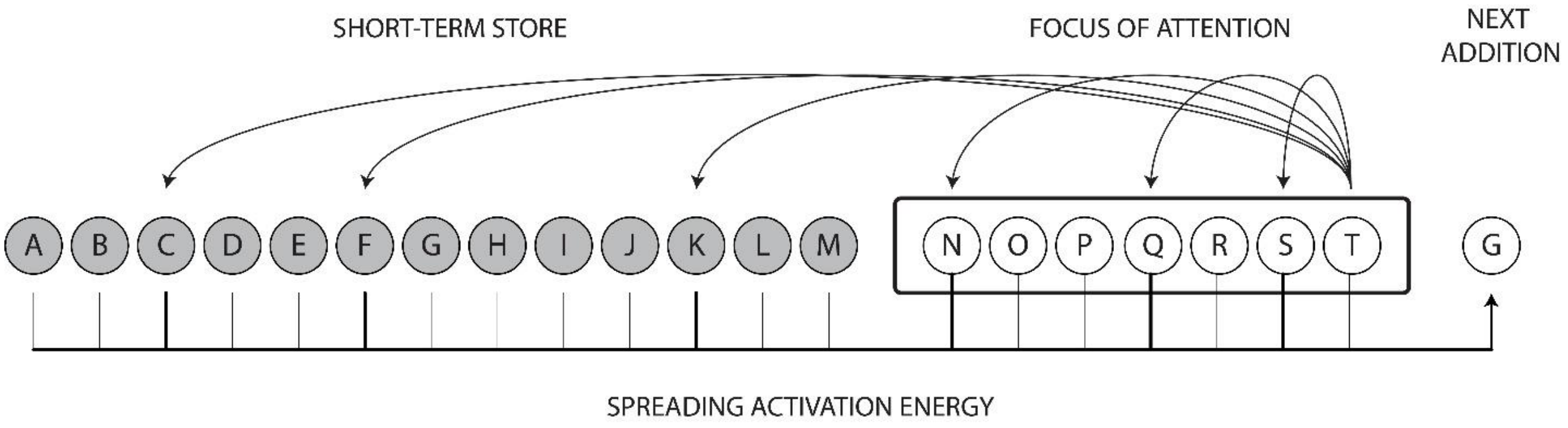

**Fig. 28.** Spreading Activation Weights are Redistributed Across the Sequence

*This figure illustrates the process introduced by step four of multiassociative search above. Each new addition to the FoA (item T) changes the distribution of activation energy. It spreads more activation energy to certain items in the focus of attention and short-term store than others (shown by the arcs). This differentially increases these items' capacity to contribute to the selection of the next item (indicated by line width below the items).*

Undoubtedly, many canonical information processing algorithms not mentioned here (e.g., Miller et al., 2018; Sreenivasan, 2019) also contribute to this search and play causal roles in this process. However, it may be parsimonious to assume that the subsymbolic components of the symbolic items in working memory work synergistically and in parallel to search for the updates to working memory in the way described. In other words, the production sequence of thought is not determined by semantic dependencies between symbols (e.g., rules, utilities, predicates, conditionals, functions, etc.) as in other cognitive architectures (e.g., ACT-R, Soar, Sigma, etc.). Instead, it is determined by syntactic dependencies among subsymbols. These dependencies may reconcile with declarative, symbolic knowledge at the psychological level. Nonetheless, they operate unconsciously below it. In other words, the outcomes of these "blind" statistical searches only appear rational because they are based on a history of structured learning from orderly environmental patterns.

Note that, in the present model, the cortical assemblies constituting items currently in the FoA are not the only contributors to selecting the next item(s). Rather, all firing neurons that participate in the spreading of activation in the cortical workspace contribute definitions to this global search. Potentiated neurons in the short-term store—as well as active neurons in sensory and motor cortex (semantic), hippocampus (episodic), basal ganglia (procedural), and other cortically connected subcortical neurons—all contribute to multiassociative search. Figure 29 depicts this situation, in which a working memory store characterized by iterative updating selects its updates using spreading activation generated by several different neuroanatomical systems. The next section will discuss how iterative updating and multiassociative search work together to formulate not only associations but also predictions.

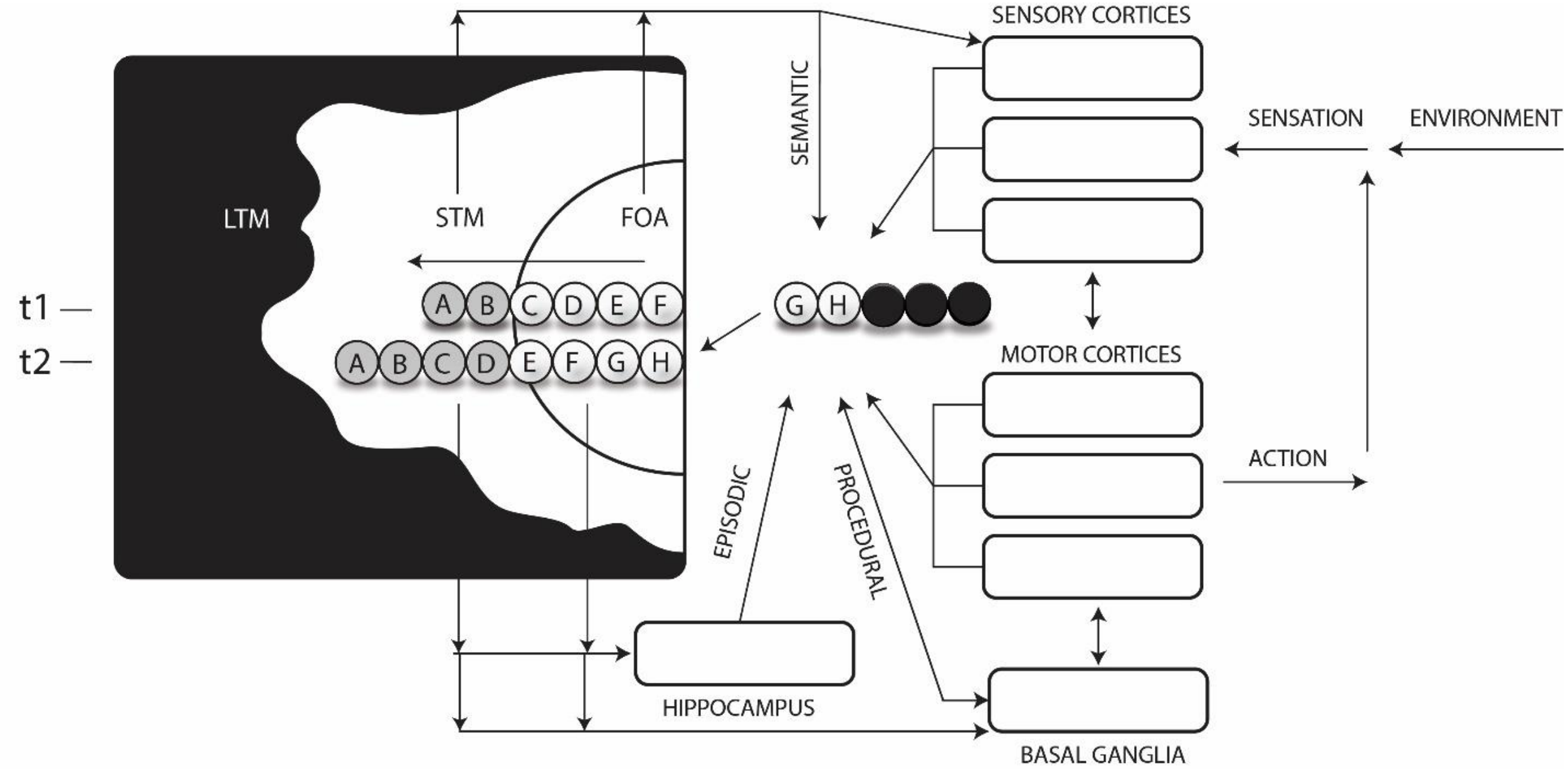

**Fig. 29.** A Single Cycle of the Iterative Updating / Multiassociative Search Procedure

*The FoA, the short-term store, as well as active neurons in the hippocampus, basal ganglia, sensory and motor cortices all contribute to the spreading activation that will select the next item(s) to be added to working memory. At time 1, two (G and H) of a potential five items are converged on, and these update the FoA in time 2.*

### 4.3 States Updated by the Products of Search Are Predictions

The meaning of an event is determined by the events that came before it and by those that will come after it. Claude Shannon, the founder of information theory, knew this and was interested in predicting events based on their foregoing context. He introduced a hypothetical situation in which a person is tasked with guessing a randomly selected letter from a book (Shannon, 1951). Because there is no contextual information available, any response would be highly uninformed and should resort to choosing the most frequently occurring letter in that language, which in the English language is "E". However, if this person is given the letter that comes before the unknown letter, a more informed guess can be made. The more previous letters known, the better the guess (Stone, 2015). For instance, if you knew the sequence of letters that precede an unknown letter was “T,” “H,” “I,” and “N,” then you would know that there is a high probability that the next letter is either “K” or “G.”

As with letters in a word or words in a sentence, events occurring along a timeline in a natural environment are not independent or equiprobable. Rather, there are correlations and conditional dependencies between successive events. Knowledge of conditional dependencies allows us to predict what other people will do next and, sometimes, to even finish their sentences for them. Working memory enables the mind to capture and record long-term dependencies. This, in turn, permits animals to treat separate events as causally related variables that can be used to predict future events. By capturing the statistical structure of a sequence of recent occurrences (including rewards and punishments), working memory provides animals with a way to form an autoregressive interpretation of an unfolding scenario, forming associative expectations of it and responses to it.

The interaction between iterative updating and multiassociative search may form the basis for prediction in the brain. Consider the case in which four environmental stimuli present themselves in quick succession. This could involve any sequence of events, such as that involved in finding food. If each stimulus is attended to and persistently activated, the items representing these stimuli and their closest associations will have the chance to comingle in the FoA. Their coactivity may cause them to become associated by activity-dependent learning even though they never actually occurred simultaneously in the environment (trace conditioning). If this sequence of four stimuli is repeated frequently (as would be expected if there were conditional dependencies between them found in nature), then they will come to be strongly associated. The next time the first three stimuli appear, their very activity may be sufficient to search for and recruit the item representing the fourth stimulus from long-term memory. Consequently, the activation of this fourth item would be a prediction. Therefore, internally generated, self-directed thought can be conceptualized as an iterative procession of concatenated, associative predictions, each predicated on the prediction before it.

In Figure 30, Diagram 1 depicts a situation in which stimulus 1 (S1) is followed by stimulus 3 (S3) and results in the selection of response 1 (R1). This can be contrasted with Diagram 2, where S3 is preceded by a different stimulus (S2), and a completely different response is selected (R2). The persistent activity of the first stimulus influences the interpretation of S3, biasing the response accordingly. That is to say, the response to S3 is conditionally dependent on the stimulus that precedes it. This situation is depicted in diagrams 1 and 2 that have been adapted from a highly popular model of PFC function (Miller & Cohen, 2001). Diagrams 3 and 4 take this idea further communicating that when the first two stimuli are the same (S1 and S2) but the subsequent stimulus differs, the responses may also differ. These diagrams underscore the hypothesis that behavior is not merely directed by the differential selection of existing neural pathways underlying stimulus-response pairings (i.e., Miller & Cohen, 2001) but rather by a series of multiassociative searches that utilize curated sets of memoranda to converge on the best response at each time step.

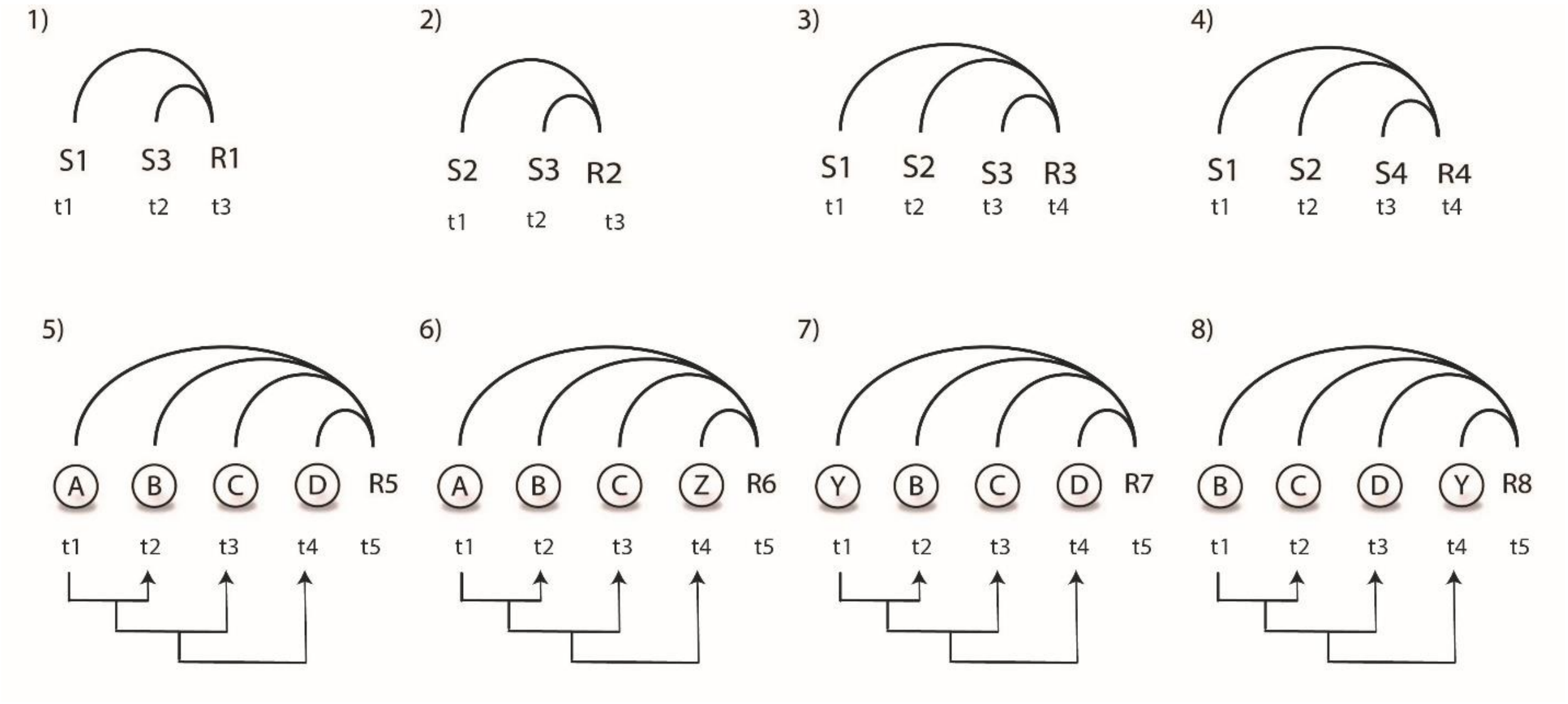

**Fig. 30.** Conditional Dependencies Between Consecutive Events

*Each arc represents the span of time since an event occurred (as in Figure 12). S represents stimuli, R represents responses, and other capital letters represent items. To provide an illustrative example, let us suppose the variables in Diagrams 1 through 4 correspond to the following events: S1 = friend, S2 = enemy, S3 = approach, S4 = depart, R1 = act friendly, R2 = act aggressive, R3 = wait, R4 = follow. In Diagrams 5 through 8 the items could correspond to: A = foraging alone, B = feel hungry, C = find berries, D = not poisonous, Z = poisonous, Y = friend approaching, R5 = eat, R6 = don't eat, R7 = share berries, R8 = eat berries before friend arrives.*

Diagrams 5 through 8 communicate that the full complement of items in working memory can be expected to show a pattern like that seen with the stimuli in Diagrams 1 through 4: each item affects the interpretation of the items after it and uniquely biases the search for a response to them. Accordingly, the arrows below Diagrams 5 through 8 indicate that, at each time step, the preceding items provide a frame of reference through which subsequent items are interpreted. Note that even though the responses in Diagrams 7 and 8 are reacting to the same four representations (B, C, D, and Y), they react to them differently because the order of items contextualizes the scenario differently. For instance, item C has a different meaning (dependency) when it follows Y versus when it precedes Y. Therefore, it elicits a different predictive response.

Consider a situation in which a person writes with a pencil, and the lead breaks. This may cause the long-term memory representations for "writing," "pencil," "lead," and "broken" to become active in the FoA. This combination of coactive items (conditioned from years of writing with a pencil in school) might result in the automatic spreading of activation to the representation for a "sharpener." During another round of updating, the representation for "writing" may exit the FoA and be replaced by the pencil sharpener's location, such as "desk drawer." In this way, sets of coactive items can prompt others in advancing sequences capable of producing not only predictions but also adaptive behavior. Especially when the items being sustained are task-relevant, this kind of iterative system should be capable of incremental advancement toward a goal.

### 4.4 Iterative Updating Allows Progressive Changes to the Contents of Thought

In addition to accounting for the serial, cyclic, continuous, narrative, and predictive functions of thought processes, iterative updating may be a fundamental feature of reasoning. This section will provide brief explanations for why this might be the case.

According to the present model, iterative updating produces sequences of interdependent states in which each state is capable of representing the current status of a problem-solving procedure and updating it with a prediction. When associative prediction is informed by meaningful causal dependencies learned from related experiences, it sets the cycle on a logical course. This makes it possible for a starting state to generate a chain of intermediate states that make progress toward a terminal solution state.

Iterative updating allows working memory to link a series of rapid, automatic associations so they can furnish a foundation for each other, resulting in the assembly of complex content. This occurs when a series of linked searches culminates in a higher-order result that any single search on its own could not otherwise attain. A prolonged stretch of tightly recursive searches (where a large proportion of items are retained throughout several states, as in Figure 20, Diagram 2) may be slower and more error-prone but can address problems too unfamiliar or complicated to be solved by less iterative, implicit processing.

Generally speaking, short bouts of iteration engage crystallized intelligence and easy-to-reach network states, whereas instances of prolonged iteration use fluid intelligence and highly processed, difficult-to-reach states. Such highly elaborated states are comprised of select subsets of previous states from various points in the recent past. This corresponds to simple thoughts building constructively “on top of” each other to form complex thoughts. William James used the term “compounding” to describe this concept:

> “…complex mental states are resultants of the self-compounding of simpler ones…. in the absence of souls, selves, or other principles of unity, primordial units of mind-stuff or mind-dust were represented as summing themselves together in successive stages of compounding and re-compounding, and thus engendering our higher and more complex states of mind.”
>
> William James (1909, p. 185)

Iterative updating may employ this compounding feature during logical or relational reasoning. The item or items that update the FoA create a context to be compared, contrasted, integrated, or otherwise reconciled with the context remaining from the previous state. This may be the same kind of reconciliation that occurs in formal logic. For instance, propositional logic combines simple statements using logical operators (subjects and predicates) and connectives (e.g., and, or, not, if, then, because, etc.) to produce complex rational statements. The operators of such a statement could be instantiated by items. Items serving as an update could form a true statement that could invoke a related premise or lead to a conclusion. By creating strings of substantiated inferences in this way, multiassociative search could permit the construction of a logical case or argument and build expectations about events that have never been encountered.

The compounding feature of iteration may also enable working memory to formulate and implement algorithms. All complex learned behaviors have algorithmic steps that must be executed in a specific sequence to reach completion (Botvinick, 2008). Activities such as hunting, tying shoes, and performing long division involve following an algorithm. Successive states of working memory could correspond to successive steps in an iterable process.

Iterative updating could be instrumental in implementing learned algorithms because virtually every step of an algorithm relates to the preceding and subsequent steps in some way. A new update could correspond to a mental operation required in the next step of the sequence of actions that need to be taken (i.e. a response, memory, or heuristic). Thus, updates represent the most applicable step at each state of solving a routine or non-routine problem. An item inhibited or allowed to subside could correspond to an operation from the algorithm that has already been executed or is no longer needed.

Once the associations relevant to an algorithm have been trained is sequence, multiassociative search can recruit the items necessary at each step (Reser, 2016). For instance, performing long division by rote requires many trials, and proficiency may only be reached when the active items in each state have been trained to converge on the items necessary to perform the operation in the next step. After the first digit of the dividend is divided by the divisor, the prevailing state of working memory automatically activates the items necessary to take the whole number result and write it above the dividend. Cognitive algorithms may be constructed in this manner during learning, as trial, error, and repetition link recursive chains of states into functional behaviors. This form of optimization may use operant conditioning to provide feedback for incremental guess refinement.

It is also conceivable that iterative updating plays a significant role in the generation of schemas and mental models. Mental models are internal representations of external systems and the relationships between their parts (Cheng & Holyoak, 2008). Iteration may afford the incremental modification of a model from its previous state, allowing capacity-limited, static models to be elaborated dynamically. Even complex systems can be modeled when their enduring features are held constant by persistent activity and the changing features are updated correspondingly. This enables tweaking the search parameters of interest to vary the simulation in stages, producing a systematic effort to investigate a structured problem space.

Solving a complex multistep problem requires the FoA and short-term store to cooperate. For instance, one line of thought developing in the FoA may be temporarily suspended in the short-term store so that the FoA can be made available to solve a related subproblem. The FoA would iterate, progressing toward the solution to this subproblem. When the subsolution is reached, it could then be merged with the pending problem to create a hybrid solution state. This interleaving and eventual merger of states of progress would facilitate the decomposition of a problem that is too computationally taxing to be processed by the FoA alone (Fig. 31).

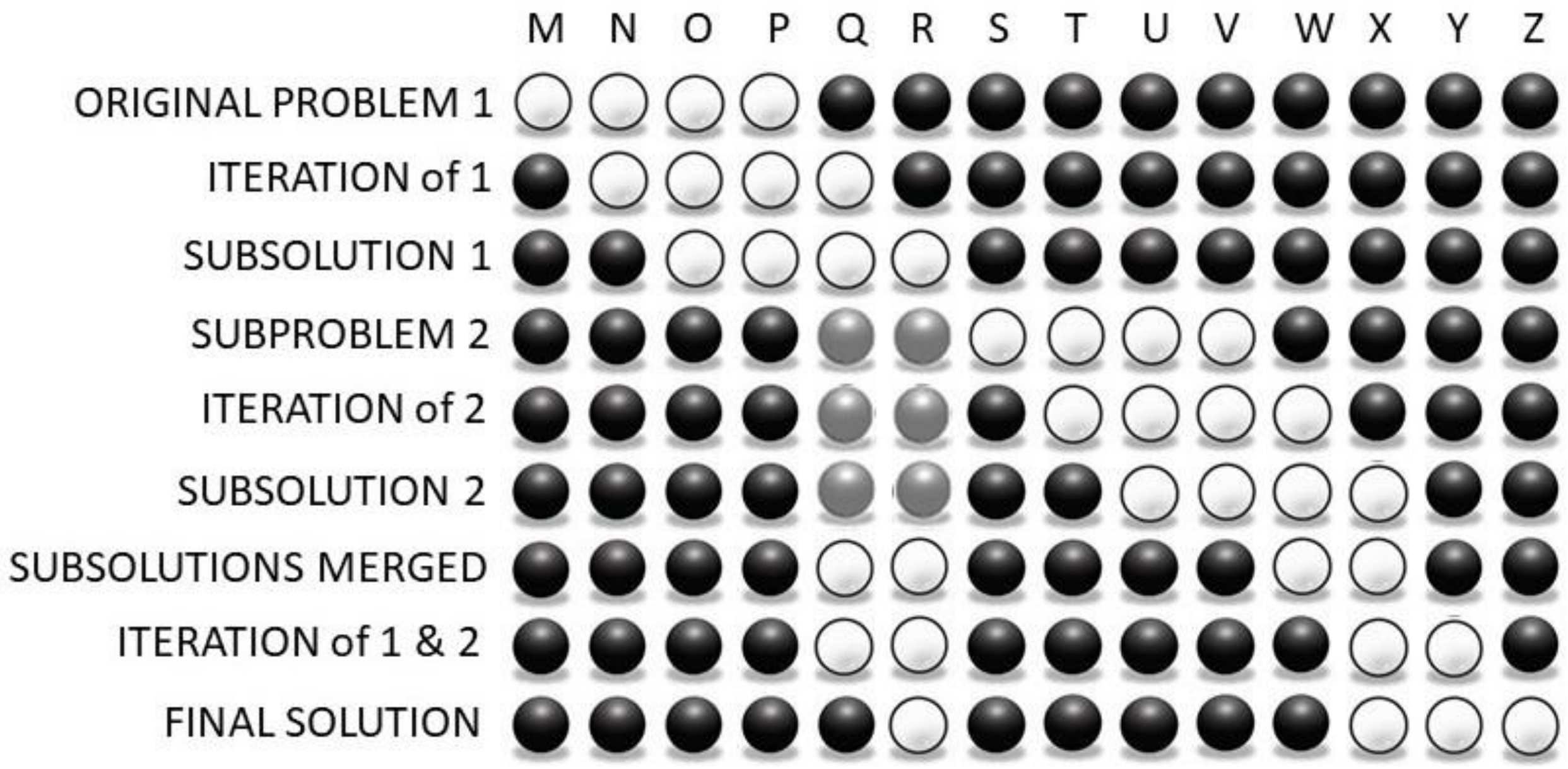

**Fig. 31.** Merging Subproblems in Working Memory

*An original problem is activated (M, N, O, P), and iterative updating is used to reach a subsolution (O, P, Q, R). This subsolution is saved in the short-term store, and a related subproblem (T, U, V, W) is introduced into the FoA. This subproblem iterates until a second subsolution is generated (U, V, W, X). Relevant items from the first subsolution (Q and R) are combined with those from the second subsolution (W and X) and iterated to generate a final solution (R, X, Y, Z). This pattern could be a fundamental aspect of human reasoning and creativity.*

According to this interpretation, the short-term store holds the agent's present objective, and the FoA produces lines of reasoning that interrogate that objective. These lines of reasoning update the objective, bringing it closer to resolution. This allows the agent to keep a present opportunity or threat in mind while considering possible responses before acting. In effect, previous threads of FoA sequences can be suspended in STM (or LTM) as interim results. These can then be retrieved rapidly if spreading activity reconverges on them. This permits working memory to deviate from its default behavior described thus far and employ a form of backward reference and conditional branching.

We have seen how continuity of thought can be established and then broken. However, Figure 31 demonstrates that instances of continuity can be reestablished, such as when one revisits a thought from the past. Other common thought patterns exemplifying related mental phenomena are illustrated in Figures 32-42.

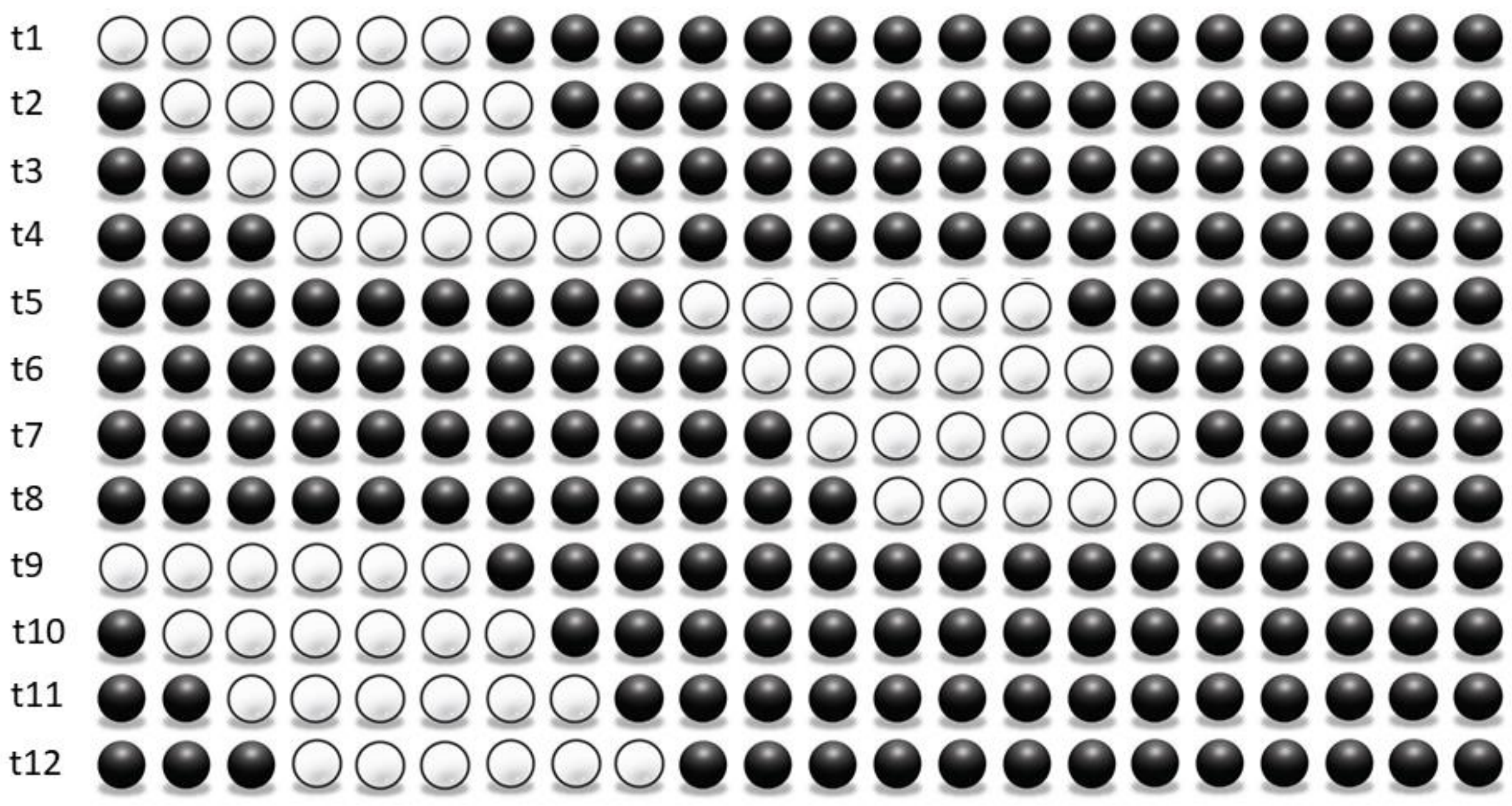

**Fig. 32.** Reiterating Through an Earlier Sequence

*A set of six items held in working memory is iteratively updated over the next three time steps, creating a self-contained thought. Starting at t5, attention shifts completely as an unrelated thought takes place using an entirely different set of items. From t9, the first sequence is reiterated as before. This might happen when someone revisits an earlier thought, such as rehashing a plan of action, retracing a set of previous steps, or retelling a story.*

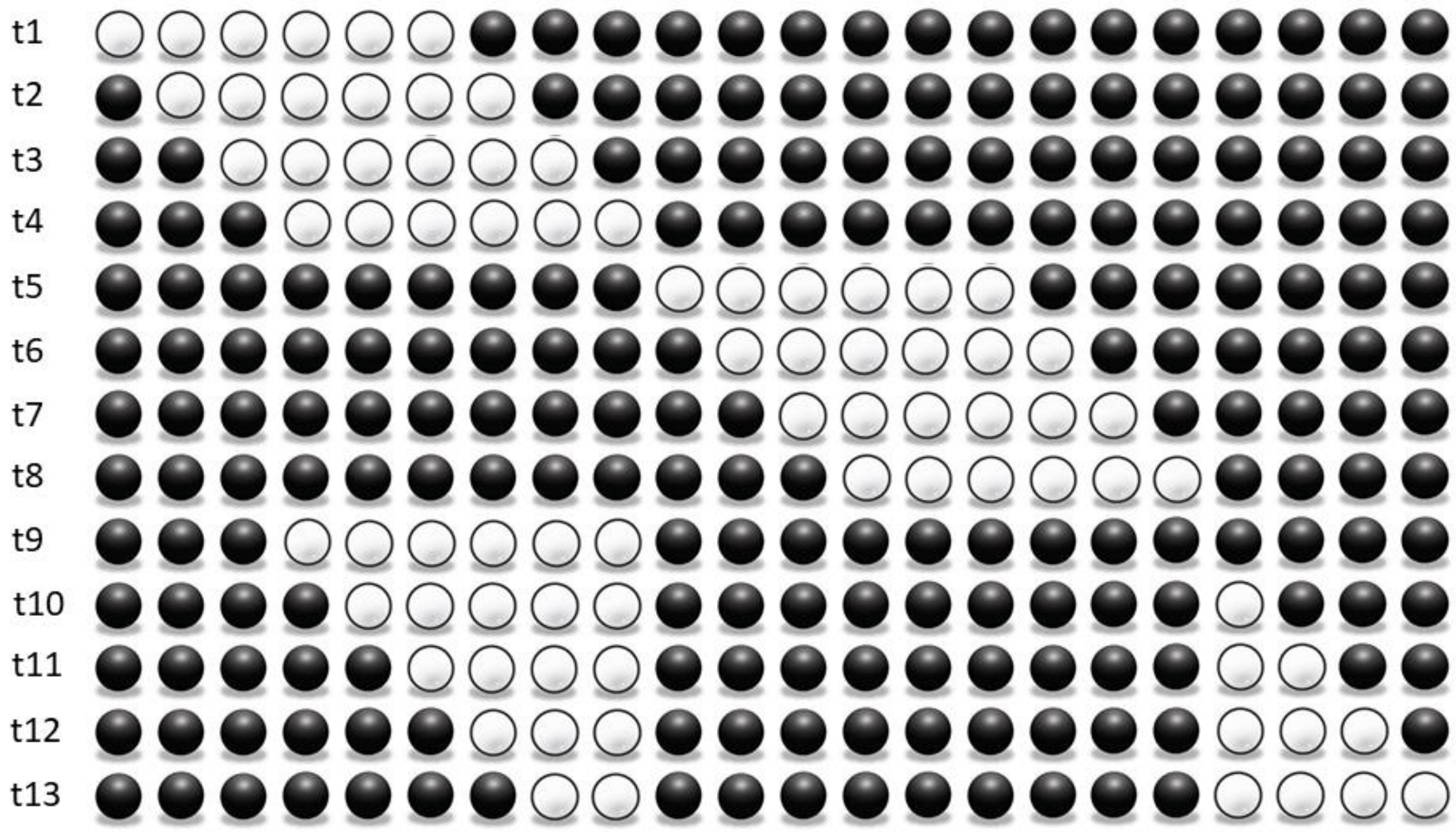

**Fig. 33.** Revisiting the Endpoint of an Earlier Iterative Sequence and Continuing It

*Six items are modified over the first three time steps, creating a line of thought composed of four related states. Attention shifts completely at t5, and an unrelated thought occurs. Starting at t9, attention shifts back to the items from t4, which are iterated without using any of the items from t5 through t8. This might happen when someone picks up a thought where it left off and continues to think about the issues from the last point at which they were considered.*

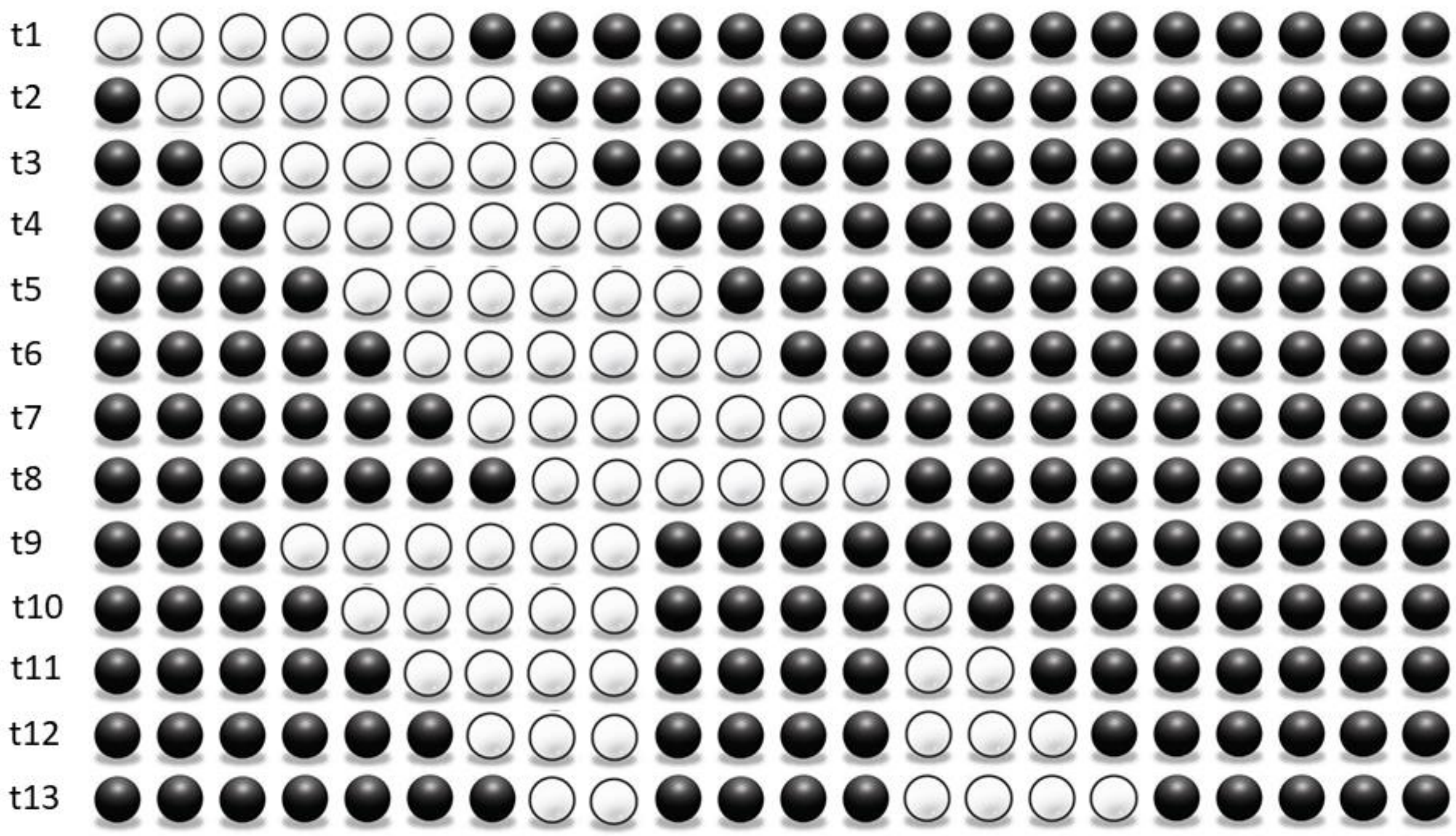

**Fig. 34.** Revisiting the Midpoint of an Earlier Iterative Sequence and Altering It

*Six items are modified over seven time steps, creating a thought composed of eight related states. At t9, attention shifts back to a point in the middle of this sequence. This set or subproblem from t4 is then iterated without including any of the items that were introduced from t5 through t8. This creates an alternate branch and a "forking" of the iterative sequence. This might happen when someone decides to assume an intermediate step in a previous problem-solving sequence and solve the problem in a different way.*

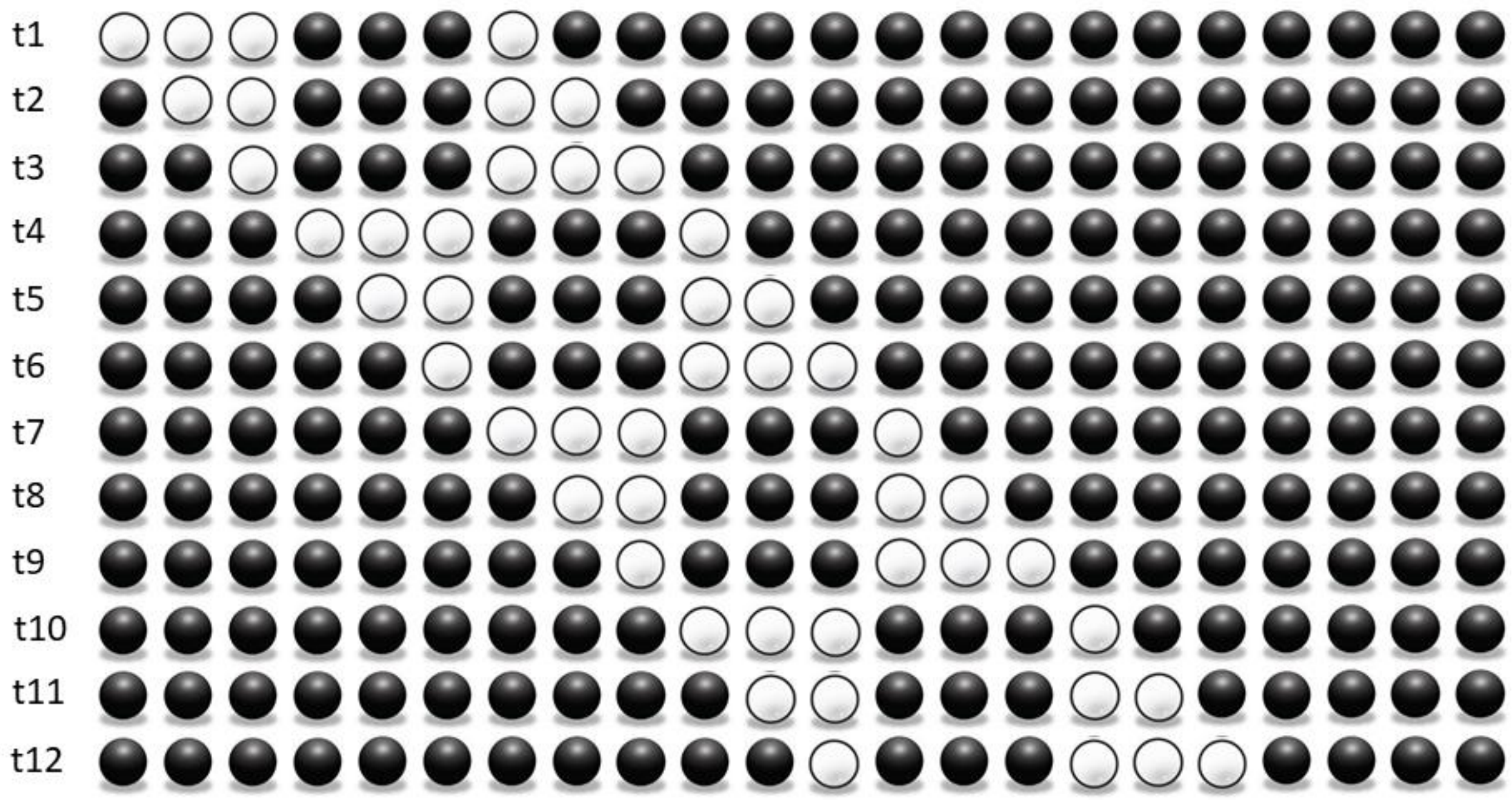

**Fig. 35.** Multitasking Occurs when Two Independent Sequences Alternate

*Two distinct threads of thought are iterated but never combined, alternating every third time step. This context switching could occur when someone is working on two separate tasks or problems simultaneously.*

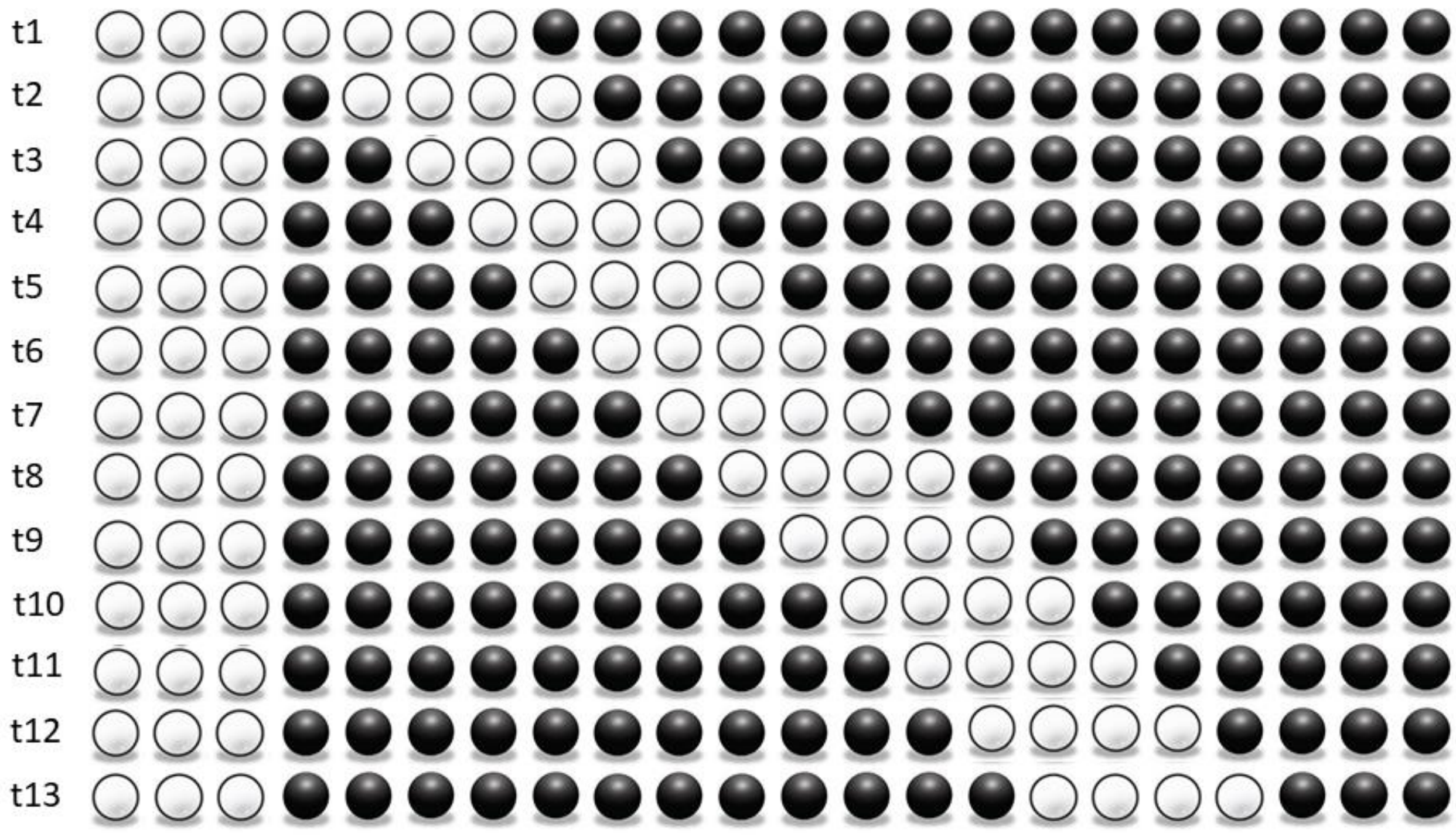

**Fig. 36.** Elaborating on a Stable Subset of Items

*Three items are held constant as iteration elaborates on their statistical relationships using related concepts. This process would strengthen the connection between these first three items and explore how they are associated within different contexts.*

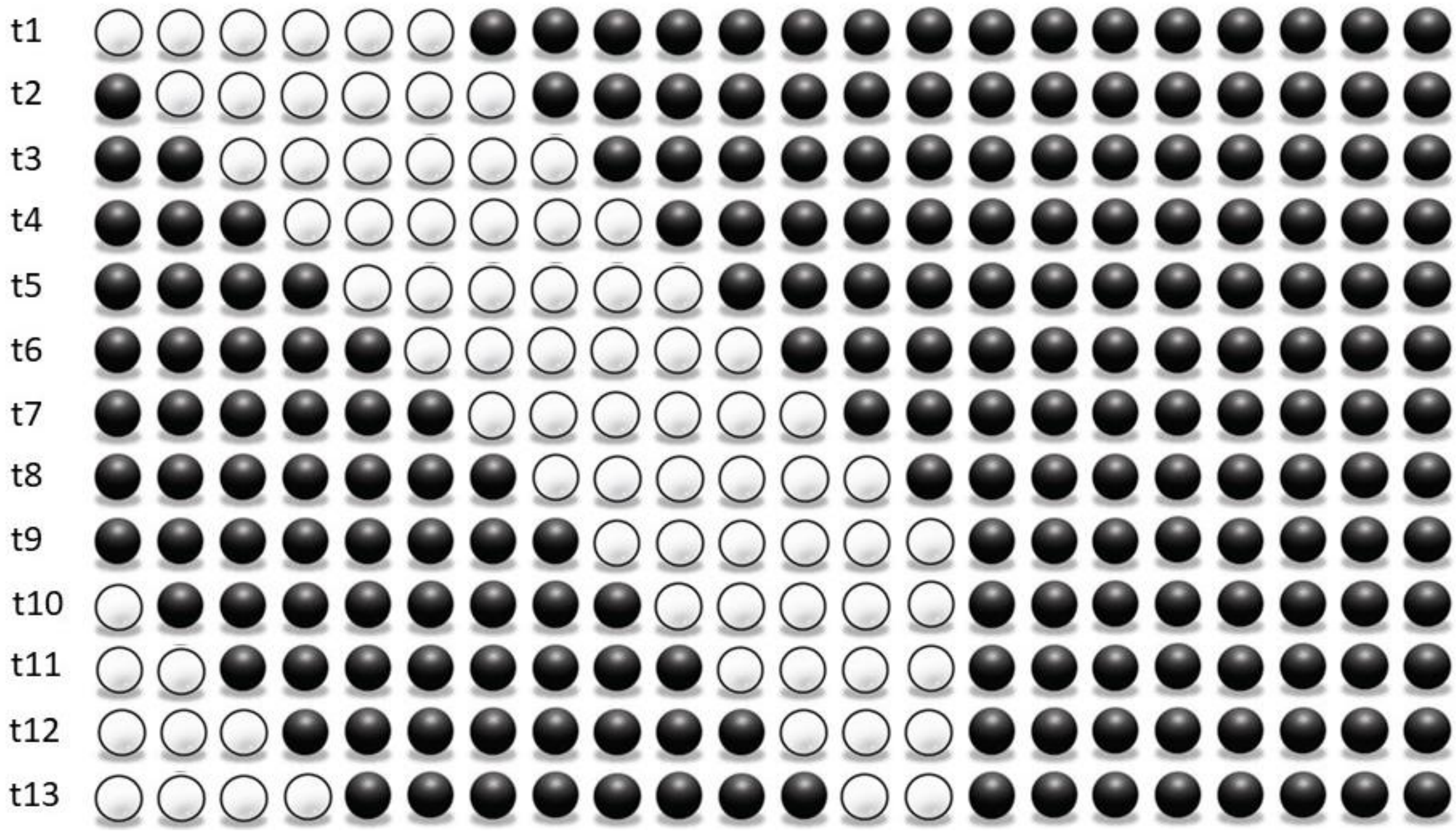

**Fig. 37.** Progressing Backward to a Subset of an Earlier State of Coactive Items

*Some lines of thought, by the end, return to the beginning. Here, the first set of coactive items is revisited and re-related to the outcome of the iterative sequence. This circularity could occur when one reconciles a predicted outcome with the original premise. This is probably a common thought pattern and can be contrasted with the previous figure (Fig. 36), where the first three items never leave coactivity.*

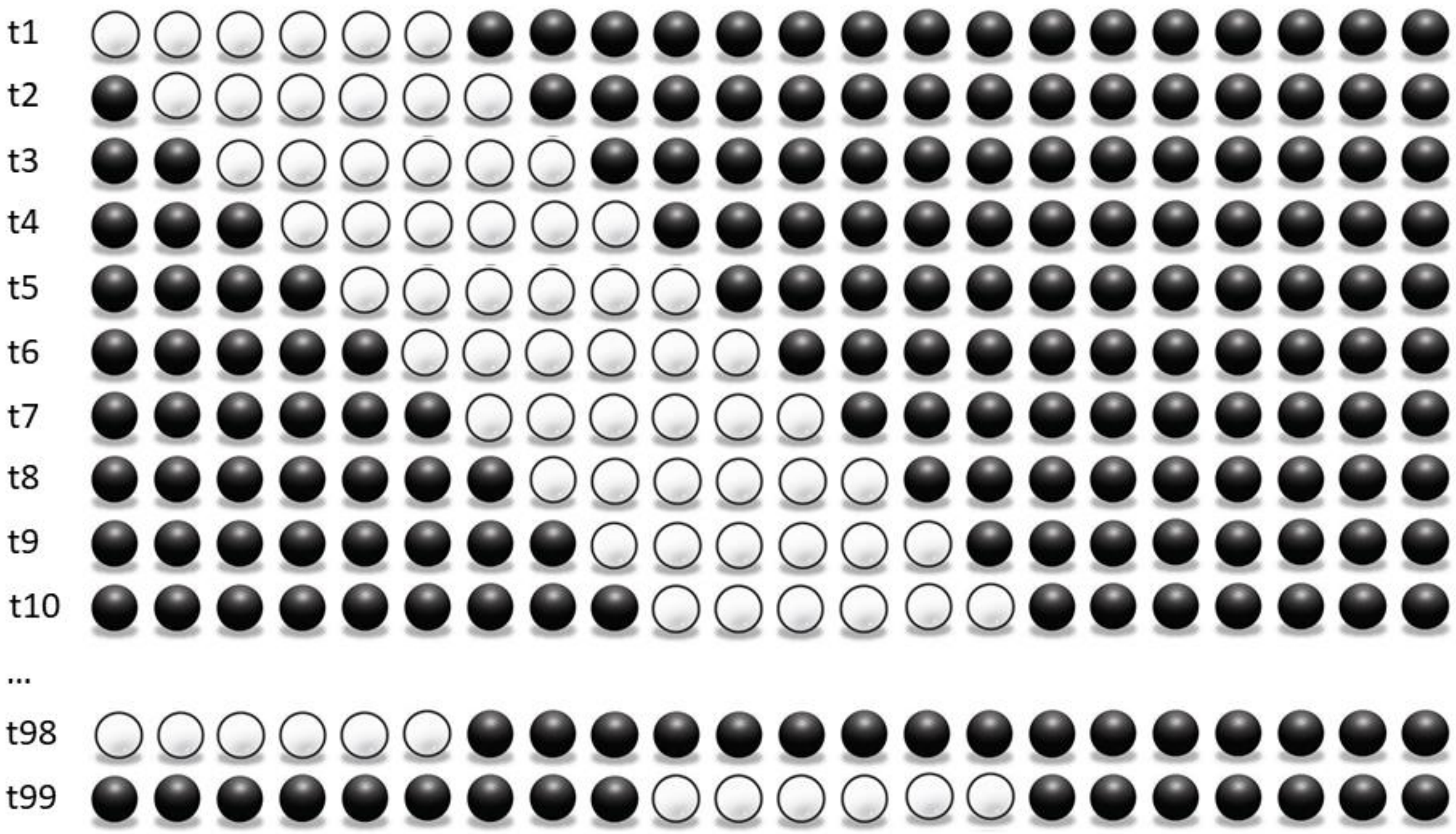

**Fig. 38.** Linking the Beginning of an Iterative Sequence with the End Makes the Intermediate Steps Implicit

*Six items are modified over nine time steps, creating a line of thought composed of ten related states. After t10, many states pass as indicated by the ellipsis. Then, later, the original six items reenter working memory. Because of Hebbian learning, these items recruit the same six-item solution reached previously without having to reiterate it. The way probability is modeled has changed due to the earlier iterative work, and multiassociative search is now capable of recruiting the final solution immediately. This is much more likely to happen when iteration involving the first six and last six items occurs, further entrenching their association. This may happen when one reflects on how their solution reconciles with the original problem state (Fig. 37).*

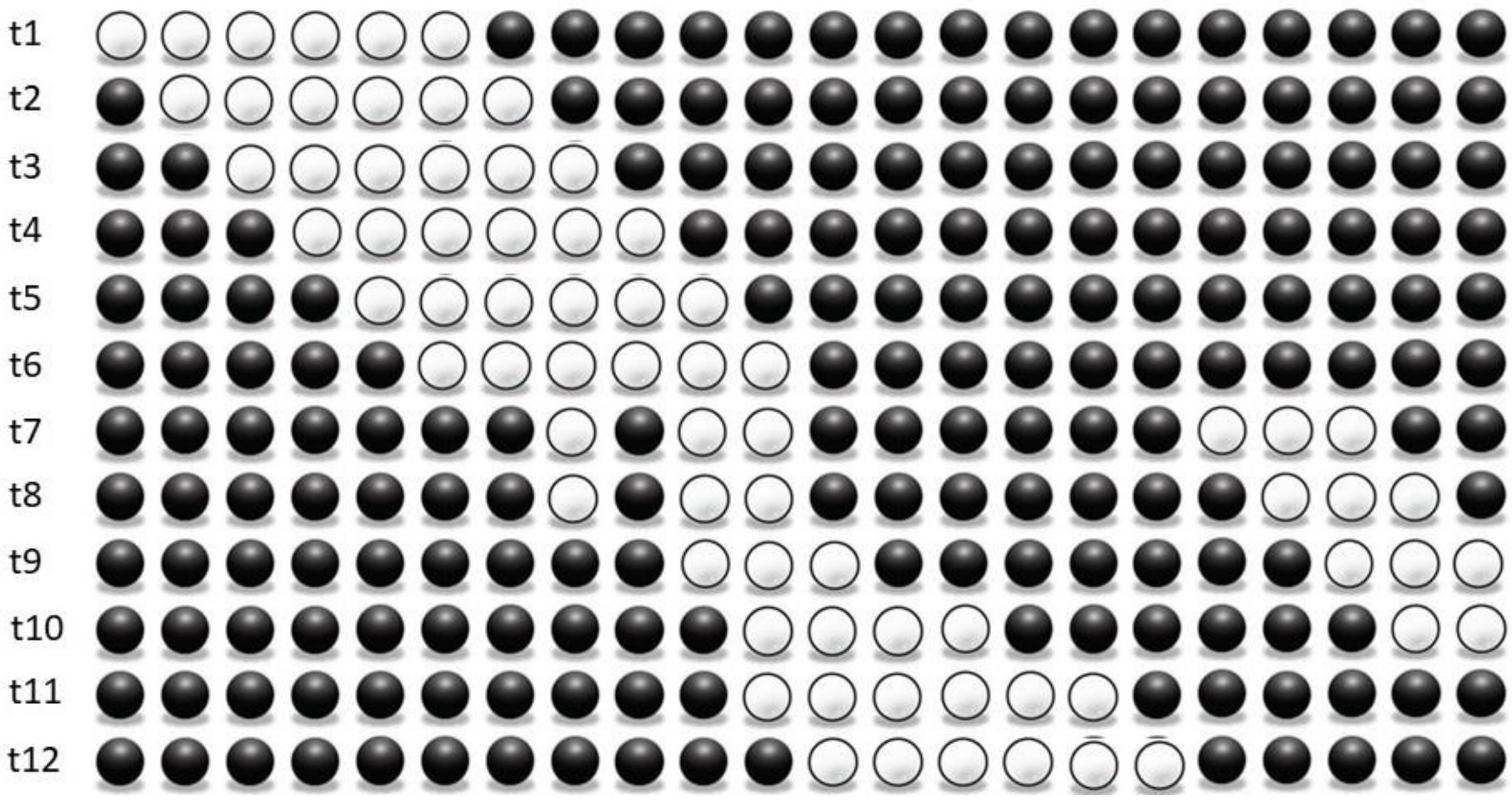

**Fig. 39**. Employing a Previously Learned Schema

*Six items are modified over five time steps, creating a line of thought composed of six related states. At time 7, a related schema composed of eleven (nonsimultaneous) items that was learned previously is recalled. This schema (a dynamic packet of information) iterates with items remaining from the ongoing line of thought. This coiteration results in the utilization of the schema as an organizing heuristic, archetype, or script to guide the thought process.*

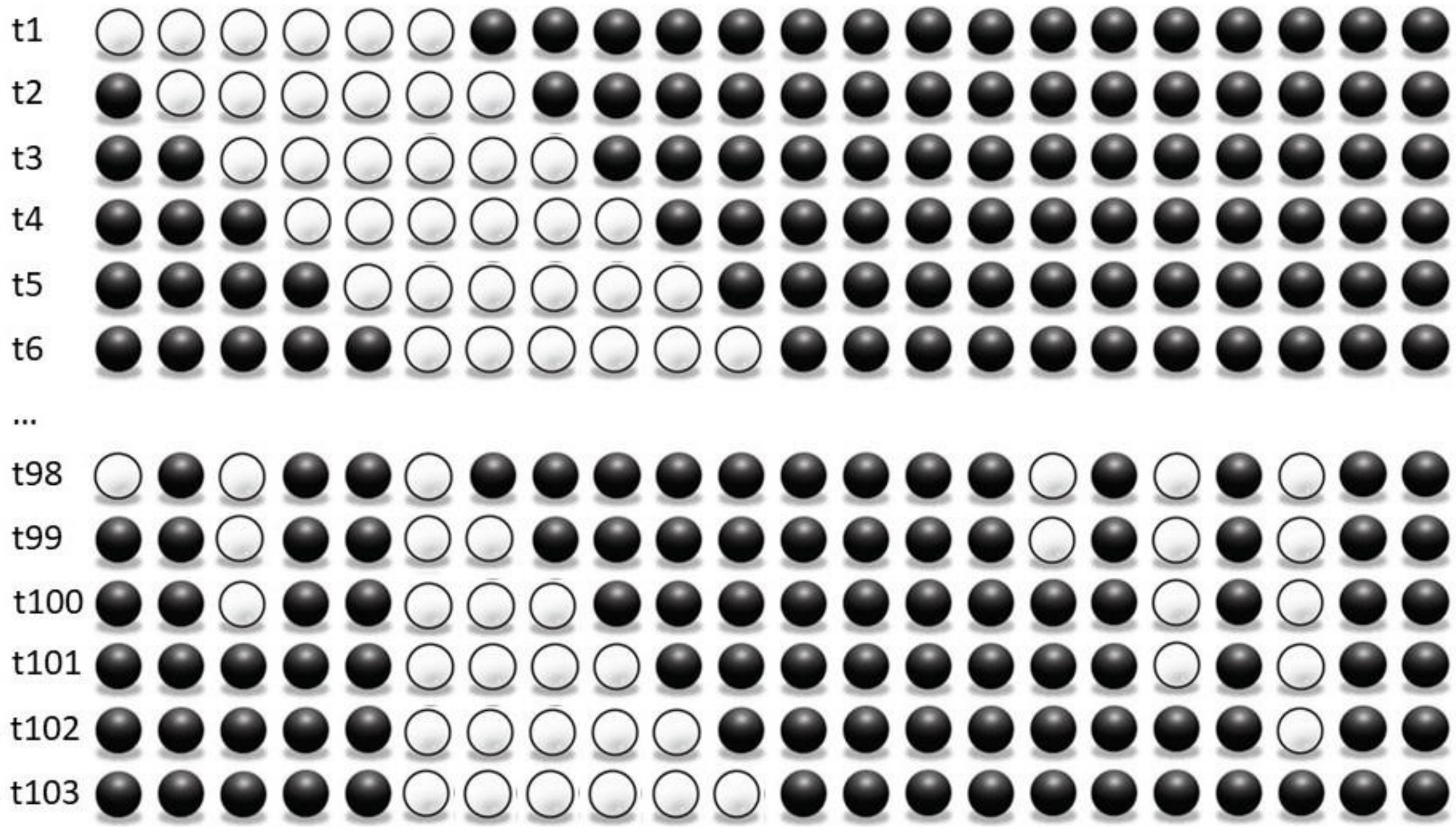

**Fig. 40**. Transfer Learning

Six items are modified over five time steps, creating a line of thought composed of six related states. After t6, many states pass as indicated by the ellipsis. Then, at time 98 an event is encountered that is similar to that at t1, containing half its items. The system recognizes this similarity and incorporates items from the previous line of thought. It thus applies information it has learned to this new context, generalizing, to arrive at a similar conclusory state.

Another commonplace pattern found in the updating of working memory may occur when an existing problem-solving process reaches an impasse. The newest addition to working memory is sometimes unhelpful or not task-relevant (e.g., because of prepotent associations formed during a similar but irrelevant task). In this case, it may be inhibited. The same items that recruited it would continue to spread activation energy without being able to reactivate it. This would force them to activate the next most pertinent item. Multiple rounds of "iterative inhibition" may be required before an appropriate item can be identified (Fig. 41). Each time a potential coactivate is vetted for exclusion, the search tree is restricted further. This situation might arise as one deliberates over different methods of completing the same task (e.g., "I should fax this letter. No, I should email it. No, I will text it instead").

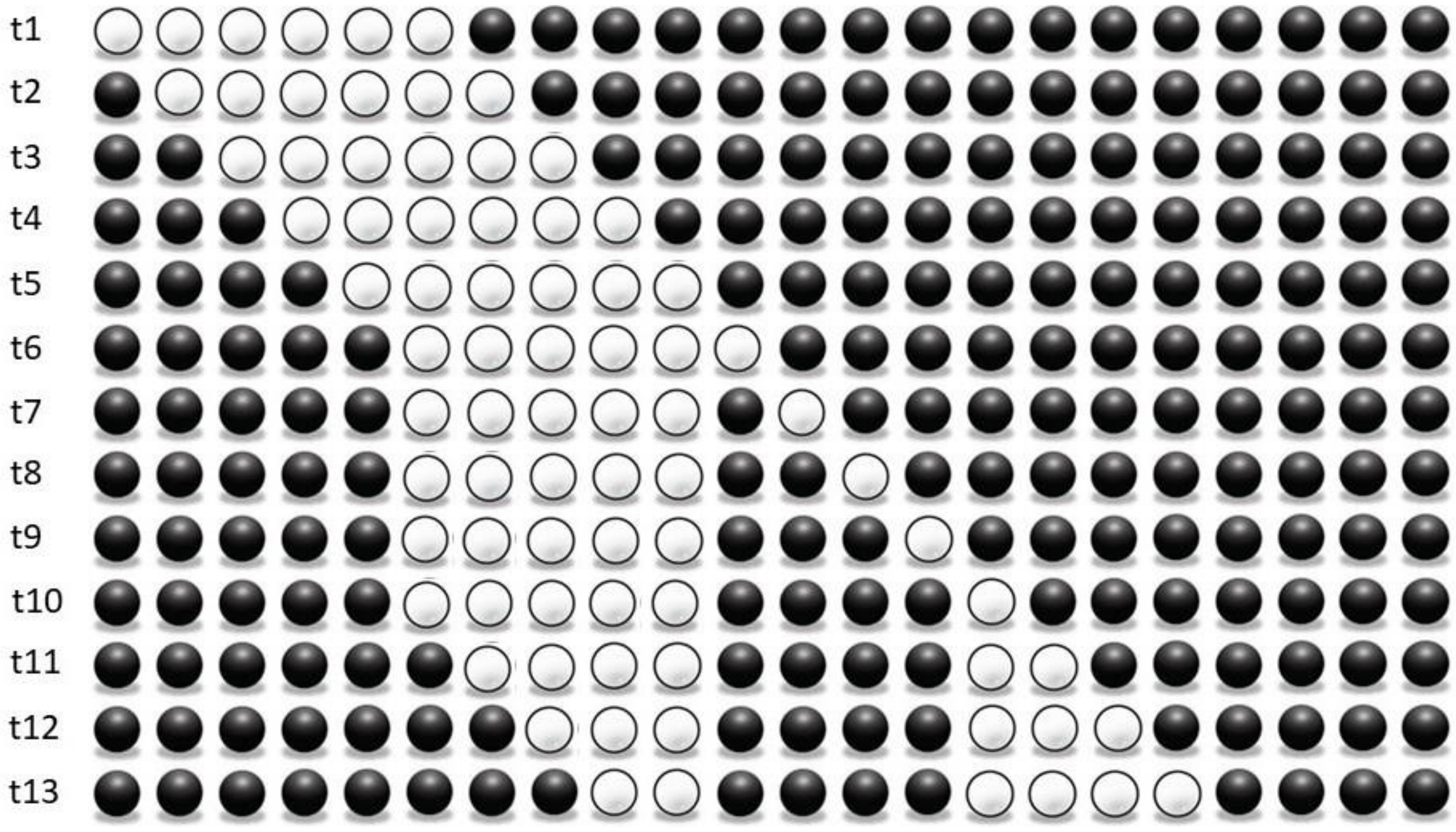

**Fig. 41.** Iterative Inhibition Excludes New Items in a Search for Something More Relevant

*An original problem is activated in time 1 creating a line of thought that continues until t6. At that point, multiassociative search determines that the most recent update is not a suitable behavioral parameter and it is*

*inhibited. With that item now unavailable, the remaining items continue to spread activation energy that converges on a different item at time 7. The same iterative inhibition occurs for three times steps until a suitable item is found and iterative updating resumes. This might happen as someone considers different solutions to a problem.*

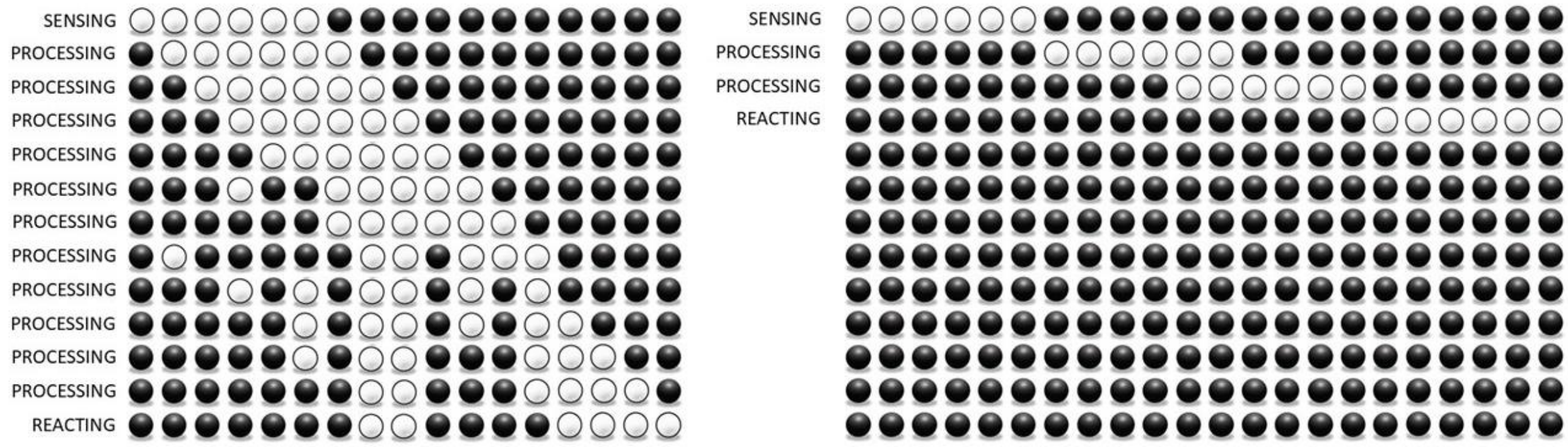

**Fig. 42.** Planning Compared to Not Planning

*The first diagram portrays densely iterative processing characteristic of planning. The second diagram portrays four mostly unrelated processing states that could not result in a cohesive plan. This comparison highlights the role of intermediate reasoning steps or chain of thought in the planning process. Planning ordinarily involves, 1) a lower rate of updating (Fig. 20), 2) a lower rate of 100% updates or new, unrelated thoughts (Fig. 22), and 3) more iterative processing steps before an action is taken.*

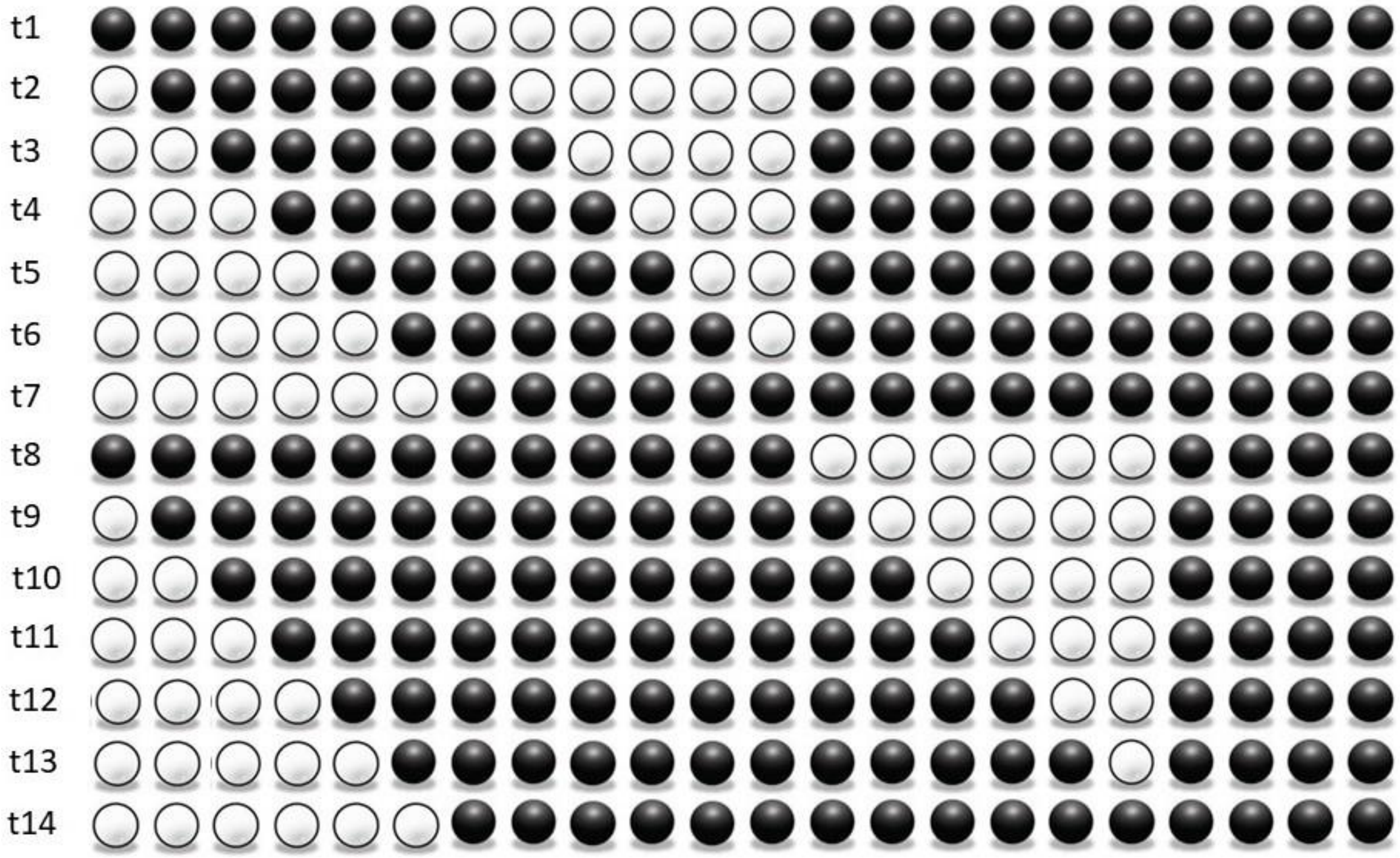

**Fig. 43**. Reconciling Disparate Situations with the Same Concept

*Two seemingly different situations from t1 and t8 iterate with the same concept (seen twice, at t7 and t14). The items at t1 and t8 independently converge toward the idea of t7/t14 as they are reconciled with its fundamental attributes. Perhaps both situations can be explained or caused by the same underlying phenomena and thus are funneled toward this specific region in the conceptual landscape.*

As the figures in this section attest, iterative updating will tend to converge toward certain stable sets of items. These are attractor states that amount to beliefs or truths about reality. Iteration then reconciles these truths with other truths. All thinking is a narrowing down (compression with dimensionality reduction) of combinations of items approaching reliable statements that can be generalized across situations. The present article itself is doing something very similar by attempting to reconcile the concept of iteration with numerous other related concepts. This section has considered how iteration of working memory content can create progress in information processing. The following section will consider how the model, as a neurobiological architecture, can be tested experimentally.

### 4.5 Testing the Neurophysiological Validity of the Model

Future research should use this framework to search for the neural signature of iteration within the brain (see Figs. 12, 13, 14 and 22). As shown in Figure 44, this search could utilize simultaneous recordings (multiple electrodes inserted into live cortical tissue) to produce time-series analyses of incremental change in populations of coactive cortical neurons.

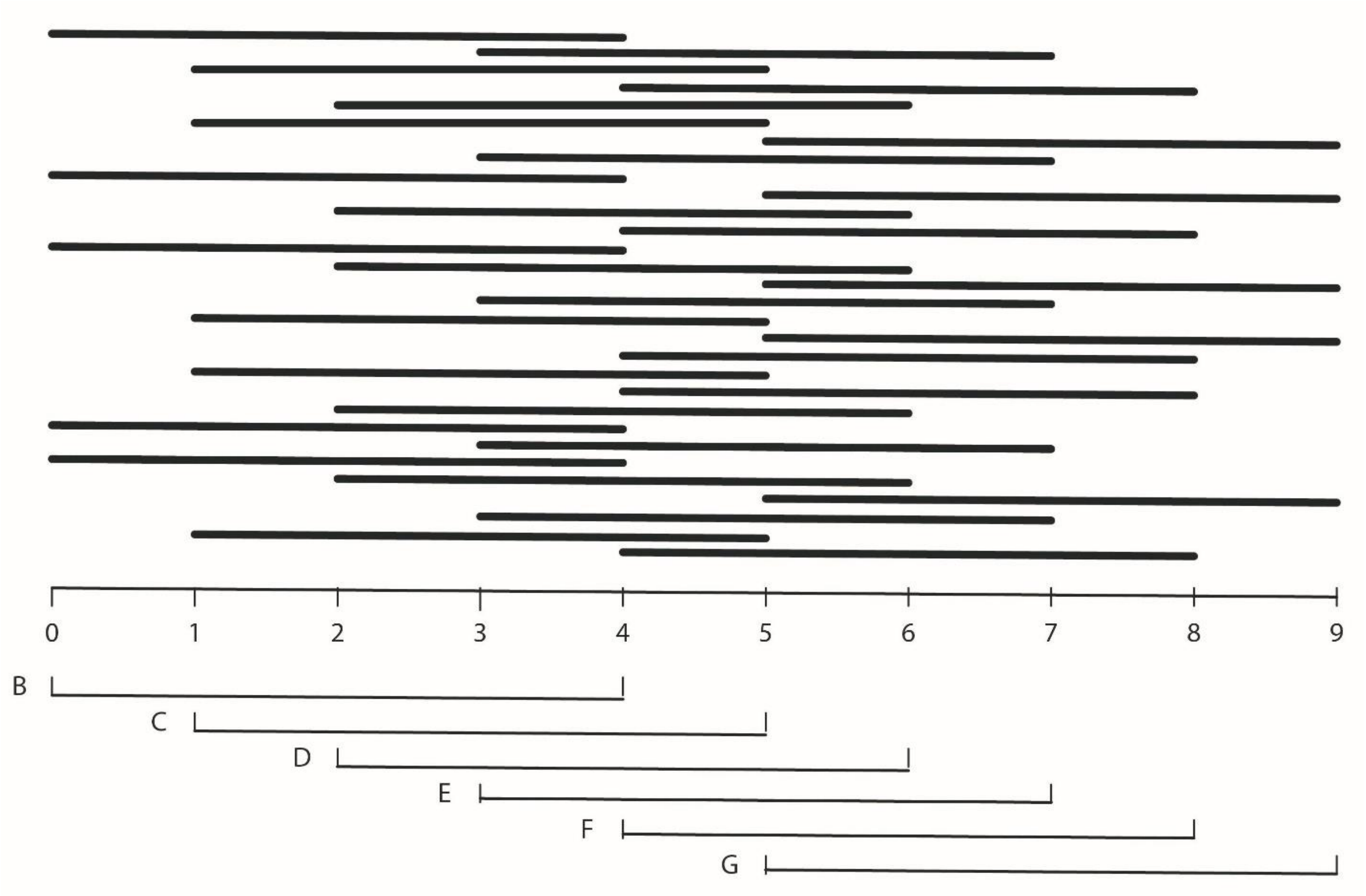

**Figure 44.** A Hypothetical Example of How Iterative Updating Could Be Found Using Electrodes

*Single-cell recording from a large number of cells in association cortex could produce an activity profile exhibiting iterative updating. In this simplified figure, the x-axis represents time in seconds while the y-axis contains the recorded activity of 30 individual neurons, each remaining active for four seconds. Five neurons become active each second. Each group of five neurons that begin and end their period of activity at the same time is assumed to belong to an individual ensemble, or item, of working memory. Brackets at the bottom of the figure indicate the item to which each group of neurons belongs. This profile coincides precisely with the pattern introduced in Figure 14 and can be meaningfully contrasted with the pattern in Figure 11. Searching new and existing data for this kind of iterative pattern could provide strong support for the present model.*

It is unclear whether it is possible to derive conclusive support for the present model using existing neuroimaging technology. Basic fMRI recording reveals the degree to which specialized brain modules exhibit involvement during a specific task but does not reveal the identity of the concepts or items involved. However, advanced recording techniques can demonstrate the onset and duration of brain responses to prepared stimuli, which could result in data similar to that in Figure 44. It should be possible to use the gain in temporal and spatial resolution to observe how the pattern of working memory activation changes over time. To that end, factorial designs that allow for the measurement of the BOLD signal for each volumetric cell should be able to test for differential activation in response to partial, as opposed to complete, updating of working memory.

Substantiating findings could be derived from neuroimaging experiments in which brain activity is recorded while participants complete a task that requires an algorithmic sequence of steps (e.g., long division). Each step of the task would need to be modeled separately. As the participant moves from one step to the next, the BOLD activity would be estimated for that particular step. A mixed model of a duration regressor covering the entire span of the problem along with individual regressors for each step would be needed to capture both the sustained attention required to solve the problem and the individual steps needed to progress from one stage to the next. It would be necessary to show that the sequence of mental representations posited as necessary to complete the task has a one-to-one correspondence with the time course of underlying cellular or hemodynamic changes. This may necessitate using multiple methods simultaneously, such as fMRI and EEG together or multivoxel pattern analysis, which has been used to resolve the addition and subtraction of individual cognitive items from working memory (Lewis-Peacock et al., 2012).

To validate the hypotheses put forth by the present model, it would be necessary to show that the activity in association areas underlying working memory contents can be partially rather than completely updated. Next, the goal would be to show that this partial updating happens constantly. Future studies should be able to resolve whether the iterative updating of cortical activity is continuous (at the level of neurons) or incremental, where entire items (and all their comprising neurons) are added or subtracted at once (Fig. 25). The line of reasoning suggested by this article predicts that the former may be true of the short-term store (i.e., Figure 13) while the latter may be true of the FoA (i.e., Figure 14).

Modern cognitive neuroscience is limited in its ability to match the components of brain states to the components of mental states. However, matching the iterative updating of ensembles to that of their corresponding items may provide a means to do so. The markers of iterative updating may establish an ordinality and translation strategy to decode the nature of the correspondence between temporary neural traces and their psychological manifestations. Machine learning may be instrumental in this work.

# **Part V:** Instantiating the Model within a Computer

### 5.1 AI Should Employ Iterative Updating

Many researchers in the field of AI expect brain science to reveal breakthroughs that will provide essential guidance for the construction of intelligent machines (Haikonen, 2012). Some have suggested that AI may not need to emulate fine-grained molecular or cellular details of the brain to create human-level intellectual function (Bostrom, 2014). Instead, they suggest simulating an abstraction of the neurological mechanisms that produce intelligence (e.g., Hassabis et al., 2017; Butlin et al., 2023). The present model introduces abstractions that could provide organizing direction to efforts at human brain emulation, integrated cognitive architectures, or the application of machine learning in commercial technology sectors.

Specifically, the present model may help close the “computational explanatory gap,” which is an effort to understand how the parallel, subsymbolic computations involved in low-level neural networks could translate into the serial, symbolic-level algorithms involved in high-level cognition (Reggia et al., 2019). The concept of multiassociative search (Figs. 26 and 27) provides a mechanistic account of how this gap could be bridged. Today, even state-of-the-art AI processing feats are typically only equivalent to brief episodes of unconscious human processing (e.g., recognizing objects in a picture) (Goodfellow et al., 2017). To create more generally intelligent AI, these short, parallel processing sessions must be chained together into iterated sequences that more closely resemble symbolic thought. Iterative updating and multiassociative search may be instrumental in realizing this. As the rest of this section will detail, even though neither are recognized by psychology or neuroscience, they are both used in computer science.

Iterative updating, on its own, is not sufficient to elevate computer information processing to the cognitive domain. In fact, updating a memory store iteratively has been commonplace in computing for several decades. All computers using the Von Neumann architecture routinely update their temporary memory stores (i.e., static RAM and dynamic RAM). These stores, known as caches, have a resemblance to working memory. They hold information that is predicted to be useful so it can be readily available to the CPU. Cached information includes intermediate results from ongoing processing, as well as data and program instructions from the storage drive. Cache stores have a limited capacity, and because they are constantly tasked with holding new information, they must evict old information. These stores are updated iteratively as the least recently used (LRU) data are replaced (Comer, 2017). For example, a computer’s RAM holds billions of bytes coactive through time, adding and subtracting from this pool in the same manner illustrated in Figure 2.

However, the next bytes of data processed by the CPU are not determined by the contents of the cache itself. Instead, the processing instruction sequence is determined by the next line of programmed, executable code. Thus, unlike the brain, computers do not make cached information globally accessible for use in multiassociative search. The various bytes of data within computer cache memory can certainly be considered coactive, but they are not “cospreading.” That is, they do not pool their activation energy to search long-term memory for relevant data as in human working memory (Fig. 27). No computer hardware or software does this as described here.

There are AI systems that employ working memory, a global workspace, recursion, and various methods of updating (e.g., Goertzel, 2016). These include cognitive architectures (Gray, 2007), evolutionary computation (Sipper et al., 2018), soft computing (Konar, 2014), and some mainstream machine learning techniques. However, such software generally utilizes either preprogrammed symbolic rules or subsymbolic ones to transform one state into the next. Because these systems are incapable of transitioning iteratively between states, they are usually restricted to formalized, narrow problem-solving domains (Haikonen, 2003).

Some contemporary artificial neural networks are capable of yoking together symbolic and subsymbolic representations. Like the brain, artificial neural networks use parallel, distributed processing to train systems of subsymbolic nodes to come to recognize complex mathematical functions. Some neural networks, such as recurrent and long short-term memory networks, have nodes capable of persistent activity that is highly analogous to sustained firing (Fig. 10). The enduring activity of these recurrent nodes permits them to cache previous inputs in the form of activated long-term memory (Sherstinsky, 2020). This working memory pool is updated iteratively as recurrent neural nodes gain and lose activation.

Similarly, attention and the context window in transformer-based large language models are updated iteratively both during training and inference. One might say that the transformer's attentional and context windows are analogous to sustained firing and synaptic potentiation, respectively. Moreover, the tokens within attention could be described as combining their spreading activation energy on each forward pass through the network in a search for the next predicted word. This behavior is similar to the present model's conception of multiassociative search. However, these AI models use symbols as their inputs and outputs (token vocabulary) and only subsymbols in the hidden layers. The human brain works in reverse with subsymbols as inputs and outputs (senses and motor commands) and symbols as hidden internal representations (see Figs. 45 and 46).

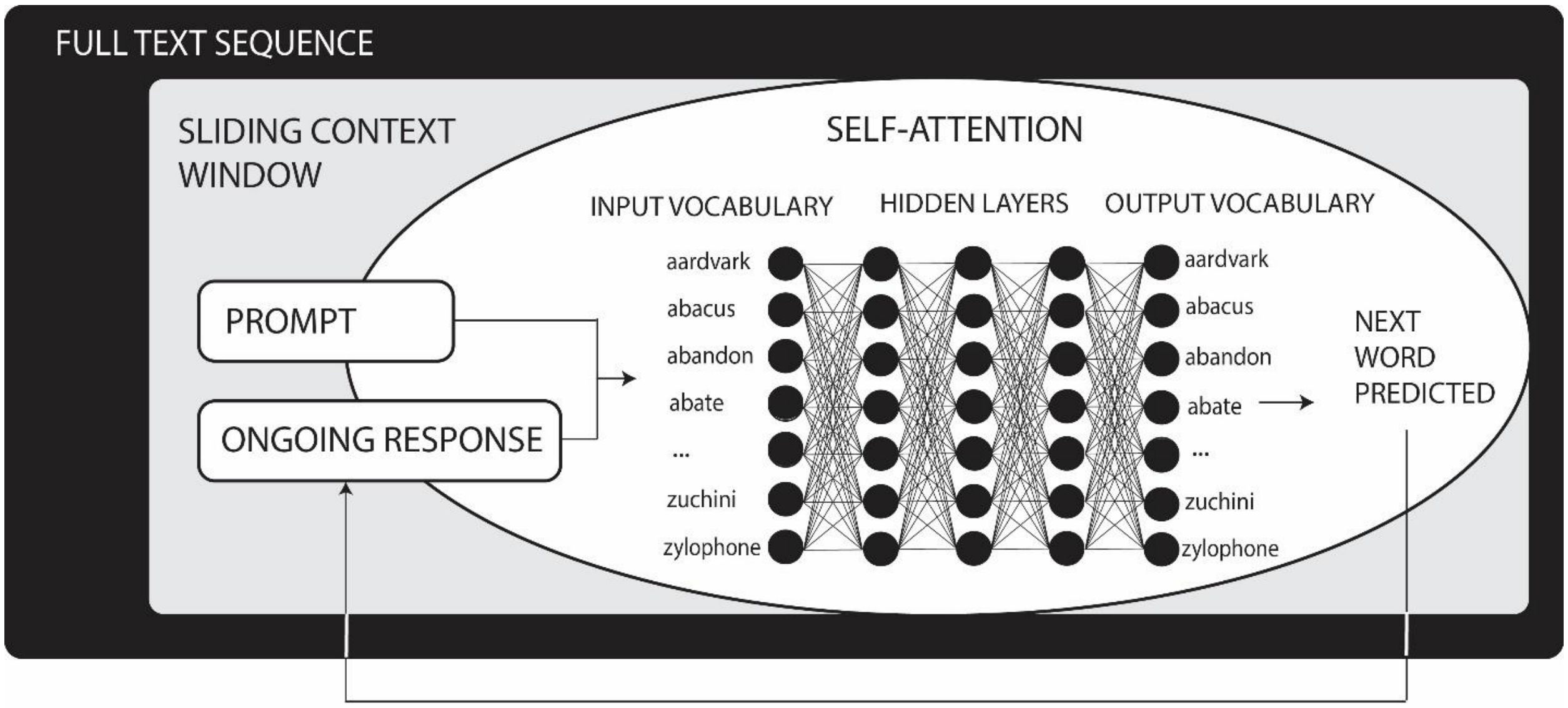

**Fig. 45.** A Caricature of a Language Model

*A simplified and abstracted representation of a transformer large language model. The prompt and ongoing response are kept active in attention along with prioritized aspects of the context (analogous to the focus of attention). Other context is held in the sliding context window (analogous to the short-term store). Symbolic tokens (words) active within self-attention pass through the neural network activating subsymbolic nodes until they converge on the next word predicted from the token output vocabulary. This network progresses from symbolic to subsymbolic to symbolic.*

Language models predict the next word in a sentence, but mammals predict the next event in a sequence of events. In other words, they predict how a situation will unfold. To do this, a cognitive agent must recognize the factors at play in each scenario, and iteratively manipulate them as internal, symbolic representations. Using these internal representations for search allows a cognitive agent to make causal predictions about how these factors will set into motion the next set of factors, an aptitude that the preceding sections have laid out a plan for re-creating.

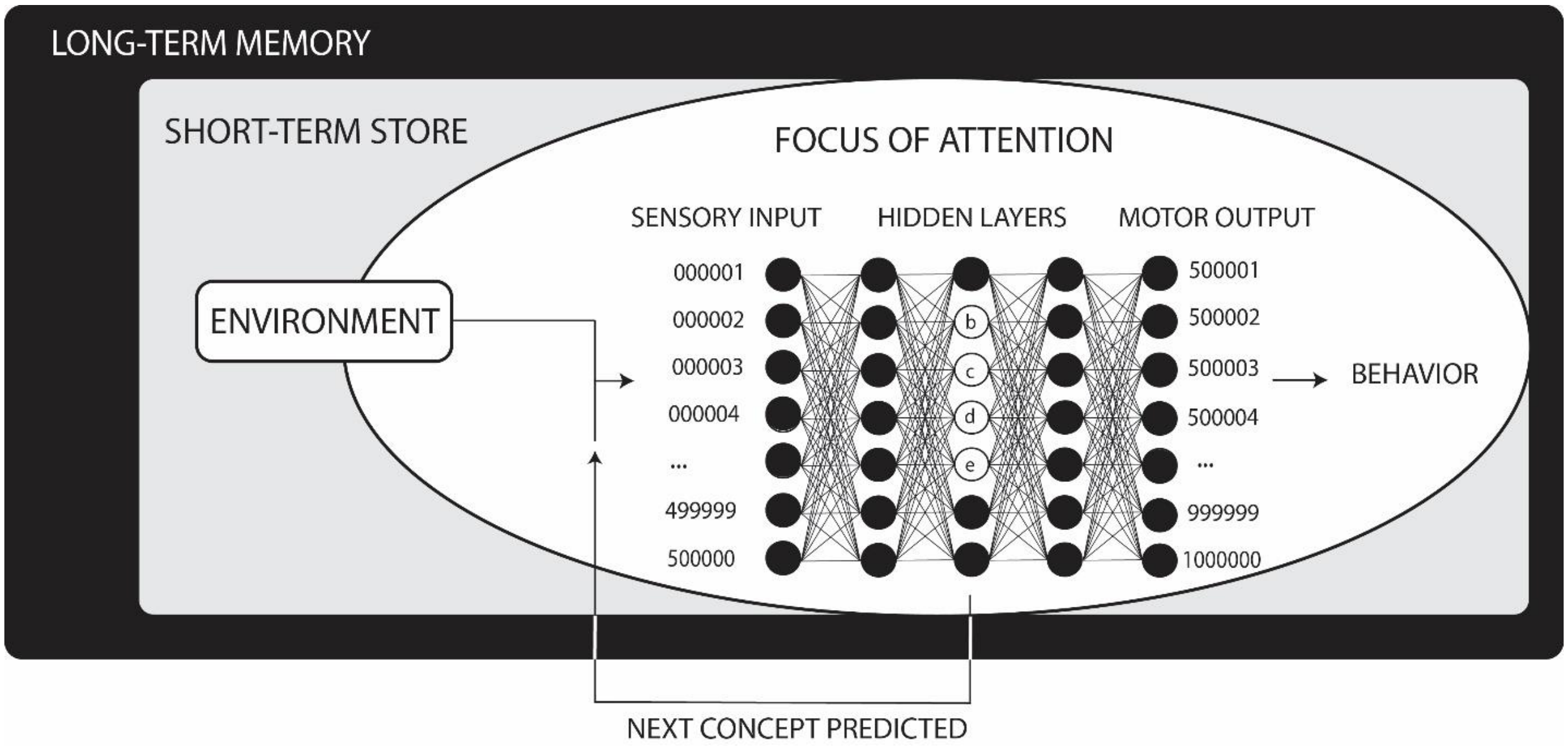

**Fig. 46.** A Caricature of a Neural Network Using Iterative Updating

*An abstracted representation of a traditional neural network implementing the iterative updating algorithm. Active sensory input nodes pass activity through the network activating items within the hidden layers. This activity then passes down toward motor output nodes guiding behavior. But it is the hidden layer representations, not the behavioral outputs that are routed back through the network cyclically to contribute to determining the next update. This is a naïve, vastly reduced implementation of the model proposed here that does not feature a global workspace, modularity, or multimodality.*

To develop and manipulate true internal representations, AI working memory should be designed to run iterative updating in lockstep with multiassociative search as outlined in the preceding pages. This is technically feasible in the near term. Because current artificial neural network technology is capable of sustained firing, synaptic potentiation, spreading activation, and Hebbian learning, everything discussed in this article thus far can be implemented by it. If an artificial neural network was engineered to do this in the manner presented in the preceding sections, the resulting system may exhibit some of the human qualities and functionality discussed thus far, including association and prediction formation, algorithm implementation, the compounding of intermediate results, progressive modification, and attentive continuity.

### 5.2 Designing an AI Capable of Iterative Updating

Implementing human-like working memory to a first approximation in an AI system would mean creating a connectionist program that spreads activity from already active information, along with incoming activity from its sensors, to search for entailed information from long-term associative memory. Using this as a partial update and then repeating this process in a cycle would structure the architecture to be self-organizing.

Iterative updating and multiassociative search may first have to be explicitly programmed into the system using rule-based code until it becomes clear how to design a system where they emerge organically as they do in the brain. Hand-coded or not, they must be defined mathematically and unambiguously to be the basis of computer software. Multiassociative search can be expressed as a function (f) that maps input variables (x) of the current state of working memory to an output variable (y) used to update them. Each network state would be a search for the update applied to the next state. As a formal algorithm, it could be modeled as a stateless Markov process in discrete time, performing non-deterministic search. As a computable function, it could be instantiated by traditional or neuromorphic computer clusters and executed using brain emulation, hierarchical hidden Markov models, stochastic grammars, probabilistic programming languages, neural networks, or others.

The rest of this section will describe how this system could be constructed using an artificial neural network architecture. It could be realized through recurrent networks or spiking ones. Either way, layers of nodes should be used to model the pattern-recognizing assemblies discussed in Section 4.1. As in the brain, each level in the hierarchy must build a statistical model of the regularities in the level below it (Eliasmith, 2013). Hierarchical pattern recognition would be achieved when primitive nodes lower in the hierarchy converge on high-order patterns in higher layers (Hawkins, 2004; Kurzweil, 2012). Figure 47 caricatures how this is actualized by nonlinear transformations.

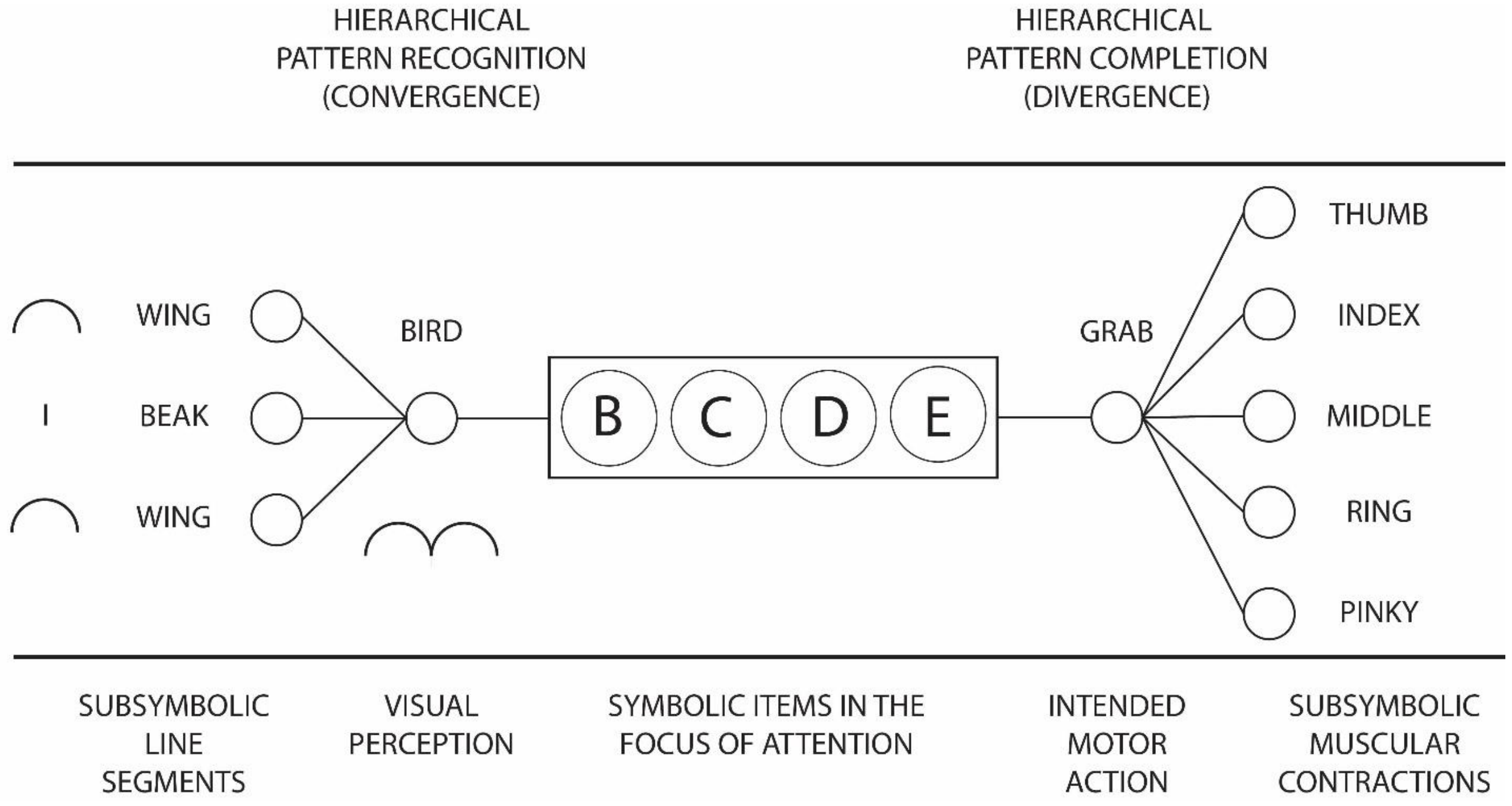

**Fig. 47**. Hierarchical Pattern Recognition and Completion

*Three subsymbolic line segments are detected by early visual cortex. These segments, corresponding to two wings and a beak, map onto three separate nodes. The nodes each fire at another node higher in the visual processing hierarchy that detects the coactivity (conjunction) of all three. In this case, that node detects the presence of a bird flying in the distance. This is an example of convergent pattern recognition made possible by hierarchical nonlinear transformations. Apparently, the prospect of a bird for dinner enters the focus of attention, creating an impulse to grab a bow and arrow. The neural engram for grabbing becomes active and fires action potentials at motor neurons responsible for the muscular contractions of each of the fingers. This is representative of divergent pattern completion. Essential to sending information to and from working memory in the mammalian brain, convergence and divergence should be emphasized in AI neural networks.*

Nodes at the top of the hierarchy would constitute high-order subsymbolic patterns due to having receptive fields composed of various inputs from multiple layers of mounting complexity. These abstract nodes should be capable of recurrent activity simulating the sustained firing of pyramidal cortical neurons. This would prolong their activity as search parameters and dependency markers, as well as contribute to contextual structuring for extended periods.

To form item-like ensembles, a Hebbian learning rule would be needed to strengthen the weights between frequently coactive nodes. This must work in such a way that groups of highly associated subsymbolic nodes can form symbolic groups (perhaps across layers). These ensembles should be sparse and fuzzy and used to represent invariant, categorical patterns (i.e. Fig 24). Such an ensemble would be equivalent to an internal mental representation and should be made capable of enduring coactivity with other items. These items should be coactive within a graph-structured global workspace. Engineering such a workspace could conceivably necessitate an analog of neural binding (i.e., Klimesch et al., 2010) and synchronized, reentrant oscillations (Edelman, 2004) to integrate (i.e., Tononi, 2004) and unify multiple ensembles into a singular situational representation. This would amount to an emulation of the FoA.

When the simulated sustained firing abates, nodes should subsequently simulate synaptic potentiation. This would enable the network to maintain pertinent information in an emulated short-term store as cached assets. Nodes potentiated in this way would continue to bias the multiassociative workspace until they are either promoted back to the FoA or demoted to inert long-term memory.

The AI's simulated FoA and short-term memory stores would undergo iterative updating such that the overlap of persistent information is congruent with Figure 13 and information replacement is congruent with Figure 29. It is imperative that the FoA be designed to cache not only external stimuli but also internal concepts as in Figure 18. Information selection should be guided by multiassociative search as in Figure 27. Each update would amount to a truth-preserving associative transition in the processing stream underwritten by structural properties of the network, which in turn are based on past statistical analyses of reliable patterns from the physical world.

### 5.3 Modularity, Modality, and Imagery in AI

An implementation of this system would necessitate modular specialization. Each module would correspond to a compartmentalized neural network meant to simulate a different cortical or subcortical area of the mammalian brain. These separate networks would interconnect to form a single dynamical system. Coordinating this assemblage to implement the multiassociative algorithm would be a considerable engineering problem. Given that the human brain accomplishes this task, human neuroscience should be used as an archetype. Thus, the system could be constructed biomimetically and inspired by general neuroanatomical connectivity.

Not only would the nodes of each modular network be organized hierarchically, but the connections between modules would establish an even larger hierarchical structure. This stratified organization, beginning from unimodal networks and progressing to densely conjunctive multimodal networks, would mirror the gradient seen from sensory cortices to association cortices in mammals. Networks higher in the hierarchy would refer to denser space-time conjunctions and multidimensional levels of abstraction. The networks could be designed to emulate specific human cortical modules if they drew inspiration from anatomical connectivity. This would emphasize intrinsic, extrinsic, short-range, and long-range connections, along with the relevant proximities and proportionalities. Multimodal areas that may be pivotal to higher-order abstractions and, therefore, in need of being reverse-engineered in this way include the angular gyrus, Wernicke's area, Broca's area, the dorsolateral PFC, the medial PFC, the supplementary motor area, and the frontal pole.

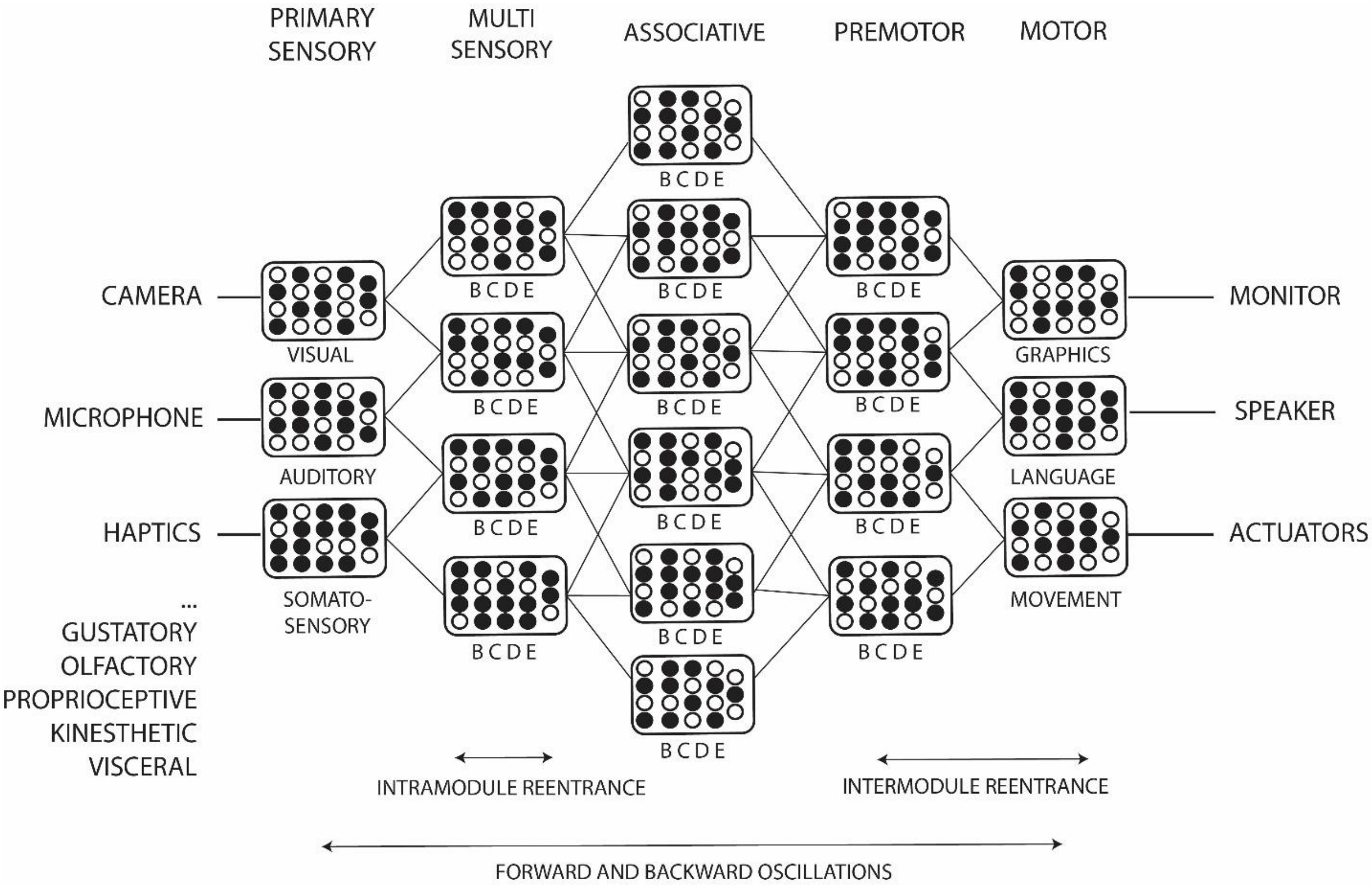

**Fig. 48.** A Hierarchical Artificial Neural Network Structured to Integrate Information Across Modules

*Each enclosed set of nodes represents a specialized neural network module wired to receive a different input modality. Networks at the bottom (left) of the hierarchy take input of a single modality from the environment. Other networks take input from multiple neural networks below them in the hierarchy. Spreading network activity would oscillate between the top and bottom of the hierarchy while allowing reentrant feedback (bidirectional signal exchange) within and between networks. This figure features 24 networks, each with 19 nodes. An actual build would necessitate dozens of networks, each with millions of nodes.*

Each modular network in the system would take inputs from associative areas (global working memory) and use them to create their own unique corresponding set of outputs with the potential to contribute to the next update. Some of these modules may produce imagery. To understand how imagery can benefit AI, let's discuss how it is formed in the brain. Neurons in sensory cortex respond to perceptual features from sensory input and fuse them into images known as topographic mappings (Moscovich et al., 2007). This imagery holds metric and compositional (precategorical) information. In addition to creating topographic mappings from patterns recognized in the external environment (bottom-up), sensory areas are thought to combine divergent (Fig. 47), top-down inputs from association cortices to generate internally derived scenery (Mellet et al., 1998; Miyashita, 2005). Generally, brain imaging research supports the idea that imagining something in the "mind's eye" activates maps in early perceptual networks (Damasio, 1989; Hasegawa et al., 1998; Ohbayashi et al., 1999).

The sensory networks of our AI system should similarly construct topographic maps (retinotopic for vision, tonotopic for sound, etc.). There are already reliable methods for using neural networks to generate such “self-organizing” maps (Hameed et al., 2019), and imagery generation by inverse neural networks is common today (Byeon et al., 2018). By creating a series of internally generated maps to match the iterative updating taking place in association areas, this system could produce iterated sequences of mental images. During this process, the topographic maps may use low-order perceptual knowledge (from prior probability) to depict associative relationships between higher-order items held in persistent activity. In so doing, the mental imagery that is formed may introduce valuable new informational content into working memory (such as features or objects incidental to the image itself). In other words, thinking and reasoning can be informed by logical information contained in visual and acoustic imagery.

As a given set of items in the FoA is updated, the set of unimodal, lower-order sensory maps held in synchrony with it would be updated correspondingly (Reser, 2011, 2012, 2013). In other words, after a mental image is formulated, it will likely be replaced by another image that uses many of the same working memory items as constraining parameters. Consecutive maps formed in this way could infuse video-like continuity into the imagery and could amount to a type of synthetic imagination. This form of hierarchical crosstalk between association and sensory areas, marked by mutual interactions (i.e., reciprocal causation), may allow an AI system to use mental imagery to see, hear, and thereby model hypothetical situations. This has been termed “progressive imagery modification” (Reser, 2016) and is depicted in Figure 49.

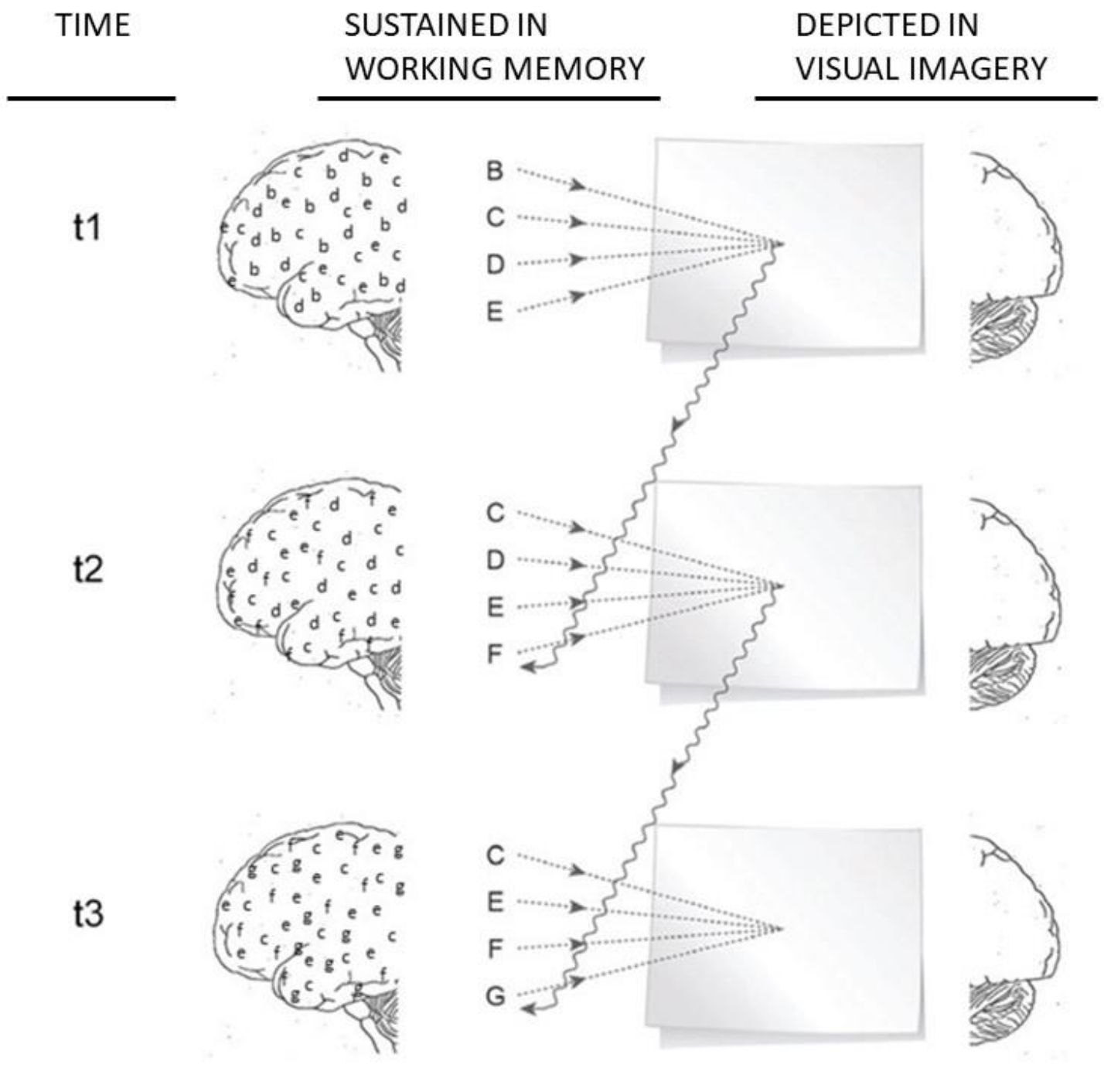

**Fig. 49.** Progressive Imagery Modification

*At time 1, items B, C, D, and E are active in association networks. The spreading activation from these items provides independent yet interactive top-down bias signals to primary visual networks where a composite*

*topographic map is built based on prior experience with these items. This gestalt sketch will introduce compatible content to working memory. Hence, at time 2, salient features created by the map from time 1 spread activation energy up the hierarchy, converging on the assemblies for item F. Item B becomes inactive while items C, D, E, and F diverge back down the hierarchy toward the primary visual network. The process repeats itself. Because of iterative updating, this process can create a logically connected series of related images.*

Brain researchers believe that sensory areas deliver information in the form of fleeting sensory maps, whereas association areas deliver lasting perceptual expectations in the form of templates and that these interact to construct higher-order cognitive processes (Hawkins, 2004; Carpenter & Grossberg, 2003). Progressive imagery modification (Fig. 49) could play an instrumental role in this reciprocal signaling between early, bottom-up sensory networks (where activity is metric, topographic, and transient) and top-down association networks (where activity is abstract, conceptual, and persistent) (Christophel et al., 2017). It could even enable an AI system to develop the kind of interplay between the central executive and the visuospatial sketchpad characteristic of the Baddeley (2000, 2007) model of working memory (Fig. 4). Further, this process of iterative modification could take place in other modules, such as language areas (where it is involved in the construction of speech), motor areas (where it is involved in action sequencing), and prefrontal areas (where it is involved in planning).

This general design could form the basis of a security precaution promoting AI safety and alignment. To human observers, the representation of knowledge in neural networks is distributed in such a complex manner that it is mostly inscrutable (Castelvecchi, 2016). This lack of transparency heightens the fear of superintelligent AI because it would be impossible to tell whether the AI was secretly harboring hostile motives (Bostrom, 2014). However, if the system was inherently obligated to build a composite topographic map of each state of working memory to initiate and inform the next state, then these maps could be displayed on a monitor for humans (or another AI) to view. A history of all visual and auditory maps could be saved to an external memory drive. This would ensure that all the AI system's mental imagery and inner speech are recorded for later inspection and interpretation. Hostile intentions would not have to be deciphered; they would be plain to see.

### 5.4 How to Train an AI that Employs Iterative Updating

The architecture described in the last three sections would not be limited to learning from discrete batches of curated input but could be exposed to continuous data streams from real-world scenarios that unfold through time. Also, the system would not suspend its activity every time it finishes a task. Rather, it would exhibit continuous, endogenous processing. The system's ontological and epistemological development could benefit from embodied, real-time, robotic interactions within physical, social, and intellectual training conditions. During exposure to these conditions, it would engage in unsupervised learning of time-series patterns from unlabeled data on a constant basis.

This would necessitate a compatible reward function to guide learning, reinforcement, and credit assignment. Said function should be based on the circuitry of the mammalian dopamine system, including the ventral tegmentum and nucleus accumbens. In mammals, novel appetitive or aversive events increase dopamine release. This ensures they are driven by predictors of food, sex, and pain. Representations of rewarding, punishing, salient, uncertain, or unpredicted events elicit dopaminergic activity in all vertebrates. This increased concentration of ambient dopamine leads to increases in sustained firing (Seamans & Yang, 2004). Thus, dopamine neuromodulation drives an animal's priorities, allowing them to prolong information about unique opportunities and threats (Seamans & Robbins, 2010).

An analog of the dopamine system's network would be needed to remember and recognize appetitively stimulating combinations of items occurring in working memory and prioritize these by sustaining their activity when they reoccur (Fig. 21). This would allow groupings with constructive incentive value to bias processing for extended periods. If we want a superintelligent AI that can further human understanding, then we should design its appetitive system to be driven to mine information from literature and databases and use it to generate new associations between ensembles. Thus, this system should latch on to unprocessed frontiers in its knowledge space (sets of unreconciled items with incentive appeal), amounting to an algorithmic form of curiosity.

Functional, preset pathways (akin to an infant's instinct to grasp something when its palm is touched) should be built into the direct connections between sensory (input) and motor (output) areas. This innate programming could come in the form of already-trained neural network modules that perform practical cognitive tasks (e.g., scene classification, paragraph comprehension, or natural language generation) embedded into the bottom of the hierarchy of this much more extensive network. For instance, a large language model could be used as a simulacrum of Broca's area (the human language area) and used to inform semantic development in the network at large. Such modules could orient the system toward effective performance on basic tasks, just as reflexes and prepared learning set developing animals on a track toward reproductive success. The machine would use operant feedback about its performance on these tasks to bootstrap learning.

Maturation of the AI's neural network should approximate that observed in the human cortex. It should start by simulating the brain of an infant (Fuster, 2015). Initially, motor output should not be driven by higher-order association areas but rather by low-order sensory and motor modules. As low-level responses are practiced and refined, and pertinent algorithms are developed through trial and error (see Section 4.4), association networks could be slowly interposed between sensory and motor networks. As in the mammalian brain (Huttenlocher & Dabholkar, 1997), sensory and motor areas should mature (myelinate) early in development, and association areas should mature late. Similarly, the capacity for persistent activity should start low but increase over developmental time.

Postponing the initialization of sustained firing would allow the formation of low-order associations between causally linked events that typically occur close together in time. This would focus the system on easy-to-predict aspects of its reality (e.g., correlations between occurrences in close temporal proximity). The consequent learning would erect a reliable scaffolding of highly predictable associations that could be used to substantiate higher-order, time-delayed associations later in development (Reser, 2016). In other words, the proportion of updating from one state to the next (Fig. 19) would start very high. This would be reversed over weeks to years as an increasing capacity for persistent working memory activity would be folded into the system.

A working memory store that uses iterative updating would be used to establish associations between related clusters of stimuli that appear close together in time from books, articles, lectures, speeches, videos, and experiences. Then, as the length of sustained firing is increased, temporally proximate contextual representations could be coactivated with other less proximate ones when multiassociative search deems them to be highly probabilistically related (i.e., they share a logical or analogical connection). Thus, two events that were never temporally local in the environment could be selected for joint iterative processing within the FoA (Fig. 31). This kind of reconciliation between separate (previously discrete) iterative threads could build and constantly retune a dynamic knowledge base of interconnected representations. After adequate training, the duration of persistent activity could be adjusted to outstrip that of humans, allowing the system to capture extremely long-term causal dependencies, resulting in the perception of high-order abstractions that would be imperceptible to humans.

Unlike biological brains, this system would be scalable (Fig. 50). There are straightforward ways to amplify the working memory of such a system beyond the physiological limitations of the human brain. These include:

1. Increasing the total number of nodes in LTM
2. Increasing the number of nodes capable of being coactivated in the short-term store
3. Increasing the number of items capable of being coactivated in the FoA
4. Increasing the length of time these can remain active (increasing the half-life and decreasing the rate of updating (Fig. 20)
5. Increasing the number of tightly coupled iterations (thoughts) that can occur before attention is disrupted (Fig. 22)

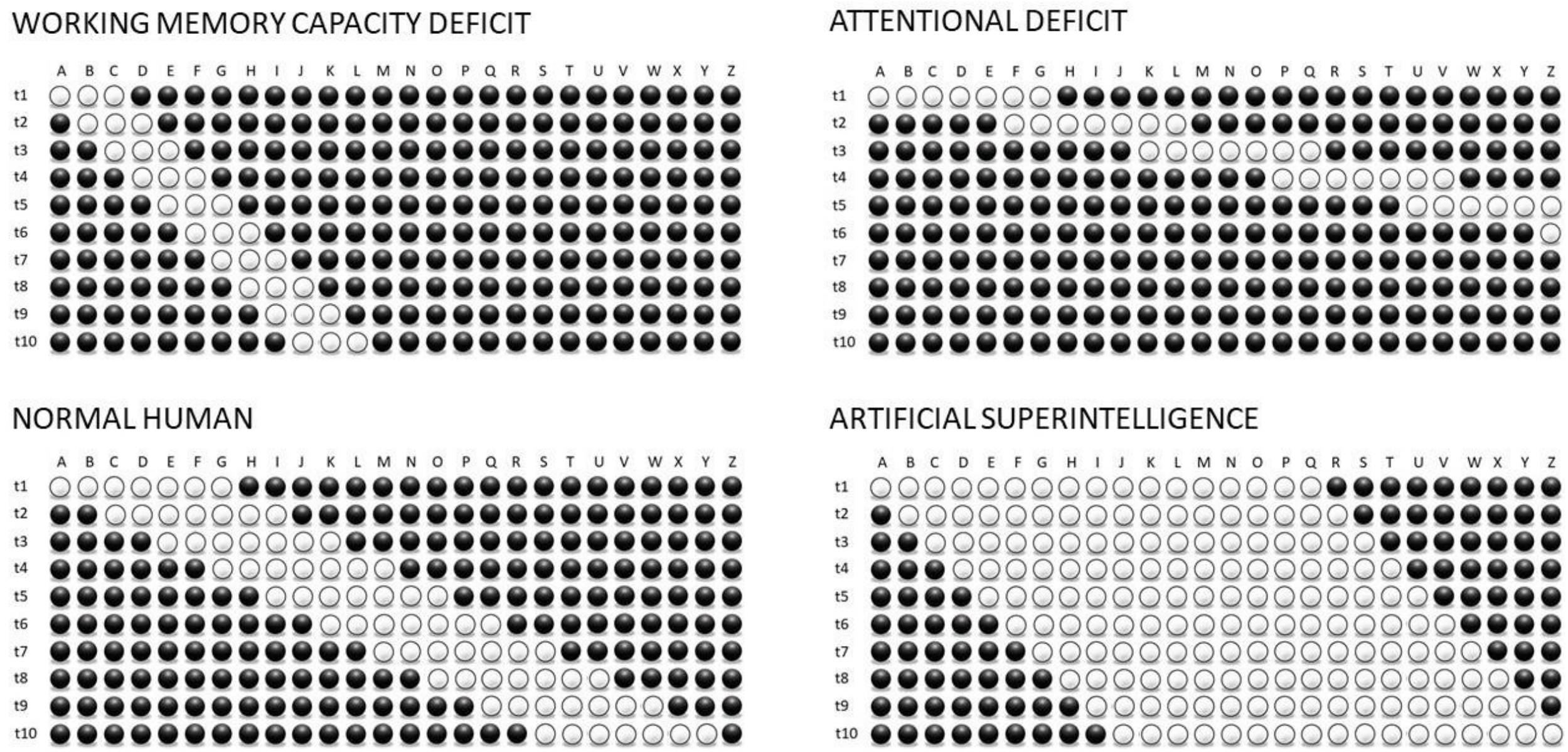

**Fig. 50.** Four Examples of Working Memory Activity Within the Focus of Attention

*The figure compares the number of items and rate of updating between a human with a limited-capacity working memory, a human with a limited attentional store, a typical human, and a superintelligent AI agent. The AI agent can maintain a larger number of nodes over a longer period, ensuring that its perceptions and actions will be informed by a larger amount of recent information.*

Under conditions of imperfect or incomplete information, the longer the backward memory span and the larger the number of related events that can be used in multiassociative search, the less uncertainty (information entropy) there is about the present state. However, in information theory, the length beyond which a backward memory span stops providing predictive information is known as the correlation length (Shannon, 1951; Stone, 2015). The working memory of a species can be seen as having a correlation length beyond which there is little predictive value to be had given its ecological niche. The long correlation length of the human FoA was likely permitted by our cognitively demanding foraging style, selection for social cognition, and the supervised learning, error feedback, and large number of training examples provided by prolonged and intensive maternal investment (Reser, 2006). However, there is no reason to believe that the length or breadth of the human FoA has been optimized for systemizing reality. It was probably constrained by several evolutionary factors that would not apply to computers.

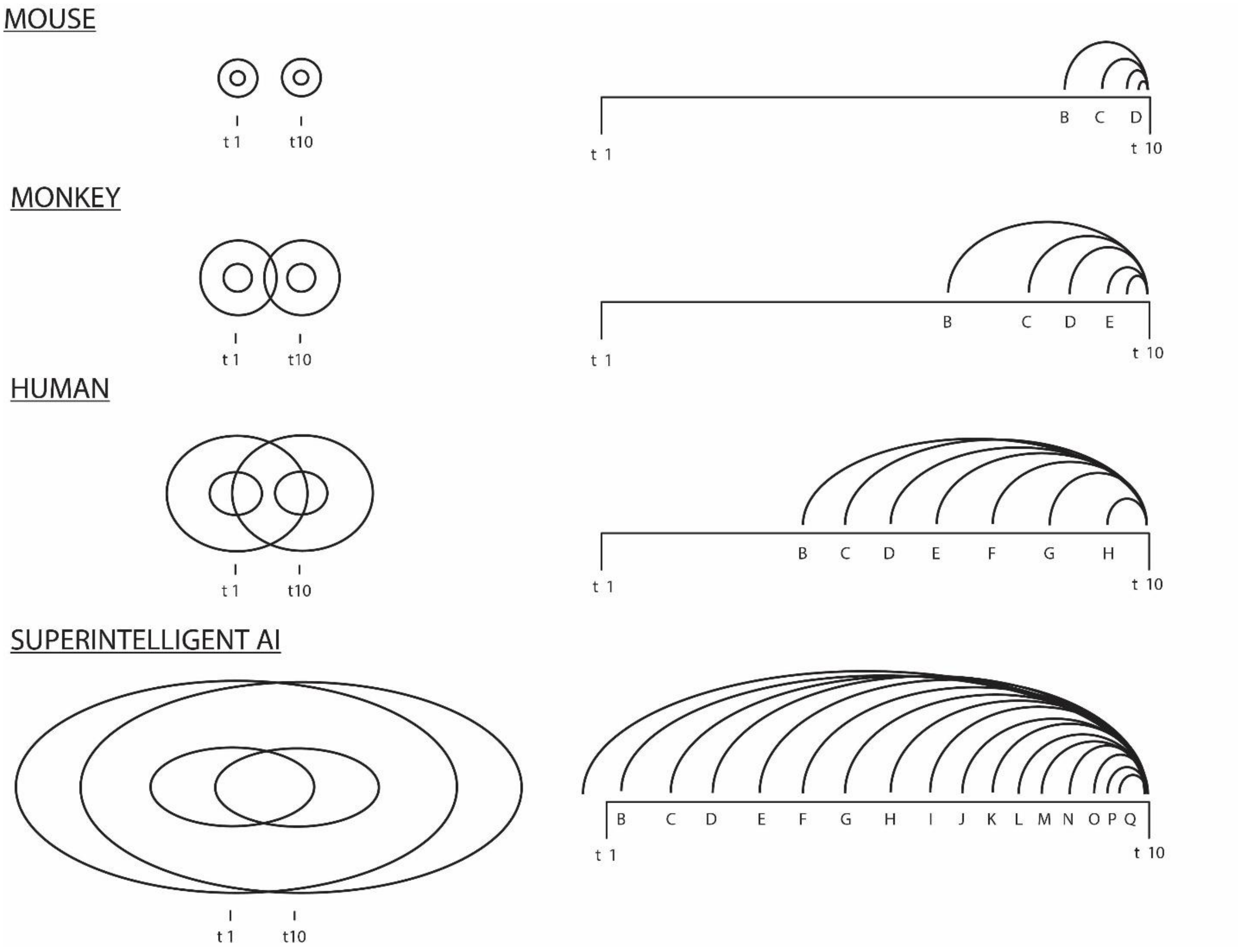

**Fig. 51**. Venn Diagrams of Working Memory in Different Systems

*These diagrams depict informational overlap between states of working memory in a span of ten seconds. The diagrams on the left use the format from Figure 13, while those on the right use the format from Figure 12. Diagram 1 shows zero overlap between working memory at times 1 and 2. This would make it more difficult for system 1, a hypothetical mouse, to make associations between events separated by the delay. For example, calling this mouse's name and feeding it 10 seconds later may not condition it to come when called, whereas feeding it two seconds later might. Training an AI should involve a maturational process where the system begins learning with a minimal working memory span (e.g., Diagram 1) before gradually developing a superhuman capacity for working memory span (Diagram 4) as formative experiences accumulate.*

Prolonging the duration of persistent activity will allow each search to be more specific and informed. This is because searches would be apprised by a larger number of specifications that stretch further back in time. It would also ensure that the system is less likely to allow crucial intermediate solutions to decay from working memory coactivity (i.e., a cache miss) before they are needed to form higher-order, compound inferences. The "thoughts" of such a system would be lengthy, highly focused, and tightly interwoven.

Now may be the time to start building large, state-of-the-art, iterative updating networks and training them as one would train a child with the expectation that aspects of intelligence will emerge. Even though they will not perform as well as leading-edge AI systems on standard benchmarks at first, they should be able to surpass them given sufficient time and training. It is hoped that through exposure to experiences with systematic patterns, a system like that described above would construct an associative network capable of producing updates to its states of working memory that build functionally on previous states. This could lead to the capacity to make valid associative connections between probabilistically related events (Fig. 31), resulting in the discovery of relationships obscured by separations in space and delays in time. Simulating iterative updating and multiassociative search and enhancing them beyond human capacities could be instrumental in the effort to construct AI capable of common sense, insight, creativity, self-directed work, long-horizon tasks, machine consciousness, and superintelligence.

### 5.5 Discussion and Conclusions

This article aimed to present an internally consistent framework for understanding how neural activity gives rise to complex thought. It is intended to inspire more detailed hypotheses, experimental tests, and machine implementations.

Previous models of working memory have attributed various high-level cognitive functions to the central executive (e.g., updating of items, coordination of modules, shifting between tasks, selective attention, gating, the construction of imagery, and others). Because the neural substrate of these advanced operations has never been delineated, the central executive remains a mysterious black box. This article has supported the case that executive functions emerge from collective processing interactions among specialized subsystems guided by iterative updating. If shown to have a tenable neural basis by future research, the concepts introduced in this article (Table 5) may amount to a viable alternative to the notion of the central executive found in other models. In so doing, these concepts may provide an organizing mechanism for self-regulating thought in AI.

| Term | Definition |
|---|---|
| **Iterative Updating** | A shift in the contents of working memory that occurs during updating as some items are added, others are removed, and still others are maintained |
| **Coactive** | A group of items that are active in the same instantaneous state |
| **Cospreading** | A group of coactive items that combine their spreading activation energy to search the same global network |
| **Multiassociative Search** | A type of search where all the coactive, cospreading contents converge in parallel on the update for the next state |
| **State-spanning Coactivity (SSC)** | Sustained coactivity exhibited by a set of two or more items that span two or more consecutive brain states |
| **Incremental Change in State-spanning Coactivity (icSSC)** | The process in which a set of items exhibiting SSC undergoes a change in group membership, where some items remain in SSC, and others are deactivated and replaced |
| **Rate of Updating** | The proportion of items updated as a function of time |
| **Mental Continuity** | The recursive interrelatedness of consecutive mental states made possible by iterative updating |
| **Iterative Compounding** | Search results from a previous state are used as an iterative update, incorporating them into the present search, thus modifying it incrementally |
| **Progressive Modification** | The logical or algorithmic progress made possible by iteration, continuity, or compounding, especially with regard to imagery updating |
| **Iterative Thread** | A chain of iteratively updated states underlying a line of thought that can be reiterated or picked up where it left off |
| **Merging of Subsolutions** | When select contents from two separate instances of iteration or separate lines of thought are coactivated in a new state and used together for multiassociative search |

**Table 5.** Definition of Terms Used and Introduced in This Article

This work reconciled iterative updating with traditional models of working memory, including those discussed in the literature review. However, it can similarly be integrated with a variety of compatible frameworks that model the dynamics of item-like constructs, including those in Table 6. These models, along with several others, provide detailed mechanistic explanations for critical neurocognitive components underspecified by the present model.

| Model / Author | Item-like Construct | Reference |
|---|---|---|
| ACT-R | symbols | Anderson (1996) |
| Adaptive Resonance Theory | templates | Grossberg (2013) |
| Global Workspace Theory | processes | Baars (2005) |
| Hierarchical Temporal Memory | time-based patterns | Hawkins et al. (2007) |
| LIDA | codelets | Baars & Franklin (2007) |
| OpenCog | atoms | Goertzel (2014) |
| Pattern Recognition Theory of Mind | pattern recognizers | Kurzweil (2013) |
| SOAR | operators | Laird (2012) |
| SPAUN | semantic pointers | Eliasmith (2013) |
| Antonio Damasio | convergent-divergent zones | Damasio (1989) |
| Gerald Edelman | neuronal groups | Edelman (2004) |
| Joaquin Fuster | cognits | Fuster (2005) |
| Douglas Hofstadter | simmballs | Hofstadter (2007) |
| Marvin Minsky | agents | Minksy (1986) |

**Table 6.** Other Models and Frameworks That Could Be Integrated with The Present Model

Working memory items in the FoA have been considered to be isomorphic with the contents of consciousness (Baars & Franklin, 2003; Buchsbaum, 2013). This suggests that the subjects of conscious thought are held in working memory and operate according to the same (or similar) rules and capacity limitations. In the classic paradigm for working memory testing, subjects can retain approximately four items in mind. However, they are holding much additional declarative content. This is because they also maintain the task requirements, active sensory perceptions, and ongoing personal thoughts (which may be limited by cognitive load). The iterative updating function applies to all this conscious content, not just to the four items described by Cowan (2017) and others. The previous figures in this article have used a small number of items and only two working memory stores (the FOA and short-term memory). Figure 52 uses an arbitrarily larger number of functionally specialized stores as an alternative to indicate that numerous items may exist along a graded continuum of activation.

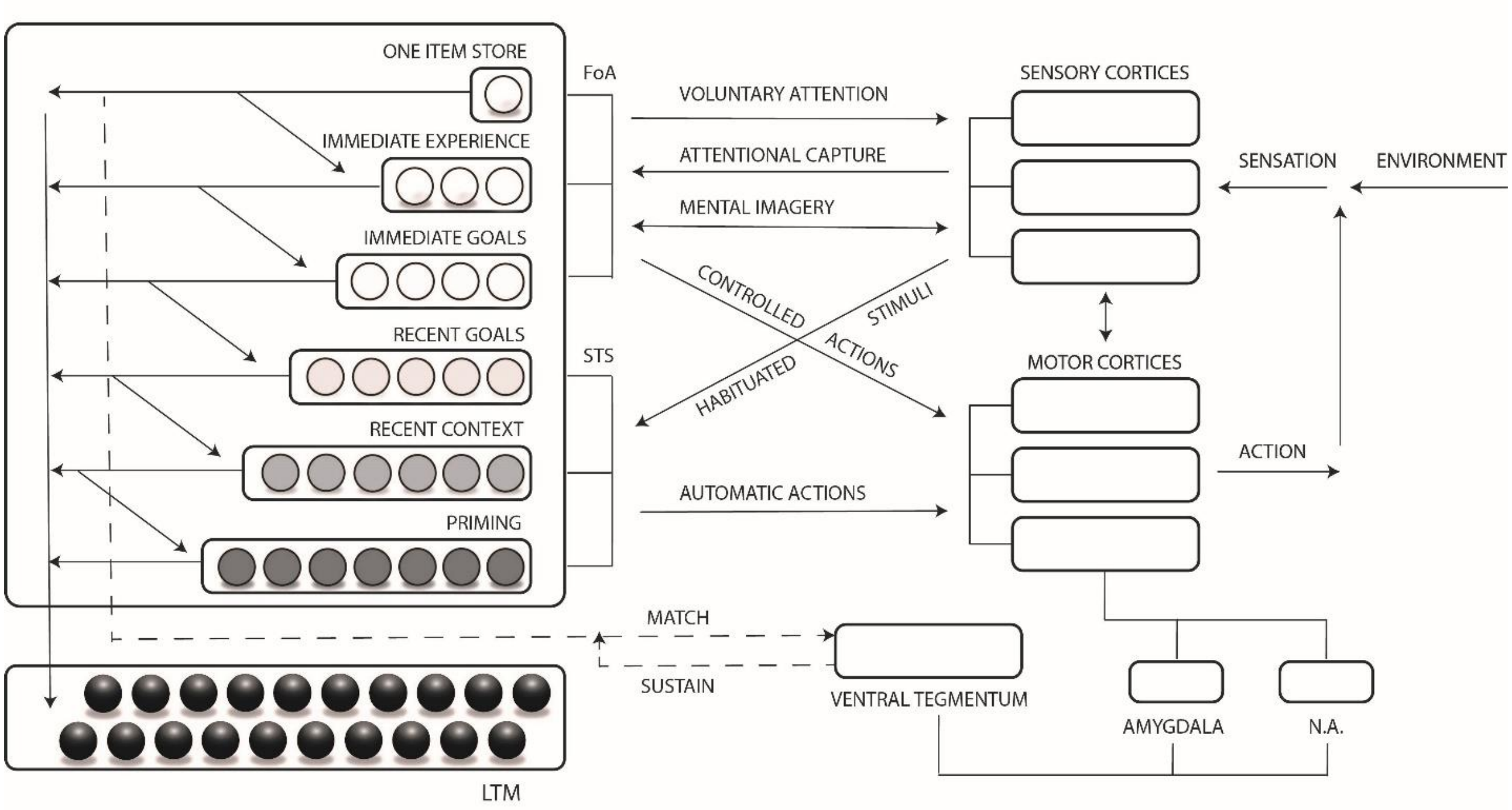

**Fig. 52.** Numerous Items Orchestrate Imagery and Behavior in the Iterative Updating Model

*Iteratively updated items in working memory interact with sensory cortices to progressively construct mental imagery. They also interact with motor cortices to progressively construct behavior. In the next state, the items in each working memory store will undergo partial replacement. The parameters used in the sensory and motor cortices will reflect this change, making their output an advancement on their previous output. This mirroring of each update will permit them to construct progressive imagery and behavior. Related cognitive processes are included as arrows. The dopamine system (ventral tegmentum) uses inputs from the amygdala and nucleus accumbens (N.A.) to determine which patterns of items match internal incentive templates and thus should be sustained.*

“Higher-order” theories of consciousness hold that conscious thought arises when a mental state is concerned with a previous mental state. This includes thoughts about perceptions and thoughts about thoughts (Rosenthal, 2004). Following this line of reasoning, thoughts that iterate from previous thoughts exhibit a backward-referential quality and could be considered "higher-order thoughts." Because of its role in generating a continual production line of higher-order thoughts, iterative updating should be considered a candidate for the neural basis of consciousness. It ensures that the train of thought does not stop and go in discrete steps but is instead propelled continuously by the items that endure through time. While individual items may exit the FoA within seconds, the content shared across successive states keeps the proverbial train on track and sustains associative connections that interlink the advancing sequence of thoughts.

| | **Features of the Present Model** |
|---|---|
| **1** | Iterative updating takes place within a cerebral cortex-like hierarchy of pattern-recognizing neural assemblies that encode subsymbolic fragments of long-term memory. |
| **2** | An item (or neural ensemble) in working memory corresponds to a persistently active group of neural assemblies that have been associated through experience. |
| **3** | Items first enter the focus of attention (FoA), which is associated with sustained firing. From there, they move toward the unattended short-term store, which is associated with synaptic potentiation. Lastly, they subside into inert long-term memory. |
| **4** | Items remain active in working memory as long as their neural assemblies demonstrate persistent activity. The activity of individual items is staggered and overlapping. Thus, the set of coactive items changes incrementally. |
| **5** | Active items in the FoA and the short-term store serve as search parameters for the next additions to working memory by spreading activation energy throughout the cortex. |
| **6** | Newly activated items are added to the set of remaining items from the previous state, completing the previous state's pattern and forming an updated set of search parameters for the next state. |
| **7** | This iterative updating process ensures that the next search is not an entirely new search but a modified version (updated iteration) of the previous search. |
| **8** | Iterative updating may play a fundamental role in event concatenation, progressive modification, learning and implementation of learned algorithms, mental modeling, inductive inference, rational thought, mental continuity, and consciousness. |

**Table 7.** Fundamental Features of the Iterative Updating Model

In his book *The River of Consciousness* (2017), Oliver Sacks asks, "But how then do our frames, our momentary moments, hold together? How, if there is only transience, do we achieve continuity?" This article postulates that our moments overlap in their set of active representations and that this ongoing confluence results in a flowing progression of states. After asking the question, Sacks quotes William James. Each thought, in James's words, is an owner of the thoughts that went before and "dies owned, transmitting whatever it realized as itself to its own later proprietor." James expounds further on this subject employing the analogy of a stream:

> "Consciousness, then, does not appear to itself chopped up in bits. Such words as 'chain' or 'train' do not describe it fitly as it presents itself in the first instance. It is nothing jointed; it flows. A 'river' or a 'stream' are the metaphors by which it is most naturally described. In talking of it hereafter let us call it the stream of thought, of consciousness, or of subjective life. [...] As the brain-changes are continuous, so do all these consciousnesses melt into each other like dissolving views. Properly they are but one protracted consciousness, one unbroken stream."

William James (1890, p. 239)

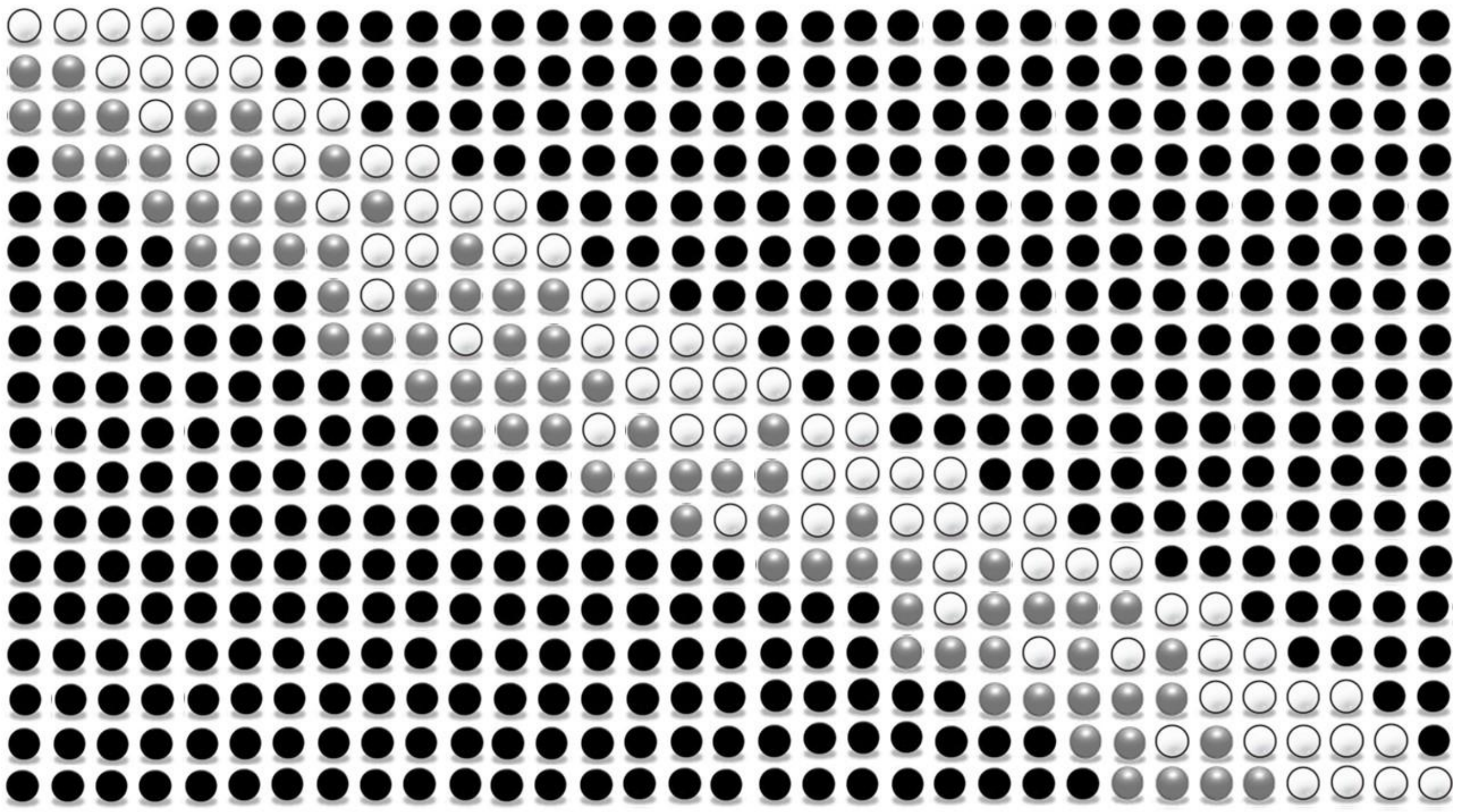

**Fig. 53.** Schematic Representation of Ongoing Iteration in the FoA and Short-term Memory Store

*This graphic expands on previous figures, incorporating a larger number of the present model's theoretical features. These include the following: (1) the number of items coactive in the FoA (white spheres) at any point in time varies between three and five, (2) the percentage of updating in the FoA varies between 25% and 100%, (3) the order of entry into the FoA does not determine the order of exit, (4) items that exit the FoA briefly enter the short-term store (gray spheres) before deactivating completely (black spheres), and (5) items that have exited the FoA are capable of reentering the FoA.*

The present model bears a resemblance to James's conception of a "stream of consciousness." A stream is a distribution of points that slides through space and time. Figure 53 extends the activity schematized in this article's other figures over 18 points in time. This results in a depiction of brain activity, working memory, and thought that, shifting gradually, appears very much like a stream.

The iterative updating of working memory sustains and shapes the flow of thought. Each new update is an influx of information that acts like a tributary, merging and interacting with the larger current of consciousness, contributing to its course. Simulating this stream within a computer could play an integral role in enabling an artificial agent to experience a cognitive continuum, traverse the expanse of consciousness, and navigate the state-space of thought.

## References

Anderson, J. R. (1983). A spreading activation theory of memory. Journal of Verbal Learning and Verbal Behavior, 22(3), 261-295.

Asok, A., Leroy, F., Rayman, J. B., & Kandel, E. R. (2019). Molecular mechanisms of the memory trace. Trends in Neurosciences, 42(1), 14-22.

Atkinson, R. C., & Shiffrin, R. M. (1968). Human memory: A proposed system and its control processes. In Spence, K. W., & Spence, J. T., The psychology of learning and motivation (Volume 2), pp. 89-195. New York: Academic Press.

Atkinson, R. C., & Shiffrin , R. M. (1969). Storage and retrieval processes in long-term memory. Psychological Review, 76(2), 179-193.

Averell, L., & Heathcote, A. (2011). The form of the forgetting curve and the fate of memories. Journal of Mathematical Psychology, 55(1), 25-25.

Baars, B. J., & Franklin, S. (2003). How conscious experience and working memory interact. Trends in Cognitive Sciences, 7, 166-172.

Baars, B. J. (2007). A framework. Baars, B. J., & Gage, N. M. (Eds.), Cognition, brain, and consciousness: Introduction to cognitive neuroscience, p. 30. London: Academic Press.

Baars, B. J., & Franklin, S. (2007). An architectural model of conscious and unconscious brain functions: Global Workspace Theory and IDA. Neural Networks, 20(9), 955-961.

Baddeley, A. (1986). Working memory. Oxford, UK: Clarendon Press.

Baddeley, A. D. (2000). The episodic buffer: A new component of working memory? Trends in Cognitive Science, 4, 417-423.

Baddeley, A. D., Hitch, G. J., & Allen, R. J. (201). From short-term store to multicomponent working memory: The role of the modal model. Memory & Cognition, 45, 575-588.

Baddeley, A. D. (2007). Working memory, thought and action. Oxford University Press.

Baddeley, A. D., & Hitch, G. J. (1994). Developments in the concept of working memory. Neuropsychology, 8(4), 485-493.

Baddeley, A. D., & Hitch, G. J. (1974). Working memory. In Bower, G. A. (Ed.), Recent advances in learning and motivation, Vol. 8, pp. 47-89. New York: Academic Press.

Baddeley, A. D. (2000). The episodic buffer: A new component of working memory? Trends in Cognitive Sciences, 4(11), 417-423.

Baddeley, A. D. (2012). Working memory: Theories, models and controversies. Annual Review of Psychology, 63, 1-29.

Bargh, J. A., & Chartrand, T. L. (2000). Studying the mind in the middle: A practical guide to priming and automaticity research. In Reis, H., & Judd, C. (Eds.), Handbook of research methods in social psychology, pp. 1-39. New York: Cambridge University Press.

Baronett, S. (2008). Logic. Upper Saddle River, NJ: Pearson Prentice Hall.

Bostrom N. (2014). Superintelligence: Paths, dangers, strategies. Oxford University Press.

Botvinick, M. M. (2008). Hierarchical models of behavior and prefrontal function. Trends in Cognitive Sciences, 12(5), 201-208.

Botvinick, M. M., & Plaut, D. C. (2006). Short-term memory for serial order: A recurrent neural network model. Psychological Review, 113(2), 201-233.

Braver, T. S., & Cohen, J. D. (2000). On the control of control: The role of dopamine in regulating prefrontal function and working memory. In Monsell, S., & Driver, J. (Eds.), Attention and performance XVIII: Control of cognitive processes, pp. 713-737. Cambridge, MA: The MIT Press.

Broadbent, D. (1958). Perception and communication. London: Pergamon Press.

Brydges, C., Gignac, G. E., & Ecker, U. K. H. (2018). Working memory capacity, short-term memory capacity, and the continued influence effect: A latent-variable analysis. Intelligence, 69, 177-122.

Buchsbaum, B. R. (2013). The role of consciousness in the phonological loop: Hidden in plain sight. Frontiers in Psychology, 4, 496.

Butlin, P., et al. (2023). Consciousness in artificial intelligence: Insights from the science of consciousness. arXiv:2308.08708

Byeon, W., Wang, Q., Srivastava, R. K., & Koumoutsakos, P. (2018). ContextVP: Fully context-aware video prediction. The European Conference on Computer Vision (ECCV), pp. 753-769.

Carpenter, G. A., & Grossberg, S. (2003). Adaptive resonance theory. In Arbib, M. A. (Ed.), The handbook of brain theory and neural networks, Second Edition, pp. 87-90. Cambridge, MA: The MIT Press.

Castelvecchi, D. (2016). Can we open the blackbox of AI: Artificial intelligence is everywhere. But before scientists trust it, they first need to understand how machines learn. Nature, 538, 7623.

Chein, J. M., & Fiez, J. A. (2010). Evaluating models of working memory through the effects of concurrent irrelevant information. Journal of Experimental Psychology: General, 139, 117-137.

Cheng, P. C., & Holyoak, K. J. (2008). Pragmatic reasoning schemas. In Adler, J. E., & Rips, L. J. (Eds.), Reasoning: Studies of human inference and its foundations, pp. 827-842. Cambridge University Press.

Chia, W. J., Hamid, A. I. A., & Abdullah, J. M. (2018). Working memory from the psychological and neurosciences perspectives: A review. Frontiers in Psychology, 27.

Christophel, T. B., Klink, P. C., Spitzer, B., Roelfsema, P. R., & Haynes, J. (2017). The distributed nature of working memory. Trends in Cognitive Sciences, 21(2), 111-124.

Cohen, G. (2000). Hierarchical models in cognition: Do they have psychological reality? European Journal of Cognitive Psychology, 12(1), 1-36.

Collins, A. M., & Loftus, E. F. (1975). A spreading-activation theory of semantic processing. Psychological Review, 82(6), 407-428.

Comer, D. (2017). Essentials of computer architecture. New York: Chapman and Hall.

Constantinidis, C., Funahashi, S., Lee, D., Murray, J. D., Qi, X., Wang, M., & Arnsten, A. F. T. (2018). Persistent spiking activity underlies working memory. Journal of Neuroscience, 38(32), 7020-7028.

Cowan, N. (1984). On short and long auditory stores. Psychological Bulletin, 96(2), 341-370.

Cowan, N. (2001). The magical number 4 in short-term memory: A reconsideration of mental storage capacity. Behavioral and Brain Sciences, 24, 87-185.

Cowan, N. (2005). Working memory capacity. New York: Psychology Press.

Cowan, N. (1988). Evolving conceptions of memory storage, selective attention, and their mutual constraints within the human information-processing system. Psychological Bulletin, 104(2), 163-191.

Cowan, N. (2009). What are the differences between long-term, short-term, and working memory? Progress in Brain Research, 169, 323-338.

Cowan, N. (2011). The focus of attention as observed in visual working memory tasks: Making sense of competing claims. Neuropsychologia, 49, 1401-1406.

Cowan, N. (2017). The many face of working memory and short-term storage. Psychonomic Bulletin & Review, 24(4), 1158-1170.

Crick, F., & Koch, C. (2003). A framework for consciousness. Nature Neuroscience, 6(2), 119-126.

D'Esposito, M., & Postle, B. R. (2015). The cognitive neuroscience of working memory. Annual Review of Psychology, 66, 115-142.

Damasio, A. R. (1989). Time-locked multiregional retroactivation: A systems level proposal for the neural substrates of recall and recognition. Cognition, 33, 25-62.

Debanne, D., Inglebert, Y., & Russier, M. (2019). Plasticity of intrinsic neuronal excitability. Current Opinion in Neurobiology, 54, 73-82.

Dehaene, S. (2020). How we learn: Why brains learn better than any machine… for now. New York: Penguin Random House.

Ecker, U. K., Oberauer, K., & Lewandowsky, S. (2014). Working memory updating involves item-specific removal. Journal of Memory and Language, 74, 1-15.

Edelman, G. (2004). Wider than the sky. Yale University Press.

Eriksson, J., Vogel, E. K., Lansner, A., Bergstrom, F., Nyberg, L. (2015). Neurocognitive architecture of working memory. Neuron, 88(1), 33-46.

Fuji, H., Ito, H., Aihara, K., Ichinose, N., & Tsukada, M. (1998). Dynamical Cell Assembly Hypothesis – Theoretical possibility of spatio-temporal coding in the cortex. Neural Networks, 9(8),1303-1350.

Funahashi, S., Bruce, C. J., Goldman-Rakic, P. S. (1993). Dorsolateral prefrontal lesions and oculomotor delayed-response performance: evidence for mnemonic 'scotomas'. Journal of Neuroscience, 13(4), 1479-1497.

Funahashi, S. (2007). The general-purpose working memory system and functions of the dorsolateral prefrontal cortex. In Osaka, N., Logie, R. H., & D'Esposito, M. (Eds.), The cognitive neuroscience of working memory, pp. 213-230. Oxford University Press.

Fuster, J. M. (2002a). Frontal lobe and cognitive development. Journal of Neurocytology, 31(3-5), 373-385.

Fuster, J. M. (2002b). Physiology of executive functions: The perception-action cycle. In Stuss, D. T., & Knight, R. T. (Eds.), Principles of frontal lobe function, pp. 96-108. Oxford University Press.

Fuster, J. M. (1973). Unit activity in prefrontal cortex during delayed-response performance: Neuronal correlates of transient memory. Journal of Neurophysiology, 36(1), 61-78.

Fuster, J. M. (2009). Cortex and Memory: Emergence of a new paradigm. Journal of Cognitive Neuroscience, 21(11), 2047-2072.

Fuster, J. (2015). The prefrontal cortex (Fifth Edition). Oxford, UK: Academic Press, Elsevier.

Glushchenko, A., et al. (2018). Unsupervised language learning in OpenCog. In: Iklé, M., Franz, A., Rzepka, R., & Goertzel B. (Eds.), Artificial general intelligence. AGI 2018. Lecture Notes in Computer Science, vol. 10999. Springer, Cham.

Goertzel, B. (2016). The AGI revolution. Humanity Press.

Goertzel, B., Pennachin, C., & Geisweiller, N. (2014). Engineering general intelligence: A path to advanced AGI via embodied learning and cognitive synergy. Atlantis Press.

Goldman-Rakic, P. S. (1987). Circuitry of the prefrontal cortex and the regulation of behavior by representational memory. In Mountcastle, V. B., Plum, F., & Geiger, S. R. (Eds.), Handbook of neurobiology, pp. 373-417. Bethesda, MD: American Physiological Society.

Goldman-Rakic, P. S. (1990). Cellular and circuit basis of working memory in prefrontal cortex of nonhuman primates. In Uylings, H. B. M., Eden, C. G. V., DeBruin, J. P. C., Corner, M. A., & Feenstra, M. G. P. (Eds.), Progress in brain research, vol. 85, pp. 325-336. Elsevier Science Publications.

Goldman-Rakic, P. S. (1995). Cellular basis of working memory. Neuron, 14(3), 447-485.

Goodfellow, I., Bengio, Y., & Courville, A. (2016). Deep learning. Cambridge, MA: The MIT Press.

Gurney, K. N. (2009). Reverse engineering the vertebrate brain: Methodological principles for a biologically grounded programme of cognitive modeling. Cognitive Computation, 1(1), 29-41.

Gray, W. D. (2007). Integrated models of cognitive systems. Oxford University Press.

Haikonen, P. O. (2003). The cognitive approach to conscious machines. Exeter, UK: Imprint Academic.

Haikonen, P. O. (2012). Consciousness and robot sentience. Hackensack, NJ: World Scientific Publishing.

Hameed, A. A., Karlik, B., Salman, M. S., & Eleyan, G. (2019). Robust adaptive learning approach to self-organizing maps. Knowledge-Based Systems, 171(1), 25-36.

Hamilton, W. (1890). In Mansel, H. L., & and Veitch, J. (Eds.), 1860 lectures on metaphysics and logic, in Two Volumes. Vol. II. Logic. Boston: Gould and Lincoln.

Hawkins, J. (2004). On intelligence. New York: Times Books.

Hasegawa, I., Fukushima, T., Ihara, T., & Miyashita, Y. (1998). Callosal window between prefronal cortices: Cognitive interaction to retrieve long-term memory. Science, 281, 814-818.

Hassabis, D., Kumaran, D., Summerfield, C., & Botvinick, M. (2017). Neuroscience-inspired artificial intelligence. Neuron, 95, 245-258.

Hebb, D. (1949). The organization of behavior. New York: Wiley.

Hofstadter, D. (2007). I am a strange loop. New York: Basic Books.

Howard, M. W., & Kahana, M. J. (2002). A distributed representation of temporal context. Journal of Mathematical Psychology, 46, 269-299.

Hummel, J. E., & Holyoak, K. J. (2003). A symbolic-connectionist theory of relational inference and generalization. Psychological Review, 110(2), 220-264.

Huttenlocher, P. R., & Dabholkar, A. S. (1997). Developmental anatomy of prefrontal cortex. In Krasnegor, N. A., Lyon, G. R., Goldman-Rakic, & P. S. (Eds.), Development of the prefrontal cortex. Baltimore, MD: Paul H. Brookes Publishing Co.

Jacob, S. N., Hähnke, D., & Nieder, A. (2018). Structuring of abstract working memory content by fronto-parietal synchrony in primate cortex. Neuron, 99(3), 588-597.

James, W. (1909). A pluralistic universe. Hibbert lectures at Manchester College on the present situation in philosophy. London: Longmans, Green, and Co.

James, W. (1890). The principles of psychology. New York: Henry Holt.

Johnson-Laird, P. N. (1998). Computer and the mind: An introduction to cognitive science. Harvard University Press.

Kaas, J. H. (1997). Topographic maps are fundamental to sensory processing. Brain Research Bulletin, 44(2), 107-112.

Kahneman, D. (2011). Thinking fast and slow. New York: Farrar, Straus, and Giroux.

Klimesch, W., Freunberger, R., & Sauseng, P. (2010). Oscillatory mechanisms of process binding in memory. Neuroscience and Biobehavioral Reviews, 34(7), 1002-1014.

Konar A. (2014). Artificial intelligence and soft computing: Behavior and cognitive modeling of the human brain. Boca Raton, FL: CRC Press.

Kounatidou, P., Richter, M., & Schöner, G.. (2018). A neural dynamic architecture that autonomously builds mental models. In Rogers, T. T. Rau, M., Zhu, X., & Kalish, C. W. (Eds.), Proceedings of the 40th Annual Conference of the Cognitive Science Society, pp. 643-648.

Kurzweil, R. (2012). How to create a mind. New York: Penguin Group.

Laird, J. E. (2012). The soar cognitive architecture. Cambridge, MA: The MIT Press.

Lansner, A. (2009). Associative memory models: From the cell-assembly theory to biophysically detailed cortex simulations. Trends in Neurosciences, 32(3),179-186.

LaRocque, J. J., Lewis-Peacock, J. A., & Postle, B. R. (2014). Multiple neural states of representation in short-term memory? It’s a matter of attention. Frontiers in Human Neuroscience, 8, 1-14.

Lewis-Peacock, J. A., Drysdale, A. T., Oberauer, K., & Postle, B. R. (2012). Neural evidence for a distinction between short-term memory and the focus of attention. Journal of Cognitive Neuroscience, 24(1), 61-79.

Manohar, S. G., Zokaei, N., Fallon, S. J., Vogels, T. P., & Husain, M. (2019). Neural mechanisms of attending to items in working memory. Neuroscience and Biobehavioral Reviews, 101, 1-12.

Mellet, E., Petit, L., Mazoyer, B., Denis, M., & Tzourio, N. (1998). Reopening the mental imagery debate: Lessons from functional anatomy. Nueroimage, 8(2),129-139.

Meyer, K., Damasio, A. (2009). Convergence and divergence in a neural architecture for recognition and memory. Trends in Neurosciences, 32(7), 376-382.

Myers, N. E., Stokes, M. G., & Nobre, A. C. (2017). Prioritizing information during working memory: Beyond sustained internal attention. Trends in Cognitive Sciences, 21(6), 449-461.

Miller, G. A. (1956). The magical number seven, plus or minus two: Some limits on our capacity for processing information. Psychological Review, 63(2), 81-97.

Miller, E. K., & Cohen, J. D. (2001). An integrative theory of prefrontal cortex function. Annual Review of Neuroscience, 24, 167-202.

Miller, E. K., Lundqvist, M., & Bastos, A. M. (2018). Working memory 2.0. Neuron, 100, 463-475.

Mongillo, G., Barak, O., & Tsodyks, M. (2008). Synaptic theory of working memory. Science, 319, 1543-1546.

Moscovich, M. (1992). Memory and working-with-memory: A component process model based on modules and central systems. Journal of Cognitive Neuroscience, 4(3),257-267.

Moscovitch, M., Chein, J. M., Talmi, D., & Cohn, M. (2007). Learning and memory. In Baars, B. J., & Gage, N. M. (Eds.), Cognition, brain, and consciousness: Introduction to cognitive neuroscience, p. 234. London: Academic Press.

Miyashita, Y. (2005). Cognitive memory: cellular and network machineries and their top-down control. Science, 306, 435-440.

Nairne, J. S. (2002). Remembering over the short-term: The case against the standard model. Annual Review of Psychology, 53, 53-81.

Newell, A., & Simon, H. A. Computer simulation of human thinking. Science, 134, 2011-2017.

Niklaus, M., Singmann, H., & Oberauer, K. (2019). Two distinct mechanisms of selection in working memory: Additive last-item and retro-cue benefits. Cognition, 183, 282-302.

Norman, D. A. (1968). Toward a theory of memory and attention. Psychological Review, 75(6), 522-536.

Nyberg, L., Eriksson, J. (2016). Working memory: maintenance, updating, and the realization of intentions. Cold Spring Harbor Perspectives in Biology, 8(2), a021816.

Oberauer, K. (2002). Access to information in working memory: Exploring the focus of attention. Journal of Experimental Psychology: Learning, Memory, and Cognition, 28(3), 411-421.

Opitz B. (2010). Neural binding mechanisms in learning and memory. Neuroscience and Biobehavioral Reviews, 34(7), 1036-1046.

Panichello, M. F., & Buschman, T. J. (2021). Shared mechanisms underlie the control of working memory and attention. Nature, 592, 601-605.

Pina, J. E., Bodner, M., & Ermentrout, B. (2018). Oscillations in working memory and neural binding: a mechanism for multiple memories and their interactions. PLOS Computational Biology, 14(11), e1006517.

Postle, B. R. (2007). Activated long-term memory? The bases of representation in working memory. In Osaka, N., Logie, R. H., & D'Esposito, M. (Eds.), The cognitive neuroscience of working memory. Oxford University Press.

Postle, B., et al. (2006). Repetitive transcranial magnetic stimulation dissociates working memory manipulation from retention functions in the prefrontal, but not posterior parietal, cortex. Journal of Cognitive Neuroscience, 18, 1712-1722.

Reggia, J. A., Katz, G. E., & Davis, G. P. (2019). Modeling working memory to identify computational correlates of consciousness. Open Philosophy, 2, 252-269.

Reser, J. E. (2006). Evolutionary neuropathology & congenital mental retardation: Environmental cues predictive of maternal deprivation influence the fetus to minimize cerebral metabolism in order to express bioenergetic thrift. Medical Hypotheses, 67(3), 529-544.

Reser, J. E. (2011). What determines belief: The philosophy, psychology and neuroscience of belief formation and change. Saarbrucken, Germany: Verlag Dr. Muller.

Reser, J. E. (2012). Assessing the psychological correlates of belief strength: Contributing factors and role in behavior. (Doctoral Dissertation). Retrieved from University of Southern California. Usctheses-m2627.

Reser, J. E. (2022). Artificial intelligence software structured to simulate human working memory, mental imagery, and mental continuity. arXiv:2204.05138

Reser, J. E. (2013). The neurological process responsible for mental continuity: Reciprocating transformations between a working memory updating function and an imagery generation system. Association for the Scientific Study of Consciousness Conference. San Diego CA, July 12-15.

Reser, J. E. (2016). Incremental change in the set of coactive cortical assemblies enables mental continuity. Physiology and Behavior, 167(1), 222-237.

Reisberg, D. (2010). Cognition: Exploring the science of the mind. New York: W. W. Norton & Co.

Rose, N. S., LaFocque, J. J., Riggall, A. C., Gosseries, O., Starrett, M. J., & Meyering, E. E. (2016). Reactivation of latent working memories with transcranial magnetic stimulation. Science, 354(6316), 1136-1139.

Rosenthal, D. M. (2004). Varieties of higher-order theory. In Gennaro, R. (Ed.), Higher-order theories of consciousness, pp. 17-44. Amsterdam: John Benjamins.

Ruchkin, D. S., Grafman, J., Cameron, K., & Berndt, R. S. (2003). Working memory retention systems: A state of activated long-term memory. Behavioral and Brain Sciences, 26, 709-777.

Rushworth, M. F., Nixon, P. D., Eacott, M. J., & Passingham, R. E. (1997). Ventral prefrontal cortex is not essential for working memory. Journal of Neuroscience, 17(12), 4829-4838.

Ryan, K., Agrawal, P., & Franklin, S. (2019). The pattern theory of self in artificial general intelligence: A theoretical framework for modeling self in biologically inspired cognitive architectures. Cognitive Systems Research. In press.

Rypma, B., Berger, J. S., & D'Esposito, M. (2002). The influence of working-memory demand and subject performance on prefrontal cortical activity. Journal of Cognitive Science, 14(5), 721-731.

Sacks, O. (2017). The river of consciousness. New York: Vintage Books.

Salmon, M. (2012). Arguments from analogy. Introduction to logic and critical thinking, pp. 132-142. Cengage Learning.

Sarter, M., Givens, B., & Bruno, J. P. (2001). The cognitive neuroscience of sustained attention: where top-down meets bottom-up. Brain Research Reviews, 35(2), 146-160.

Schvaneveldt, R. W., & Meyer, D.E. (1973). Retrieval and comparison processes in semantic memory. In Kornblum, S., Attention and performance: IV, pp. 395-409. New York: Academic Press.

Shanks, D. (2010). Learning: From association to cognition. Annual Review of Psychology, 1, 273-301.

Shannon, C. (1951). Prediction and entropy of printed English. Bell System Technical Journal, 30, 47-51.

Shastri, L. (1999). Advances in Shruti—A neurally motivated model of relational knowledge representation and rapid inference using temporal synchrony. Applied Intelligence, 11(1), 79-108.

Sherstinsky A. (2020). Fundamentals of recurrent neural network (rnn) and long short-term memory (lstm) network. Physica D: Nonlinear Phenomena, 404, 132306.

Shipstead, Z., Harrison, T. L., & Engle, R. W. (2015). Working memory capacity and the scope and control of attention. Attention, Perception, & Psychophysics, 77(6), 1863-1880.

Seamans, J. K., & Robbins, T. W. (2010). Dopamine modulation of the prefrontal cortex and cognitive function. The Dopamine Receptors, 373-398.

Seamans, J. K., & Yang, C. R. (2004). The principal features and mechanisms of dopamine modulation in the prefrontal cortex. Progress in Neurobiology, 74(1), 1-58.

Silvanto, J. (2017). Working memory maintenance: Sustained firing or synaptic mechanisms? Trends in Cognitive Sciences, 21(3), 152-154.

Sipper, M., Fu, W., Ahuja, K., & Moore, J. H. (2018). Investigating the parameter space of evolutionary algorithms. BioData Mining, 11(2).

Sreenivasan, K. K., & D’Esposito, M. (2019). The what, where and how of delay activity. Nature Reviews Neuroscience. May 13.

Sousa, A. M. M., Meyer, K. A., Santpere, G., Gulden, F. O., & Sestan, N. (2017). Evolution of the human nervous system function, structure, and development. Cell, 170(2), 226-247.

Sperling, G. (1960). The information available in brief visual representations. Psychological Monographs, 74, 1-29.

Stokes, M. G. (2015). ‘Activity-silent’ working memory in prefrontal cortex: A dynamic coding framework. Trends in Cognitive Science, 19(7), 395-405.

Stone, J. V. (2015). Information theory: A tutorial introduction. Sebtel Press.

Striedter, G. (2005). Principles of brain evolution. Sunderland, MA: Sinauer Associates.

Tomita, H., Ohbayashi, M., Nakahara, K., Hasegawa, I., & Miyashita, Y. (1999). Top-down signalfrom prefrontal cortex in executive control of memory retrieval. Nature, 401, 699-703.

Tononi, G. (2010). An information integration theory of consciousness. BMC Neuroscience, 5, 42.

Treisman, A. M. (1964). Selective attention in man. British Medical Bulletin, 20, 12 16.

von der Malsburg, C. (1999). The what and why of binding: The modeler’s perspective. Neuron, 24, 95-104.

Weger, U., Wagemann, J., & Meyer, A. (2018). Introspection in psychology: Its contribution to theory and method in memory research. European Psychologist, 23, 206-216.

Zanto, T. P., Rubens, M. T., Thangavel, A., & Gazzaley, A. (2011). Causal role of the prefrontal cortex in top-down modulation of visual processing and working memory. Nature Neuroscience, 14, 656-661.